\documentclass[a4paper,usenames,dvipsnames,11pt]{article}
\pdfoutput=1

\usepackage{jheppub}
\usepackage{slashed}
\usepackage{mathrsfs,booktabs,multirow,tabularx}
\usepackage{stmaryrd}
\usepackage{xspace}
\usepackage{fancyvrb}
\usepackage[makeroom]{cancel}
\usepackage{amsmath}    
\usepackage{amssymb}    %
\usepackage{graphicx}   
\usepackage{verbatim}   
\usepackage{lscape}
\usepackage{subfig}
\usepackage{listings}

\def\beq{\begin{equation}}
\def\beqn{\begin{eqnarray}}
\def\eeq{\end{equation}}
\def\eeqn{\end{eqnarray}}
\def\abs#1{\left|#1\right|}

\def\pdistr#1#2{\left(\frac{1}{#1}\right)_{\!#2}}

\def\EWeq#1{eq.~({\bf I}.#1)}

\def\PDF#1#2{\Gamma_{\!#1/#2}}

\newcommand\sss{\scriptscriptstyle}

\newcommand\mydot{\!\cdot\!}
\newcommand\ep{\epsilon}
\newcommand\half{\frac{1}{2}}

\newcommand\aem{\alpha}
\newcommand\aemotpi{\frac{\aem}{2\pi}}

\newcommand{\bt}{\bar{t}}

\newcommand{\bk}{\bar{k}}

\newcommand{\mpmm}{\ell^+\ell^-}
\newcommand{\lp}{\ell^+}
\newcommand{\lm}{\ell^-}

\newcommand{\lep}{\ell}

\newcommand{\pb}{\bar{p}}

\newcommand{\ord}{{\cal O}}

\newcommand\hsig{\hat{\sigma}}
\newcommand\bsig{\bar{\sigma}}

\newcommand\vet{\vec{e}}
\newcommand\amp{{\cal A}}

\newcommand\ampsq{{\cal M}}

\newcommand\bampsq{\overline{\cal M}}

\newcommand\MSb{\overline{\rm MS}}
\newcommand\xib{\bar{\xi}}
\newcommand\yb{\bar{y}}

\newcommand\mtos{\frac{m^2}{s}}

\newcommand\stepf{\Theta}
\newcommand\mZ{m_{\sss Z}}

\newcommand{\muz}{\mu_0}

\newcommand{\minf}{M_{\rm\sss INF}}
\newcommand{\hminf}{\hat{M}_{\rm\sss INF}}
\newcommand{\pt}{p_{\sss T}}

\newcommand{\deltaI}{\delta_{\sss\rm I}}

\newcommand{\bdeltaI}{\bar{\delta}_{\sss\rm I}}

\newcommand{\GeV}{{\rm GeV}}

\newcommand{\LOG}{{\rm LO}_\Gamma}
\newcommand{\NLOG}{{\rm NLO}_\Gamma}
\newcommand{\NNLOG}{{\rm NNLO}_\Gamma}
\newcommand{\dNLOG}{\delta{\rm NLO}_\Gamma}
\newcommand{\dNNLOG}{\delta{\rm NNLO}_\Gamma}
\newcommand{\dNLO}{\delta{\rm NLO}}
\newcommand{\dNNLO}{\delta{\rm NNLO}}

\title{Double neutral-current corrections to NLO electroweak leptonic 
cross sections}

\author[a]{Stefano Frixione,}
\author[b,c,d]{Fabio Maltoni,}
\author[d]{Davide Pagani,}
\author[e,f]{Marco Zaro}

\affiliation[a]{INFN, Sezione di Genova, Via Dodecaneso 33, Genoa, 
I-16146 Italy}

\affiliation[b]{
Centre for Cosmology, Particle Physics and Phenomenology (CP3),
Universit\'e Catholique de\\ Louvain, 
Louvain-la-Neuve, 
B-1348 Belgium
}

\affiliation[c]{
Dipartimento di Fisica e Astronomia, Universit\`a di Bologna, 
Via Irnerio 46,\\
Bologna,
I-40126 Italy
 } 

\affiliation[d]{
INFN, Sezione di Bologna,
Via Irnerio 46, 
Bologna,
I-40126 Italy
 }

\affiliation[e]{
INFN, Sezione di Milano,
Via Celoria 16, 
Milano,
I-20133 Italy
 }

\affiliation[f]{
TIFLab, Universit\`a degli Studi di Milano,
Via Celoria 16, 
Milano, 
I-20133 Italy
 }

\emailAdd{Stefano.Frixione@cern.ch}
\emailAdd{fabio.maltoni@uclouvain.be}
\emailAdd{davide.pagani@bo.infn.it}
\emailAdd{marco.zaro@unimi.it}

\abstract{
We present a method for improving next-to-leading order electroweak (EW)
predictions for lepton-scattering processes by consistently including
double neutral-current corrections arising from vector-boson-fusion
topologies, which are formally of higher order. By combining, in a
process-independent manner, exact fixed-order results, collinear resummation
of QED radiation, and a subtraction procedure, we obtain results which are
gauge invariant and valid in the entire phase space, retain any dependence 
on the masses of electroweak bosons, and can be systematically improved, while
avoiding the need for complete next-to-next-to-leading order calculations.
This paper is devoted to the development and validation of the formalism;
phenomenological applications are presented in a companion study, where we
also discuss and motivate why our approach is superior to the one based on
EW parton distribution functions for targeting percent-level precision at
multi-TeV lepton colliders.
}


\preprint{
\begin{flushright}
TIF-UNIMI-2025-11\\
IRMP-CP3-25-22\\
COMETA-2025-20\\
\today
\end{flushright}
}

\begin{document}
\maketitle
\flushbottom

\section{Introduction\label{sec:intro}}
Lepton colliders have always been regarded as precision machines,
where the discovery of new phenomena may happen only through a
very careful comparison between measurements and the corresponding
theoretical predictions. This viewpoint is informed by the fact that,
historically, all lepton machines ever built have been essentially
low-energy colliders, whose main characteristic is that the overwhelming 
majority of the collisions are initiated by the annihilation of the two
incoming leptons, and produce low-multiplicity final states.

In view of the fact that the high-energy physics community is presently
considering the post-LHC strategy, it is useful to regard the behaviour
of lepton colliders that we have just mentioned as an almost accidental
by-product of two facts, in the context of a description that can be
applied to both hadron and lepton collisions. Such a description is
given by the factorisation theorem\footnote{For a recent review on this
and its alternatives which focuses on $e^+e^-$ physics, see 
e.g.~\cite{Frixione:2022ofv}.}, which represents each incoming 
particle as a set of collinear partons, which predominantly
collide in pairs in an incoherent manner, and whose longitudinal
momentum fraction ($z$) distributions are given by pre-computed 
(or measured) functions, the Parton Distribution Function (PDFs). The two 
facts we have alluded to before are: {\em a)} the set of partons within 
a lepton is dominated by a lepton of the same flavour as that of the 
incoming one, with its PDF almost equal to a $\delta(1-z)$ (i.e.~that
parton carries away almost all of the momentum of the incoming particle);
{\em b)}~lepton collisions are initiated by QED and weak interactions; hence,
they tend to favour low-multiplicity final states. In hadronic collisions,
the almost polar opposite happens: PDFs are peaked at small-$z$ values;
and QCD drives the hard reactions, thus rendering it easy to find
high-multiplicity final states.

As was mentioned before, this dichotomy between lepton and hadron colliders
is an accident due to the low-energy nature of the former. In fact, by 
increasing the c.m.~energy, a lepton collider gradually acquires the
characteristics typical of a hadron collider. In particular,
one starts to see that the peaks of the transverse momenta tend to
move from the high-end of the spectrum towards small values; likewise
for the invariant masses of the systems of objects tagged in the final state.
Technically, this happens because large energies mean the possibility of
probing parton contents at smaller $z$ values, where one is not
dominated any longer by lepton PDFs, but rather by those of the photon
and, increasingly, of the gluon and of the quarks. Furthermore, phase 
spaces associated with larger multiplicities become sizeable, hence in part
offsetting the suppression due to the QED/weak coupling constant factors.

The increase of the collider energy therefore results in a richer set of
phenomena, and in new discovery strategies\footnote{It is important to
note that, in large part, the problems and opportunities associated
with an energy increase are the same as those one is confronted with
when the goal is to reach extremely high precision. In other words,
the discussions about a 10~TeV muon colliders with a percent-level program 
will touch many of the issues relevant to a 500~GeV electron machine which 
aims at measuring a few observables with a relative precision of, say, 
$10^{-5}$ -- an example of particular relevance for this paper is
the impact of the $\gamma\gamma$-fusion channel.}. 
In particular, in the context of muon-collider studies, the class of
VBF processes has received a significant amount of attention, in view
of its becoming one of the dominant production mechanisms at high 
c.m.~energies; this fits nicely with the fact that, theoretically,
the vector bosons exchanged in the $t$ channels can effectively be treated 
as partons within the incoming leptons, i.e.~as {\em massless} objects
associated with a PDF. While one can prove that all of the contributions
neglected by this approximation vanish with the c.m.~energy that goes
to infinity, the speed at which one approaches such a limit is
logarithmic; therefore, for realistic collider configurations, with 
energies in the range of a few TeVs or even tens of TeVs, the neglected
contributions may still be important, and even dominant. When this is the 
case, one sees that VBF processes are not even well-defined quantities, 
but merely denote a subset of the Feynman diagrams that contribute to the
production of the system of interest (e.g.~a Higgs, a $W^+W^-$ pair,
and so forth). This implies that, in order to carry out studies which
are guaranteed to be unbiased theoretically, one must have predictions 
that are correct in the whole phase space, thus being able to describe both 
the regions where VBF kinematic configurations are important, and those where 
they are negligible, with a seamless transition between the two; an important
component of this is the ability to control the theoretical uncertainties.

The approach where EW vector bosons are regarded as massless 
partons within the incoming lepton does not fulfill this
requirement. The solution, in itself or as a starting point
for further refinements, is the perturbative computation of the cross
section of interest that features all of the relevant contributions,
including but not limited to VBF topologies. The problem with such a
solution is that calculations increase sharply in complexity with the 
perturbative order. While this is directly relevant to VBF topologies, 
since these are suppressed by the square of the coupling constant 
w.r.t.~Born contributions\footnote{Barring accidental zeros at the Born 
level; however, when such zeros are present, the cross section remains small 
in  absolute value, still owing to the suppression by the coupling constant.}
(i.e.~are of next-to-next-to-leading order (NNLO)), it is fortunately the
case that VBF graphs are simple to compute -- they are tree-level
ones, with a straightforward singularity structure.

This observation is the motivation behind this work, whose goal is
the inclusion of terms of NNLO, and specifically those which stem from
VBF topologies, in perturbative predictions whose accuracy is of 
next-to-leading order (NLO), without performing complete NNLO calculations, 
but in a gauge-invariant manner, and by retaining the exact dependences on 
vector boson masses and the capability of firmly assessing the 
theoretical systematics.

This paper is organised as follows. We introduce the problem and
discuss some general issues in sect.~\ref{sec:gen}. We present the
core part of the procedure in sect.~\ref{sec:VBF}. The resulting
formula cannot be used as is in order to obtain physical predictions, and we 
expose its limitations and scope in sect.~\ref{sec:meaning}. We then proceed
to lift such limitations in sect.~\ref{sec:tech}, and arrive at the
final result in sect.~\ref{sec:eq35}. We conclude in sect.~\ref{sec:concl},
and collect additional material in the appendices.

This is a technical paper, whose goal is achieved by means of a sequence of 
formal manipulations. The reader who is interested only in the underpinning 
logic and in the final result can find a summary of the former in
sect.~\ref{sec:gen}, and the latter in eqs.~(\ref{xsNNLOimprep4}),
(\ref{dsig2Gdef2}), and~(\ref{final00NNLO2rep}).
Some immediate phenomenological consequences of our work are presented
in sect.~\ref{sec:rescuts}, sect.~\ref{sec:resfact}, and more extensively
in a companion paper~\cite{Frixione:2025xxx}.

\section{Generalities\label{sec:gen}}
We are interested in studying the properties
of a system \mbox{$T=\{T_1,\ldots T_m\}$} of $m$ particles produced
in $\mpmm$ collisions, with $\ell=e$ or $\ell=\mu$. We shall loosely 
identify the $T_i$'s as ``tagged'' objects, regardless of whether they 
can actually be directly tagged in a detector (this happens e.g.~when 
\mbox{$T=\{e^+,e^-\}$} or \mbox{$T=\{\gamma,\gamma,\gamma,\gamma\}$}), or 
only indirectly, for example through their decay products (this is the 
case e.g.~when \mbox{$T=\{t,\bt\}$} or \mbox{$T=\{W^+,W^-\}$}). Indeed,
from a theoretical viewpoint the direct or indirect nature of the
tagging is irrelevant; what matters is the ability to calculate,
in a manner which we assume to be perturbative, the cross sections
associated with the processes:
\beq
\lm+\lp\;\;\longrightarrow\;\;T\cup U_q\,,\;\;\;\;\;\;
T\cup U_q\,\equiv\,T_1+\ldots T_m+u_1+\ldots u_q\,,
\label{proc0}
\eeq
for any set \mbox{$U_q=\{u_1,\ldots u_q\}$} of $q$ particles that may
be taggable or not, but that we generally regard as {\em not} tagged
in the context of our experiment: thus, we call them ``untagged''. 
Roughly speaking, the cross section associated with eq.~(\ref{proc0}) 
will be proportional, in a fixed-order perturbative approach, to a 
coupling-constant factor\footnote{It is always possible to use
eq.~(\ref{coupls}) when working at the tree level, while it becomes
ambiguous or outright impossible when virtual corrections are involved.
Here, we employ that equation {\em only} in order to help one
understand the arguments of this section in an intuitive manner.
Furthermore, we ignore the contributions due to strongly-interacting
particles in either the initial or the final state; that said, our
formalism will require no modifications should one wish to include them.}
\beq
\aem^b\aem^q\,,
\label{coupls}
\eeq
having denoted by $\aem$ the coupling constant of the theory we are
working with (i.e.~typically the EW sector of the SM), and by $b$ an
integer characteristic of the production of the sole system $T$.

\vskip -0.5truecm
\begin{figure}[htb]
  \begin{center}
  \includegraphics[scale=1.0,width=0.99\textwidth]{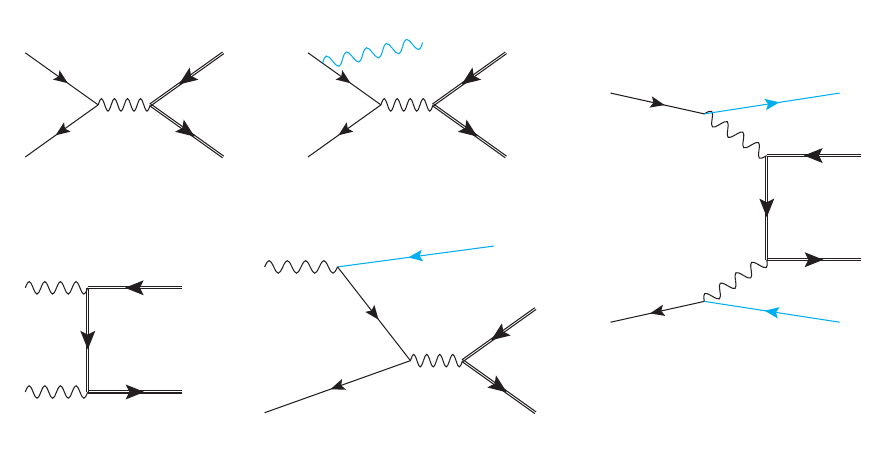}
\caption{\label{fig:ttdiag1}
Representative tree-level $t\bt$-production diagrams at the level of
short-distance cross sections, of order $\aem^2\aem^0$ (left panel,
$\mpmm$ and $\gamma\gamma$ channels), $\aem^2\aem^1$ (middle panel,
$\mpmm$ and $\gamma\ell$ channels), and $\aem^2\aem^2$ (right panel,
$\mpmm$ channel only). 
}
  \end{center}
\end{figure}

\vskip -0.15truecm
\begin{figure}[htb]
  \begin{center}
  \includegraphics[scale=1.0,width=0.9\textwidth]{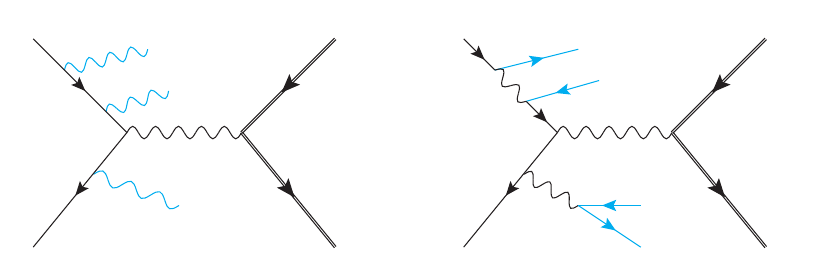}
\caption{\label{fig:ttdiag2}
Representative tree-level $t\bt$-production diagrams of order $\aem^2\aem^3$
(left panel) and $\aem^2\aem^4$ (right panel). 
}
  \end{center}
\end{figure}
In order to be definite, let us consider the case of $t\bt$ production. Sample 
perturbative diagrams for the production of this tagged system are depicted
in figs.~\ref{fig:ttdiag1} and~\ref{fig:ttdiag2}. There, we have represented
as thick black lines the $t$ and the $\bt$ quarks, and as blue lines the
untagged particles. In reference to eq.~(\ref{coupls}), we have $b=2$,
whilst the relevant value of $q$ is obtained by counting the number
of final-state untagged particles. The wiggly lines represent 
electroweak vector bosons, including $W$'s in the case of the rightmost 
panel of fig.~\ref{fig:ttdiag1} (when this happens, the $t$-channel
fermion line actually represents a $b$ quark, whereas the two outgoing
fermion lines represent neutrinos).

The calculation of the matrix elements that correspond to the graphs
above increase in complexity with the overall power of $\aem$; there is
both a technical complexity stemming from each graph, and a bookkeeping
complexity due to the growth of the number of diagrams. The
problem becomes rapidly so severe that one must resort to some kind of
approximation -- for example, the diagrams of fig.~\ref{fig:ttdiag2}
are typically replaced by the diagrams in the leftmost panel of 
fig.~\ref{fig:ttdiag1}, whose matrix elements are then convoluted with
the so-called Parton Distribution Functions (PDFs), which account for
any number of {\em collinear} emissions of any types of particles (whence
the necessity of including contributions where the initial state is
{\em not} an $\mpmm$ one\footnote{While the identities of the
initial-state particles at the short-distance level may be different from 
those of the beam particles, we always regard the mechanism which produces
them as perturbative, as opposed to multi-particle phenomena such as
beamstrahlung, with which we are not concerned in this work.}, as in the 
graphs at the bottom of the left and middle panels of that figure). In the 
notation introduced before, the PDFs are an effective way to sum over 
{\em all} sets $U_q$, with \mbox{$0\le q\le\infty$}, provided that all 
of the particles in such sets are emitted collinearly to the incoming lines.

An unavoidable consequence of employing an approximation based on PDFs
is that any information on the masses of the particles involved in the
corresponding collinear process is solely constituted by the arguments
of logarithms, which are equal to the ratios of such masses over a hard 
scale that characterises the process (e.g.~the invariant mass of the
tagged system $T$). Specifically, the information is lost on the behaviour
of the cross section as a function with polynomial- or square-root-type 
dependence on these masses (henceforth, we shall call this behaviour 
``power-suppressed effects''). 
This is not a particularly steep price to pay if the particles in 
question are photons or light leptons, whose masses are already null or 
negligible, and the use of PDFs is in fact helpful in keeping the logarithmic 
growth of the cross section under control. The situation is manifestly very 
different if one considers $Z$ and $W$ bosons, whose masses are hardly 
negligible but at asymptotically-high energies. By far and large, this 
strongly hints at the fact that the uncertainties due to neglecting such 
masses in simulations relevant to {\em any collider configurations ever 
considered to be realistic} are possibly very significant. In particular, 
power-suppressed effects are expected to be dominant or fairly large in a 
sizeable part of the phase space. In general, the reasons for questioning
the accuracy of PDF approaches that involve weak bosons (called EW PDFs
henceforth), such as those of refs.~\cite{Han:2020uid,Garosi:2023bvq}, 
are related to the following issues.
\begin{itemize}
\item
Results obtained by means of exact matrix element computations exhibit
large differences w.r.t.~those emerging from the so-called Effective 
Weak-boson Approximation (EWA), in the absence of observable-specific 
fine tunings of the scale parameters that enter the latter (see 
ref.~\cite{Frixione:2025xxx}). One needs to bear in mind that an
EWA approach relies on functions with can be regarded as the first-order
expansions\footnote{\label{ft:pcEWA}This condition may be relaxed, and 
thus {\em some} power-suppressed effects can be included in the EWA 
even at the LO. While this improves the agreement with matrix-element
predictions, it further degrades the control on the theoretical
uncertainties associated with the EWA.} of the corresponding EW PDFs 
(or, equivalently, as the initial conditions for their evolution).
\item
The differences between EW-PDF- and EWA-induced results are smaller than 
the (very large) uncertainties associated with the approximations that 
these methods entail. This fact, and the previous item, imply that all-order 
resummation effects, included in the EW PDFs but absent in the EWA, are 
subdominant w.r.t.~to the differences that either of these predictions 
(if not fine-tuned) have in comparison to matrix-element-based results,
which are relevant precisely because resummation is not necessary.
\item
The uncertainties inherent in the EW PDFs and EWA approaches are in part 
not parametric in nature (i.e.~they do not scale with powers of the vector 
boson masses or the coupling constants), and thus cannot be assessed in a 
fully reliable manner.
\item
Being equivalent to a massless approach, EW-PDF- and EWA-based computations 
lack any description of power-suppressed effects$^{\ref{ft:pcEWA}}$;
therefore, we expect them to be unreliable in particular at and near the 
threshold region, where any agreement with the correct result is accidental, 
except in the cases where the typical scale of the produced system is much 
larger than the EW scale.
\item
Current EW PDFs are leading-logarithmic (LL) and leading-order (LO) accurate 
(contrary to their light-particle counterparts). As a consequence of this, 
short-distance cross section predictions are beyond accuracy if they include 
contributions of next-to-leading order (NLO) or higher.
\item
It follows from the previous item that the usage of (LL and LO) EW
PDFs implies that only a subset of the diagrams relevant to a given process 
can be considered: the effects of the remaining diagrams are neglected, and 
their impact cannot be estimated.
\item
It then further follows that, since simulations with EW PDFs are 
fully inclusive in the particles that emerge from the boson branchings and
one can only use LO matrix elements, such simulations implement a much simpler
kinematic structure than that of the processes they are meant to approximate.
This may pose serious difficulties when one tries to reproduce the typical 
experimental setup; for example, certain observables are impossible to
define, trivial, or over-simplified.
\end{itemize}
For these reasons, it seems wise to develop a rigorous approach based on 
exact matrix element calculations whenever massive vector bosons are 
involved. In this context, there are two main obstacles, namely: {\em a)} the
computations are technically demanding; {\em b)} a strict matrix element
approach cannot work well when photons and light leptons are also present,
since logarithmic effects associated with such particles are not
resummed, and they must~be.

A situation where either or both of these drawbacks are relevant is
that represented by the VBF-like graph in the rightmost panel of
fig.~\ref{fig:ttdiag1}, which is indeed the case most often cited
as {\em the} reason for employing an EW-PDF approach. Preliminarily, let us 
remark that the set of such graphs constitutes, from a perturbative viewpoint,
a single class of the much larger set of all of the NNLO diagrams, 
which are suppressed by two powers of the coupling
constant w.r.t.~the leading-order (LO) contributions. Therefore, for
the VBF-like topologies to be dominant, one must be in a rather
special kinematical configuration where both the coupling-constant
suppression and the competition from other NNLO graphs are overcome.
Clearly, complete NNLO results would render both points moot, but such
results are indeed very difficult to obtain at present. One can therefore limit 
oneself to considering all double-real tree-level graphs. This bypasses
a significant fraction of the technical difficulties, but the one associated
with the fixed-order treatment of massless-particle branching remains,
for those processes that feature both $Z$ and $\gamma$ exchanges.

The formalism we shall derive in the remainder of this paper is such that:
\begin{itemize}
\item
It uses all of the exact matrix-elements of LO and NLO, as well as those
associated with all of the $\ell^+\ell^-$-initiated double-real graphs of 
NNLO, for all particles involved, be them massive, nearly-massless, 
or massless.
\item
It includes the resummation of small-mass effects by using lepton,
photon and, if necessary, quark and gluon PDFs, at the accuracy at
which such PDFs are available (presently, next-to-leading 
logarithmic~\cite{Bertone:2019hks,Frixione:2023gmf,Bonvini:2025xxx}).
\item 
It smoothly matches the two (i.e.~hard and collinear) regimes, so that no 
information is lost anywhere in the phase space, and the energy range can be 
scanned from the threshold region to the collider energy at the same level 
of accuracy.
\item
It allows one to include further improvements, such as the inclusion of
Sudakov-resummed results and of higher-order corrections for those channels
which open up at the NNLO.
\end{itemize}
The key for achieving these objectives is a careful treatment of
VBF-like diagrams, which we shall discuss in the next section.

In conclusion, a remark about notation.
In view of the fact that this point is generally
ambiguous, we specify that by NLO and NNLO we understand a prediction
at  that level of accuracy -- that is, in reference to eq.~(\ref{coupls}),
NLO corresponds the {\em sum} of the $\ord(\aem^b)$ and the $\ord(\aem^{b+1})$
contributions, while NNLO is the NLO result {\em plus} the $\ord(\aem^{b+2})$
contributions. Conversely, by $\dNLO$ and $\dNNLO$ we understand
the sole {\em corrections}, i.e.~the $\ord(\aem^{b+1})$ and $\ord(\aem^{b+2})$
contributions, respectively. In addition to that, the LO, $(\delta)$NLO, and 
$(\delta)$NNLO contributions that we specifically address in this paper are 
denoted by $\LOG$, $(\delta)$NLO$_\Gamma$, and $(\delta)$NNLO$_\Gamma$, 
respectively. In the case of the LO, $\LOG$ coincides with the results
of the $\gamma\gamma$-initiated process, and can thus be denoted by
LO$_{\gamma\gamma}$ as well.\label{page:names}

\section{VBF-like configurations with electroweak vector bosons
\label{sec:VBF}}
We shall assign the kinematics of any $2\to 2+m$ process thus including,
but being not limited to, a VBF-like one, as follows:
\beq
\lm(p_1)+\lp(p_2)\;\longrightarrow T+\lm(k_1)+\lp(k_2)\,,
\label{VBFproc}
\eeq
having used the same notation as in eq.~(\ref{proc0}), with the untagged 
particles explicitly chosen to be leptons of the same flavour as
that of the incoming ones. As was said in sect.~\ref{sec:gen},
we are specifically interested in the contributions to the process of 
eq.~(\ref{VBFproc}) due to VBF topologies that feature both $Z$ and $\gamma$
exchanges. Absent hard cuts on the outgoing leptons, there will be
a phase-space region where the cross section is dominated by kinematical 
configurations with such leptons collinear to the incoming ones. In fact, if 
the leptons are assumed to be massless, the matrix elements that feature 
photon exchanges are divergent in the collinear limit; with massive leptons, 
the matrix elements are dominated by $\log Q^2/m^2$ terms, with $Q$ the 
typical hard scale of the process and $m$ the lepton mass; in all relevant
scenarios, power-suppressed effects associated with the lepton mass
($(m^2/Q^2)^c$ for some $c>0$) are negligible. Conversely, both 
logarithmic ($\log Q^2/\mZ^2$) and power-suppressed ($(\mZ^2/Q^2)^d$
for some $d>0$, not necessarily equal to $c$) effects due to the $Z$ 
boson need to be retained.

\begin{figure}[htb]
  \begin{center}
  \includegraphics[scale=1.0,width=0.2\textwidth,angle=270]{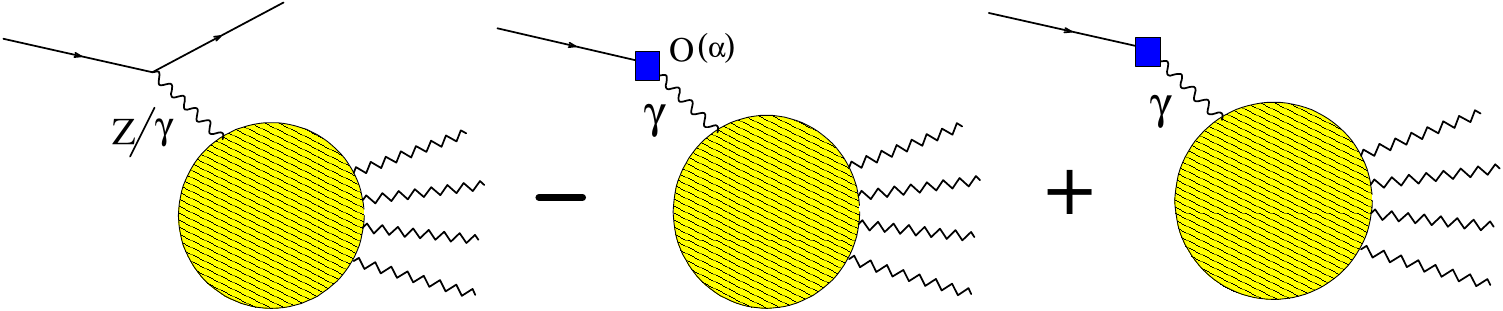}
\caption{\label{fig:hvbf} 
Upper half of a VBF configuration in $\mpmm$ collisions; for brevity, 
only one side of the Cutkosky diagrams is shown. Left panel: exact matrix 
elements. Middle panel: with $\ord(\aem)$ expansion of the $\gamma$ 
PDF. Right panel: with the full $\gamma$ PDF.
}
  \end{center}
\end{figure}
We start by pointing out that the procedure we are going to define
works also for one ``VBF vertex'' at a time; thus, we can concentrate, say,
on the upper half of the VBF diagram, and subsequently apply our results
to the bottom half. This has the advantage that the procedure can be applied
to any kind of process that features at least one $Z$-boson $t$-channel
exchange, a class larger than, but which includes, the VBF processes.
Thus, we begin by considering the process\footnote{Here and elsewhere,
the intermediate $(Z/\gamma)^*$ notation reminds one that such vector
bosons {\em may} be exchanged in the $t$ channel(s); however, the matrix 
elements always include all contributions.}
\beq
\lm(p_1)+a(p_2)\;\longrightarrow(Z/\gamma)^*\longrightarrow\;
\lm(k_1)+X
\label{VBFprocred}
\eeq
rather than that of eq.~(\ref{VBFproc}); this is symbolically
depicted in the left panel of fig.~\ref{fig:hvbf} (where we omit
particle $a$). By assuming the system $X$ to be made of $n$
particles\footnote{While from eq.~(\ref{VBFproc}) we have $a=\lp$,
$X=\{\lp(k_2),T\}$ and $n=m+1$, eq.~(\ref{VBFprocred}) 
and its subsequent manipulations do not rely on these relationships.
\label{ft:one}}, we write the cross section for the process of 
eq.~(\ref{VBFprocred}) as follows:
\beq
d\bsig_{\lep}^{(n+1)}=\bampsq_{\lep}^{(n+1)} d\phi_{n+1}\,,
\label{xsecm}
\eeq
with $\bampsq_{\lep}^{(n+1)}$ the matrix element computed with
massive leptons (hence the overline, following the notational conventions
of ref.~\cite{Frixione:2019lga}). The basic idea of the procedure that
follows is this:
\label{page:one}
\begin{itemize}
\item Add to the matrix element of eq.~(\ref{xsecm}) 
(the square of the graph in the leftmost panel of fig.~\ref{fig:hvbf})
a quantity which is equal to zero at $\ord(\aem)$, given by the difference
of the reduced cross section convoluted with the photon PDF
(the square of the graph in the rightmost panel of fig.~\ref{fig:hvbf})
minus the same thing expanded at $\ord(\aem)$, supplemented by
collinear remainders of matrix-element origin (the square of the graph 
in the middle panel of fig.~\ref{fig:hvbf}).
\end{itemize}
In order to do this, we must compute the
leading behaviour of eq.~(\ref{xsecm}) in the collinear configuration 
$p_1\!\parallel\! k_1$. From \EWeq{5.15}\footnote{Henceforth, eq.~(x.y)
of  ref.~\cite{Frixione:2019lga} will be denoted by \EWeq{x.y}.}:
\beq
\bampsq_{\lep}^{(n+1)}\;\stackrel{p_1\parallel k_1}{\longrightarrow}\;
\frac{e^2}{p_1\mydot k_1-m^2}\,P_{\gamma^\star\lep}^{<}\left(z\right)\,
\ampsq_{\gamma}^{(n)}(zp_1)\,.
\label{reMEcll}
\eeq
The massive-lepton Altarelli-Parisi kernel can be found in app.~B
of ref.~\cite{Frixione:2019lga}:
\beq
P_{\gamma^\star\lep}^{<}(z)=\frac{1+(1-z)^2}{z}-
\frac{z\,m^2}{p_1\mydot k_1-m^2}\,.
\label{Pgstq}
\eeq
The matrix element $\ampsq_{\gamma}^{(n)}(zp_1)$ is the one relevant to 
the process:
\beq
\gamma(zp_1)+a(p_2)\;\longrightarrow\;X\,,
\label{VBFprocga}
\eeq
where all lepton-mass effects are neglected (i.e., $p_1^2=0$ in
eq.~(\ref{VBFprocga}): the photon is on-shell), and eq.~(\ref{reMEcll}) 
is understood to be valid up to terms suppressed by powers of $m^2/Q^2$.
Equation~(\ref{reMEcll}) follows from the usual argument of collinear
dominance; the reduced $n$-body matrix element that factorises features
only the photon (and not the $Z$), since a collinearly-divergent
Feynman graph (which is only associated with photon exchange) must appear
on both sides of the cut to give a leading $1/k_\perp^2$ behaviour.

For the explicit computation of eq.~(\ref{reMEcll}) we parametrise
the relevant momenta as follows:
\beqn
p_1&=&(E_1,0,0,p)\,,
\;\;\;\;\;\;\;\;
E_1=\sqrt{m^2+p^2}\,,
\\
k_1&=&\frac{\sqrt{s}}{2}\,\xi\left(1,\vet\,\sqrt{1-y^2}\,\beta_k,
y\,\beta_k\right)\,,
\label{knpocll}
\eeqn
where, having defined $s=(p_1+p_2)^2$ and $m_2^2=p_2^2$:
\beqn
E_1&=&\frac{\sqrt{s}}{2}\left(1+\frac{m^2-m_2^2}{s}\right)\,,
\\
p&=&\frac{\sqrt{s}}{2}\,\sqrt{1-\frac{(m+m_2)^2}{s}}\,
\sqrt{1-\frac{(m-m_2)^2}{s}}\,,
\\
\beta_k&=&\sqrt{1-\frac{4m^2}{s\,\xi^2}}\,.
\eeqn
This parametrisation is the standard FKS~\cite{Frixione:1995ms,Frixione:1997np}
one, where one works in the 
c.m.~frame of the incoming particles. As usual, we identify $\xi=1-z$, 
$z$ being that which appears in eq.~(\ref{reMEcll}). In practice, we
are interested in the two cases $m_2=m$ and $m_2=0$, which correspond
to $a\equiv\lp$ and $a=\gamma$, respectively; we shall show that
these lead to the same result for eq.~(\ref{reMEcll}). The above 
momentum parametrisation corresponds to the following phase space (see 
\EWeq{5.21}--\EWeq{5.23}):
\beqn
&&d\phi_{n+1}(p_1,p_2;k_1,X_1\ldots X_n)=
d\tilde{\phi}_n(p_1,p_2;k_1,X_1\ldots X_n)\,d\phi_1(k_1)\,,
\label{phinpo}
\\
&&d\tilde{\phi}_n(p_1,p_2;k_1,X_1\ldots X_n)=
\nonumber\\*&&\phantom{aaa}
(2\pi)^4\delta\left(p_1+p_2-k_1-\sum_{i=1}^n X_i\right)
\prod_{i=1}^n\frac{d^3 X_i}{(2\pi)^3 2X_i^0}\,,
\phantom{aaaaaa}
\label{tphin}
\\
&&d\phi_1(k_1)=\frac{d^3 k_1}{(2\pi)^3 2k_1^0}=
\frac{1}{2(2\pi)^3}\,\frac{s\beta_k^3}{4}\,\xi d\xi\,dy\,d\varphi\,.
\label{phi1}
\eeqn
By means of direct computations we obtain what follows:
\beq
p_1\mydot k_1-m^2=\frac{s}{4}f\left(1-\frac{y}{1+\rho}\right)\,,
\eeq
where, if $m_2=m$:
\beqn
f&=&\xi-\frac{4m^2}{s}\,,
\label{fcase1}
\\
\frac{1}{1+\rho}&=&\frac{\xi\beta_k}{f}\sqrt{1-\frac{4m^2}{s}}
\;\;\;\;\Longrightarrow\;\;\;\;
\rho=2\frac{(1-\xi)^2}{\xi^2}\frac{m^2}{s}+
\ord\left(\frac{m^4}{s^2}\right),\phantom{aa}
\label{rocase1}
\eeqn
while if $m_2=0$:
\beqn
f&=&\xi\left(1+\frac{m^2}{s}\right)-\frac{4m^2}{s}\,,
\label{fcase2}
\\
\frac{1}{1+\rho}&=&\frac{\xi\beta_k}{f}\left(1-\frac{m^2}{s}\right)
\;\;\;\;\Longrightarrow\;\;\;\;
\rho=2\frac{(1-\xi)^2}{\xi^2}\frac{m^2}{s}+
\ord\left(\frac{m^4}{s^2}\right).\phantom{aa}
\label{rocase2}
\eeqn
Importantly, this shows that the leading behaviour in $m^2/s$ of the $\rho$ 
parameter is the same in the two cases of interest; ultimately, this is
due to the invariance under longitudinal boosts of massless-lepton results.
By using these formulae and the distributions identities in \EWeq{4.108}
and \EWeq{4.182}, namely:
\beqn
\frac{1}{\left(1\pm\frac{y}{1+\rho}\right)^2}&=&
\left(\frac{1}{\rho}+\frac{3}{2}\right)\delta(1\pm y)
+\left(\log\frac{\rho}{2}+1\right)\delta^\prime(1\pm y)
\nonumber\\*&+&
\sum_{j=2}^\infty \frac{(-1)^j\,2^{j-1}}{(j-1)\,j!}\,\delta^{(j)}(1-y)
+\ord(\rho)\,,\phantom{aaaa}
\label{ybe2}
\\
\frac{1}{1-\frac{y}{1+\rho}}&=&
-\log\frac{\rho}{2}\,\,\delta(1-y)
+\pdistr{1-y}{+}+\ord(\rho)\,,
\label{ybe3}
\eeqn
and by discarding terms suppressed by powers of $m^2/s$, we arrive at:
\beqn
d\bsig_{\lep}^{(n+1)}&\stackrel{p_1\parallel k_1}{\longrightarrow}&
\aemotpi\,
{\cal Q}_{\gamma\lep}(1-\xi)\ampsq_{\gamma}^{(n)}\big((1-\xi)p_1\big)\,
\delta(1-y)\,d\xi\,dy\,d\tilde{\phi}_n
\nonumber\\*&+&
\frac{4e^2}{s\xi}\,\frac{1+\xi^2}{1-\xi}\left(\frac{1}{1-y}\right)_+
\ampsq_{\gamma}^{(n)}\big((1-\xi)p_1\big)\,d\phi_{n+1}\,,
\label{xsecmlim}
\eeqn
with (see \EWeq{4.187}):
\beq
{\cal Q}_{\gamma\lep}(1-\xi)=-\left[\frac{1+\xi^2}{1-\xi}
\left(2\log\frac{1-\xi}{\xi}+\log\mtos\right)+\frac{2\xi}{1-\xi}\right]\,.
\label{Qres}
\eeq
From eqs.~(\ref{reMEcll}) and~(\ref{Pgstq}), we see that the expression 
on the second line of eq.~(\ref{xsecmlim}) is, apart from the endpoint
contribution of the plus distribution, the collinear limit of the
massless-lepton matrix elements, $\ampsq_{\lep}^{(n+1)}$. In the first line,
the exact $n$-body phase space appears, since from \EWeq{5.24}:
\beq
d\tilde{\phi}_n(p_1,p_2;k_1,X_1\ldots X_n)\,\delta(1-y)
\;\stackrel{m\to 0}{\longrightarrow}\;
d\phi_n\big((1-\xi)p_1,X_1\ldots X_n\big)\,\delta(1-y)\,,
\label{phinlim}
\eeq
At the same level of accuracy as that of eq.~(\ref{xsecmlim}), we can 
therefore write:
\beqn
d\bsig_{\lep}^{(n+1)}&\stackrel{p_1\parallel k_1}{\longrightarrow}&
\aemotpi\,
{\cal Q}_{\gamma\lep}(z)\ampsq_{\gamma}^{(n)}\big(zp_1\big)\,
d\phi_n\big(zp_1\big)\,dz
\nonumber\\*&+&
\left(\frac{1}{1-y}\right)_+
\left((1-y)\ampsq_{\lep}^{(n+1)}\right)
d\phi_{n+1}\,.
\label{xsecmlim2}
\eeqn
Equation~(\ref{xsecmlim2}) gives the collinear limit of the massive-lepton
cross section in terms of the subtracted {\em massless}-lepton one (that
features both $\gamma$ and $Z$ $t$-channel exchanges, as well as any other
tree-level diagram of the same order), plus an $n$-body quantity
that is the convolution of a universal kernel ${\cal Q}_{\gamma\lep}$
with the cross section for the $\gamma$-initiated process of
eq.~(\ref{VBFprocga}). Note that, owing to the subtraction, the
massless-lepton cross section is finite in the collinear limit:
the role of its singularity is played by the $\log m^2/s$ term which
appears in ${\cal Q}_{\gamma\lep}$, and that corresponds to the leading
behaviour of the massive-lepton cross section.

The $n$-body $\gamma$-initiated cross section is also what appears
in the expression for the process of eq.~(\ref{VBFprocred}) that
emerges from collinear factorisation. This, we write as follows:
\beqn
d\hsig_\gamma^{(n)}&=&\PDF{\gamma}{\lep}(z)\,
\ampsq_{\gamma}^{(n)}\big(zp_1\big)\,
d\phi_n\big(zp_1\big)\,dz
\label{fact0}
\\&=&
\left(\aemotpi\PDF{\gamma}{\lep}^{[1]}(z)+\ord(\aem^2)\right)
\ampsq_{\gamma}^{(n)}\big(zp_1\big)\,
d\phi_n\big(zp_1\big)\,dz
\nonumber\\*&\equiv&
d\hsig_{\gamma\ord(\aem)}^{(n)}+\ord(\aem^2)\,,
\label{fact1}
\eeqn
where eq.~(\ref{fact1}) is simply eq.~(\ref{fact0}) after expanding the 
photon PDF at $\ord(\aem)$ . Equations~(\ref{fact0}) and~(\ref{fact1}) 
are depicted in the right and middle panels, respectively, of
fig.~\ref{fig:hvbf}. $\PDF{\gamma}{\lep}^{[1]}$ can be found in
\EWeq{4.189}:
\beq
\PDF{\gamma}{\lep}^{[1]}(z)=\frac{1+(1-z)^2}{z}\left(
\log\frac{\mu^2}{m^2}-2\log z-1\right) + K_{\gamma\lep}(z)\,,
\label{Ggesol2}
\eeq
where $\mu^2\sim s$ is the mass-scale squared at which the PDF is evaluated,
and $K_{\gamma\lep}(z)$ is a function that defines the factorisation 
scheme (see refs.~\cite{Frixione:2021wzh,Bertone:2022ktl}
for more details on its definition, for both LL and NLL PDFs).
It is now apparent that the leading $m\to 0$ behaviours of
eqs.~(\ref{xsecmlim2}) and~(\ref{fact1}) are identical (as they
should by construction). This implies that:
\beqn
d\bsig_{\lep}^{(n+1)}&\!\!=\!\!&
\aemotpi\,
{\cal Q}_{\gamma\lep}(z)\ampsq_{\gamma}^{(n)}\big(zp_1\big)\,
d\phi_n\big(zp_1\big)\,dz-d\hsig_{\gamma\ord(\aem)}^{(n)}
\label{master}
\\*&+&
\left(\frac{1}{1-y}\right)_+
\left((1-y)\ampsq_{\lep}^{(n+1)}\right)
d\phi_{n+1}+d\hsig_\gamma^{(n)}+
\ord\left(\aem^2,\mtos\right)\,.\phantom{a}
\nonumber
\eeqn
The r.h.s.~of this equation is depicted symbolically in fig.~\ref{fig:hvbf}:
the sought collinear-improved $(n+1)$-body cross section which features 
$Z/\gamma$ interference is obtained as the sum of the subtracted 
$(n+1)$-body massless-lepton cross section, plus the convolution 
of the $\gamma$ PDF with the $n$-body $\gamma$-initiated one, plus the 
convolution of the latter with a collinear-finite kernel, whose expression 
is obtained from the first line of eq.~(\ref{master}), namely:
\beqn
&&\aemotpi\,
{\cal Q}_{\gamma\lep}(z)\ampsq_{\gamma}^{(n)}\big(zp_1\big)\,
d\phi_n\big(zp_1\big)\,dz-d\hsig_{\gamma\ord(\aem)}^{(n)}
\\*&&\phantom{aaaaaaaa}
=\aemotpi\,
{\cal Q}_{\gamma\lep}^\prime(z)\ampsq_{\gamma}^{(n)}\big(zp_1\big)\,
d\phi_n\big(zp_1\big)\,dz\,,
\eeqn
with:
\beq
{\cal Q}_{\gamma\lep}^\prime(z)=
\frac{1+(1-z)^2}{z}\left(
\log\frac{s}{\mu^2}+2\log(1-z)\right) +z - K_{\gamma\lep}(z)\,.
\label{Qpres}
\eeq
Thus finally the collinear-improved expression of the original
cross section is:
\beqn
\bampsq_{\lep}^{(n+1)} d\phi_{n+1}&\longrightarrow&
\left(\frac{1}{1-y}\right)_+\!
\left((1-y)\ampsq_{\lep}^{(n+1)}\right)
d\phi_{n+1}
\nonumber\\*&&+
\aemotpi\,{\cal Q}_{\gamma\lep}^\prime(z)\ampsq_{\gamma}^{(n)}\big(zp_1\big)\,
d\phi_n\big(zp_1\big)\,dz\,+\,d\hsig_\gamma^{(n)}\,,
\label{final0}
\eeqn
with $d\hsig_\gamma^{(n)}$ given in eq.~(\ref{fact0}). Equation~(\ref{final0})
achieves formally the goal stated in the bullet point at 
page~\pageref{page:one} and symbolically\footnote{Among other things,
this means that the figure stands for both sides of the Cutkosky diagram,
although only one side is shown -- one must bear in mind that the
procedure uses matrix elements, i.e.~squared amplitudes.} depicted in 
fig.~\ref{fig:hvbf}. Loosely speaking, the first and second terms of 
eq.~(\ref{final0}) correspond to the two leftmost panels of the figure, 
whereas the third term corresponds to the rightmost panel. 

Now we observe that each of the cross sections that appear in 
eq.~(\ref{final0}) may feature the same structure as the cross section 
we have started from, i.e.~that for eq.~(\ref{VBFprocred}) -- in other words, 
they may correspond to the lower half of a VBF diagram. In that case, the 
procedure above should be iterated; this explains why it has been carried 
out for the process in eq.~(\ref{VBFprocred}), rather than for the original 
one of eq.~(\ref{VBFproc}). In order to carry out such an 
iteration\footnote{The procedure henceforth is easy to understand, but
not necessarily technically rigorous. A more careful approach will
be discussed in sect.~\ref{sec:eq35}, which of course will not change
the conclusions we shall reach here.}, we start by re-writing 
eq.~(\ref{final0}) in a more explicit manner
as far as parton indices are concerned, thus:
\beqn
&&\bampsq_{\lep\lep}^{(m+2)} d\phi_{m+2}\longrightarrow
\left(\frac{1}{1-y_1}\right)_+\!
\left((1-y_1)\ampsq_{\lep\lep}^{(m+2)}\right)
d\phi_{m+2}\big(p_1,p_2\big)
\nonumber\\*&&\phantom{aaaaa}+
\left(\PDF{\gamma}{\lep}(z_1)+
\aemotpi\,{\cal Q}_{\gamma\lep}^\prime(z_1)\right)
\ampsq_{\gamma\lep}^{(m+1)}\,
d\phi_{m+1}\big(z_1p_1,p_2\big)\,dz_1\,,\phantom{aaa}
\label{final0A}
\eeqn
having renamed $z\to z_1$ and $y\to y_1$, and having employed the
relationships in footnote~\ref{ft:one}. By applying the replacement
of eq.~(\ref{final0A}) to the rightmost lepton indices that appear in
each of the matrix elements in its r.h.s., we obtain what follows:
\beqn
&&\bampsq_{\lep\lep}^{(m+2)} d\phi_{m+2}\longrightarrow
\nonumber\\*&&\phantom{aaa}
\left(\frac{1}{1-y_1}\right)_+\!
\left(\frac{1}{1-y_2}\right)_+\!
\left((1-y_1)(1-y_2)\ampsq_{\lep\lep}^{(m+2)}\right)
d\phi_{m+2}\big(p_1,p_2\big)
\nonumber\\*&&\phantom{a}+
\left(\PDF{\gamma}{\lep}(z_2)+
\aemotpi\,{\cal Q}_{\gamma\lep}^\prime(z_2)\right)
\left(\frac{1}{1-y_1}\right)_+\!
\nonumber\\*&&\phantom{aaaaaaaaa}\times
\left((1-y_1)\ampsq_{\lep\gamma}^{(m+1)}\right)
d\phi_{m+1}\big(p_1,z_2p_2\big)\,dz_2
\nonumber\\*&&\phantom{a}+
\left(\PDF{\gamma}{\lep}(z_1)+
\aemotpi\,{\cal Q}_{\gamma\lep}^\prime(z_1)\right)
\left(\frac{1}{1-y_2}\right)_+\!
\nonumber\\*&&\phantom{aaaaaaaaa}\times
\left((1-y_2)\ampsq_{\gamma\lep}^{(m+1)}\right)
d\phi_{m+1}\big(z_1p_1,p_2\big)\,dz_1\,\phantom{aaa}
\nonumber\\*&&\phantom{a}+
\left(\PDF{\gamma}{\lep}(z_1)+
\aemotpi\,{\cal Q}_{\gamma\lep}^\prime(z_1)\right)
\left(\PDF{\gamma}{\lep}(z_2)+
\aemotpi\,{\cal Q}_{\gamma\lep}^\prime(z_2)\right)
\nonumber\\*&&\phantom{aaaaaaaaa}\times
\ampsq_{\gamma\gamma}^{(m)}\,
d\phi_{m}\big(z_1p_1,z_2p_2\big)\,dz_1\,dz_2\,,\phantom{aaa}
\label{final00}
\eeqn
which is the two-VBF-side version of eq.~(\ref{final0}). As it should
be clear by construction, $\ampsq_{\lep\lep}^{(m+2)}(p_1,p_2)$ is the matrix
element associated with eq.~(\ref{VBFproc}), 
$\ampsq_{\lep\gamma}^{(m+1)}(p_1,p_2)$ and 
$\ampsq_{\gamma\lep}^{(m+1)}(p_1,p_2)$ are the matrix elements associated 
with
\beqn
\lm(p_1)+\gamma(p_2)&\longrightarrow&\lm(k_1)+T\,,
\label{mugaWW}
\\
\gamma(p_1)+\lp(p_2)&\longrightarrow&\lp(k_2)+T\,,
\label{gamuWW}
\eeqn
respectively, and $\ampsq_{\gamma\gamma}^{(m)}$ is the matrix
element associated with
\beq
\gamma(p_1)+\gamma(p_2)\;\longrightarrow\;T\,.
\label{gagaWW}
\eeq
In all of these matrix elements, the leptons are massless.

In summary, the steps that we have taken are these: starting from a 
$\mpmm\to \mpmm+m$-body process with massive leptons which features 
VBF-like configurations, by means of unitary operations
(amounting to adding and subtracting the same quantities), augmented by
the resummation of large collinear logarithms due to initial-state branchings 
of light objects, we have re-expressed its tree-level matrix element in 
terms of massless-leptons matrix elements of different multiplicities, that
are usually associated with LO, $\dNLO$, and $\dNNLO$ contributions. 

Ultimately, we want to exploit this fact for replacing the LO- and $\dNLO$-like 
terms with their exact, complete, counterparts, while keeping the $\dNNLO$-like 
term as a relative $\ord(\aem^2)$ {\em improvement} of the underlying NLO
prediction. Thus, this is not equivalent to performing a complete NNLO
calculation and, while technically easier, its work-flow is much less well
established than that of the latter; specifically, one needs to be careful 
to not double count (with either sign) some contributions.

More in detail, we shall arrive at the improved prediction we are
seeking by considering the following issues, in turn:
\label{page:two}
\begin{enumerate}
\item Our starting point is eq.~(\ref{final00}), which shows how
to express a massive-lepton matrix element in terms of massless-lepton
matrix elements, plus resummation effects -- we discuss the physics contents, 
limitations, and scope of that equation in sect.~\ref{sec:meaning}.
\item One can relate massive- and massless-lepton matrix elements
only by introducing some arbitrary quantities (among which, the 
factorisation scheme, and the way to subtract and/or cutoff singularities)
relevant to the latter. We discuss the nature of the massive vs massless 
description (specifically, the emergence of the mass logarithms by means
of the factorisation theorem) in sect.~\ref{sec:tech}, as well as a 
tailored cutoff strategy in sect.~\ref{sec:cuts}, and the 
factorisation-scheme dependence of physical predictions in
sect.~\ref{sec:scheme}.
\item We have simply stated that eq.~(\ref{final00}) follows from
eq.~(\ref{final0}), essentially by flipping fig.~\ref{fig:hvbf} w.r.t.~an
horizontal axis. In sect.~\ref{sec:sides} we provide the reader with 
more formal arguments.
\item Finite terms can be exchanged among different contributions to
a fixed-order cross section without changing its physical predictions.
Such finite terms are typically associated with the remainders of the
subtractions of the singularities emerging from the collinear and soft
emissions of massless particles. In view of the fact that we shall
replace {\em part of} eq.~(\ref{final00}) with a complete NLO calculation,
it is of crucial importance that the subtractions in the former induce the
{\em same} finite parts as in the latter. Our complete NLO calculation
will be based on the FKS formalism; therefore, we shall re-write the
$\dNLO$-like parts of eq.~(\ref{final00}) exactly as in the context of
the FKS procedure -- this is discussed in sect.~\ref{sec:eq35}.
\end{enumerate}

\section{Interpretation\label{sec:meaning}}
Since the physics contents of eqs.~(\ref{final0}) and~(\ref{final00})
are identical, it is easier to start discussing the former, which is
less complicated from a technical viewpoint.

The crucial observation is that the kernel 
${\cal Q}_{\gamma\lep}^\prime$ of eq.~(\ref{Qpres}) {\em coincides}
with the kernel relevant to the degenerate $(n+1)$-body contributions
in FKS in the case of $\lep\to\gamma\lep$ branchings (see e.g.~the formulae 
in sect.~4.3 and appendix~D of ref.~\cite{Frederix:2009yq}). Likewise, the
first term of eq.~(\ref{final0}) is the $(n+1)$-body contribution to
an FKS cross section, for those cases where the soft subtraction is 
trivial (see sect.~4.2 of ref.~\cite{Frederix:2009yq}), and when the
singularity structure is such that it does not require the use
of ${\cal S}$ functions (as is the case of eq.~(\ref{VBFprocred})).
This is important, because it shows that the one-leg procedure we have 
followed here, which deals solely with tree-level matrix elements, is 
consistent with that relevant to a fully-fledged subtracted NLO computation.
However, this does {\em not} imply that eq.~(\ref{final0}) exactly coincides
with part of the corresponding $\dNLO$ result: to mention just the most trivial 
aspect, for this to happen the first and second terms of eq.~(\ref{final0}) 
would need to be further convoluted with $\PDF{\lep}{\lep}$\footnote{One must 
bear in mind that the initial-state parton that comes from the right 
($a(p_2)$ in eq.~(\ref{VBFprocred})) has an idle role in the procedure 
that leads to eq.~(\ref{final0}). Therefore, we may or may not (for
testing purposes) further convolute with the PDF relevant to that parton, 
but if we do, we must do so for all of the terms in eq.~(\ref{final0}).}, 
while the third term is already in the form that we usually associate 
with a Born contribution (i.e.~an $n$-body matrix element convoluted 
with PDFs, per eq.~(\ref{fact0})). While the convolution with the PDFs
is a trivial operation, much more crucially and technically difficult
is to make sure that the subtraction in eq.~(\ref{final0}) exactly 
(lest one double count) matches the corresponding one of the $\dNLO$ 
computation that will eventually replace it -- this point is discussed
throughout the following sections, and will finally be addressed in 
sect.~\ref{sec:eq35}.

We remark that in the context of QED computations, where PDFs have
a well defined perturbative expansion, for ``Born'' to be unambiguously
identified one needs some qualifications. In particular, while just
above we have used a criterion based on final-state multiplicities
(i.e.~$n$- vs $(n+1)$-body, as was done in sect.~\ref{sec:gen}), we may as 
well employ a perturbative counting\footnote{The two coincide at the tree
level and in the absence of PDFs; this is not necessarily true if either
of these conditions is relaxed, as is shown by the discussion in the
main text.}. In doing so, and assuming that Born-level contributions are 
those of $\ord(\aem^b)$ associated with \mbox{$\gamma a\to X$}, then all 
terms in eq.~(\ref{final0}) (on both sides) are of $\ord(\aem^{b+1})$,
and this regardless of whether the convolution by $\PDF{\lep}{\lep}$
is carried out in the first and second terms on the r.h.s.~(since
\mbox{$\PDF{\lep}{\lep}=\ord(\aem^0)$} while
\mbox{$\PDF{\gamma}{\lep}=\ord(\aem)$}).

Two further observations are necessary. Firstly, we point out again that 
the $Z$ boson plays no role in the manipulations that lead us
to eq.~(\ref{final0}); its effects, including the interference with
the photon, are included {\em exactly} by the matrix element 
$\ampsq_{\lep}^{(n+1)}$, whereby \mbox{$\log Q^2/\mZ^2$} terms are not
resummed (as we have argued, this is an operation which is actually
not necessary). Secondly, the role of the second term in eq.~(\ref{final0})
is that of removing the double counting that stems from the sum of the first
and last terms. For this to happen, one needs to bear in mind that the
mass-scale squared $\mu^2$ employed in the evaluation of 
${\cal Q}_{\gamma\lep}^\prime(z)$ must be the same as that used as an 
argument of the photon PDFs. We also point out that, if one uses the 
NLL PDFs defined in the $\Delta$ scheme~\cite{Frixione:2021wzh}, by means
of a direct computation one finds:
\beq
\left.{\cal Q}_{\gamma\lep}^\prime(z)\right|_
{K_{\gamma\lep}(z)=K_{\gamma\lep}^{(\Delta)}(z)}=
{\cal Q}_{\gamma\lep}(z)\left[{\rm eq}.~(\protect\ref{Qres})\,,
\log\frac{s}{m^2}\longrightarrow\log\frac{s}{\mu^2}\right],
\label{QpDelta}
\eeq
with $K_{\gamma\lep}^{(\Delta)}(z)$ given in eq.~(3.6) of
ref.~\cite{Frixione:2021wzh}. This is interesting, since as is
discussed in sect.~4.2.3 of ref.~\cite{Frixione:2019lga} the
quantity ${\cal Q}_{\gamma\lep}(z)$ is one of the possible 
forms~\cite{Frixione:1993yw} of the Weizsaecker-Williams (WW) function.
However, this does {\em not} mean that in the context of the
$\Delta$ scheme ${\cal Q}_{\gamma\lep}^\prime(z)$ coincides with the WW
function because, crucially, of the replacement of the logarithmic
term in eq.~(\ref{QpDelta}): while the WW function is a collinear-divergent
quantity, ${\cal Q}_{\gamma\lep}^\prime(z)$ is collinear-finite.
It remains true that, as far as their forms in the $z$
space are concerned, the two functions are identical, and this is a
further evidence of the fact that the $\Delta$ scheme is better
suited to an intuitive physical picture than the $\MSb$ one. More extended
considerations on this point can be found in sect.~\ref{sec:scheme},
sect.~\ref{sec:resfact}, and in ref.~\cite{Frixione:2025xxx}.

All of the remarks presented above essentially apply also to 
eq.~(\ref{final00}), with some obvious differences. In particular,
if one now defines $b$ so that $\ord(\aem^b)$ is relevant to the
\mbox{$\gamma\gamma\to T$} process, then all of the
terms in eq.~(\ref{final00}) are of $\ord(\aem^{b+2})$. In the 
context of a multiplicity-based counting, the term proportional
to \mbox{$\PDF{\gamma}{\lep}(z_1)\PDF{\gamma}{\lep}(z_2)$} is a
Born-level one, while all of the others would be part of either
an $\dNLO$ computation with real processes given by eqs.~(\ref{mugaWW})
and~(\ref{gamuWW}) (these are the terms that feature a single
\mbox{${\cal Q}_{\gamma\lep}^\prime(z_{1,2})$} factor or a single $y_{1,2}$ 
subtraction -- of which there are four), or an $\dNNLO$ computation with 
a double-real process given by eq.~(\ref{VBFproc}) (these are the terms 
that feature either a factor \mbox{${\cal Q}_{\gamma\lep}^\prime(z_{1,2})$} 
or an $y_{1,2}$ subtraction for {\em each} leg -- of which there are 
four)\footnote{An equivalent way of figuring out who is who is to count the 
number of instances of $\PDF{\gamma}{\lep}(z_{1,2})$, with zero, one, and two 
corresponding to $\dNNLO$-like, $\dNLO$-like, and LO contributions, 
respectively.}, provided that each of such terms would be convoluted 
with $\PDF{\lep}{\lep}$ (which implies one or two convolution(s) for 
the $\dNLO$- or $\dNNLO$-like terms, respectively). 

\vskip -0.4truecm
\begin{figure}[htb]
  \begin{center}
  \includegraphics[scale=1.0,width=0.85\textwidth]{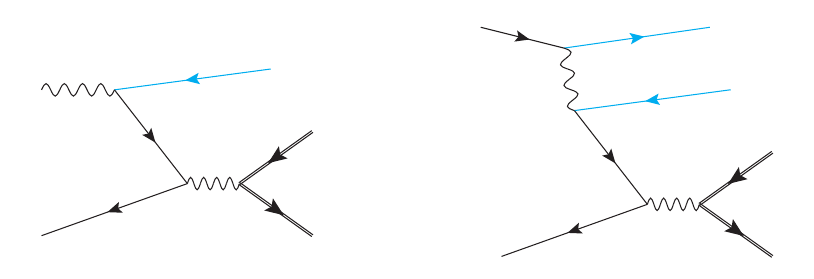}
\caption{\label{fig:div}
Divergent diagrams that contribute to eq.~(\ref{final00}); see the
text for details.
}
  \end{center}
\end{figure}
The key difference w.r.t.~what happens with a proper NLO or NNLO 
computation is that generally the contributions we are considering
here are still divergent, since there are no subtractions in some of the
singular regions, specifically those which do {\em not} correspond to
VBF-like configurations. Examples of diagrams relevant to the production of 
$T=\{t,\bt\}$ whose divergences are not subtracted in eq.~(\ref{final00}) 
are given in fig.~\ref{fig:div} -- that on the left (an $\dNLO$-type one, 
associated with an $\ord(\aem)$ $\gamma\to\lep\lep$ splitting kernel) 
contributes to  eq.~(\ref{gamuWW}), while that on the right (an $\dNNLO$-type 
one, associated with an $\ord(\aem^2)$ $\lep\to\lep\lep\lep$ splitting 
kernel) to eq.~(\ref{VBFproc}). 

In both of these examples, the unsubtracted singularities (which can 
also equivalently be seen as not damped by suitable ${\cal S}$ functions)
stem from an anti-collinear configuration, here the one in which the 
outgoing $\lp$ is collinear to the $\lm$ incoming from the left 
(the relevant reduced amplitude which is factorised contributes to 
\mbox{$\lm\lp\to t\bt$} rather than to \mbox{$\gamma\gamma\to t\bt$} 
as is the case for VBF-like diagrams). This is clearly a general characteristic 
in this matter, since anti-collinear configurations cannot be dominant 
in a VBF-like regime; nevertheless, regardless of their nature and lacking
a subtraction they must be avoided, lest eq.~(\ref{final00}) become 
meaningless, by means of suitable final-state cuts, to which we shall 
refer henceforth to as {\em technical cuts}. Note, finally, that anti-collinear 
configurations might not be the only sources of unsubtracted divergences; 
for example, eq.~(\ref{VBFproc}) receives divergent contributions when 
$k_1\!\parallel\! k_2$, that emerge from a final-state $\gamma\to\lp\lm$
branching. A more complete discussion on the issues posed by the
technical cuts will be given in sect.~\ref{sec:cuts}.

In summary, eq.~(\ref{final00}) will constitute the basis for a 
relative-$\ord(\aem^2)$ improvement of NLO predictions, the formulation
of which requires that one understands the role played by the technical
cuts, and finds a precise one-to-one correspondence between some of the 
terms of eq.~(\ref{final00}) and those in a proper NLO calculation. On 
top of that, it is helpful to keep in mind that, prior to any convolution
with PDFs, eq.~(\ref{final00}) is still very strictly related to
the $2\to 2+m$ massive-lepton matrix element it stems from, which 
offers some opportunities to test a number of technical issues that
emerge in the course of the procedure we shall follow. All of these
aspects can be discussed in the context of the factorisation theorem,
with which we begin sect.~\ref{sec:tech}.

\section{Technical aspects\label{sec:tech}}
As was said before, one key ingredient is the relationship between the 
massive- and massless-lepton matrix elements, which can be framed in 
the context of the factorisation theorem.

Before going into that, we preliminarily remind the reader that,
while featuring a more complicated divergent structure w.r.t.~their
massive counterparts, massless-lepton matrix elements have the advantage
that, at the tree and one-loop levels, they can be automatically computed 
and subsequently integrated over the phase space, whereas with a non-zero
but small lepton mass the automatic generation of virtual matrix elements is
impossible, and the phase-space integration of tree-level ones is 
difficult. Having said that, if we restrict ourselves to tree-level 
computations, and assign the lepton a mass which is not too small 
(roughly speaking, this means using a value of the order of the muon mass), 
we can still suppose that massive-lepton matrix elements also do not pose 
major problems from a numerical viewpoint. This implies that, at any fixed 
order, both its logarithmic and non-logarithmic terms can be obtained 
numerically, which is sufficient for our needs, since as we have mentioned
before we shall employ the massive-lepton matrix elements solely for
cross-checking results at the (integrated) matrix-element level.

Coming to the factorisation theorem proper, among other things one can
exploit it to study analytically the logarithms that appear in the 
massive-lepton matrix elements. By omitting parton indices in order to
simplify the notation, we write it symbolically thus\footnote{The
rightmost term in eq.~(\ref{factsimp}) reminds one of the existence
of power-suppressed effects associated with lepton masses. Since these
effects are negligible, henceforth we shall understand them.}:
\beq
d\bsig=\Gamma\star\Gamma\star d\hsig+\ord((m^2/Q^2)^c)\,,
\label{factsimp}
\eeq
where
\beq
d\bsig=\sum_{i=0}^\infty\left(\aemotpi\right)^i d\bsig^{[i]}\,,
\;\;\;\;\;\;
d\hsig=\sum_{i=0}^\infty\left(\aemotpi\right)^i d\hsig^{[i]}\,,
\;\;\;\;\;\;
\Gamma=\sum_{i=0}^\infty\left(\aemotpi\right)^i \Gamma^{[i]}\,,
\label{pexps}
\eeq
are the perturbative expansions of the complete massive-lepton cross 
sections, massless-lepton cross sections, and PDFs, respectively.
By replacing eq.~(\ref{pexps}) into eq.~(\ref{factsimp}), by expanding
the two sides in a series of $\aem$, and by equating the emerging terms with
the same power of $\aem$, one relates the $d\bsig^{[i]}$ and $d\hsig^{[i]}$
coefficients. The first two orders give (since $\Gamma^{[0]}=\delta(1-z)$ 
for the lepton, and $\Gamma^{[0]}=0$ for all of the other partons):
\beqn
d\bsig^{[0]}&=&d\hsig^{[0]}\,,
\label{fct0}
\\
d\bsig^{[1]}&=&d\hsig^{[1]}+d\delta^{[1]}\,,
\label{fct1}
\eeqn
where
\beq
d\delta^{[1]}=\Gamma^{[1]}\star\Gamma^{[0]}\star d\hsig^{[0]}+
\Gamma^{[0]}\star\Gamma^{[1]}\star d\hsig^{[0]}\,.
\label{fctdel}
\eeq
We stress again that we understand that the two sides of eqs.~(\ref{fct0}) 
and~(\ref{fct1}) generally differ by power-suppressed terms $(m^2/Q^2)^c$, 
per eq.~(\ref{factsimp}),
since these are neglected on the r.h.s.'s. By construction, $d\hsig^{[1]}$ is 
mass-independent and IR-finite, and therefore $d\delta^{[1]}$ accounts for 
the entire mass-logarithmic behaviour of $d\bsig^{[1]}$, which stem from 
$\Gamma^{[1]}$ -- see eq.~(\ref{Ggesol2}), and from its lepton counterpart, 
\EWeq{4.121}:
\beq
\PDF{\lep}{\lep}^{[1]}(z)=\left[\frac{1+z^2}{1-z}\left(
\log\frac{\mu^2}{m^2}-2\log(1-z)-1\right)\right]_+ + K_{\lep\lep}(z)\,,
\label{G1sol2}
\eeq
where the plus distribution implements soft-photon subtractions.
We point out that the previous statement holds in an {\em inclusive}
sense -- before any integration is carried out, $d\bsig^{[1]}$ and 
$d\delta^{[1]}$ live in an $(n+1)$- and $n$-body phase space, 
respectively\footnote{It is actually slightly more complicated than this.
In the context of an NLO calculation, $d\bsig^{[1]}$ features {\em also}
$n$-body-like contributions (e.g.~the soft remainders); however,
this does not change the essence of what is said here, which applies
to tree-level matrix elements that play the dominant role in the matter we
are discussing.}. In particular, for the logarithmic structure which is
explicit in the latter to appear in the former, 
an integration must be carried out over the extra degrees of freedom 
present in the $(n+1)$-body phase space w.r.t.~those of the $n$-body one. 
It is precisely this numerical integration that is problematic in the
case of massive-lepton matrix elements. However, if such an integration is 
successful, the independent calculation of the two sides of eq.~(\ref{fct1})
constitutes a powerful check of the consistency of one's results for 
the massive- and massless-lepton matrix elements, as well as for the 
analytical understanding of the mass-logarithmic structure of the former.

The lepton-mass logarithms in eqs.~(\ref{Ggesol2}) and~(\ref{G1sol2})
are in one-to-one correspondence with subtractions of initial-state
collinear singularities, examples of which are given by the plus
distributions in eqs.~(\ref{final0}) and~(\ref{final00}), in the sense
that the angular integrations that give rise to them in the context
of a massive-lepton computation correspond to the angular integrations
which are subtracted in a massless-lepton computation; in the former, the
mass screens the singularities which are subtracted in the latter. One can
also consider the screening of any singularity by means of a cut; loosely
speaking, if one denotes by $y_c$ the largest value that the angular variable
$y_i$ of eqs.~(\ref{final0}) and~(\ref{final00}) may assume, (i.e.,
\mbox{$-1\le y_i\le y_c$} rather than \mbox{$-1\le y_i\le 1$}), then the 
mass logarithm becomes:
\beq
\log\frac{\mu^2}{m^2}\;\;\longrightarrow\;\;
-\log\left(\frac{1-y_c}{2}+\frac{m^2}{\mu^2}\right)\,.
\label{lgwc}
\eeq
The r.h.s.~of eq.~(\ref{lgwc}) with $m\to 0$ shows also what happens 
if a massless calculation is not collinear-subtracted but cutoff-ed.
At $y_c\to 1$ the massive-lepton result displays again the mass logarithm,
whereas the (unsubtracted and without cutoff) massless-lepton one
diverges. Conversely, if \mbox{$m^2/\mu^2\ll 1-y_c$} the two kinds of
calculation feature the same logarithmic dependences on the cutoff $y_c$.

We shall soon show how this is relevant to our case. In order to do
that, we extend the procedure above to $\ord(\aem^{b+2})$, whereby the 
analogue of eqs.~(\ref{fct0}) and~(\ref{fct1}) reads as follows:
\beq
d\bsig^{[2]}=d\hsig^{[2]}+d\delta^{[2]}\,,
\label{fct2}
\eeq
with
\beqn
\;\;\;\;\;\;\;\;
d\delta^{[2]}&=&d\delta_D^{[2]}+d\delta_S^{[2]}\,,
\label{fctdel2}
\\
d\delta_D^{[2]}&=&\Big(
\Gamma^{[1]}\star\Gamma^{[1]}
+\Gamma^{[2]}\star\Gamma^{[0]}
+\Gamma^{[0]}\star\Gamma^{[2]}\Big)\star d\hsig^{[0]}\,,
\label{fctdel2D}
\\*
d\delta_S^{[2]}&=&\Big(
\Gamma^{[1]}\star\Gamma^{[0]}
+\Gamma^{[0]}\star\Gamma^{[1]}\Big)\star d\hsig^{[1]}.
\label{fctdel2S}
\eeqn
Equation~(\ref{fctdel2D}) is essentially a doubly-logarithmic contribution,
although single logarithms are present there as well. Conversely,
eq.~(\ref{fctdel2S}) features only single logarithms; the drawback is
that its r.h.s.~also depends on the subtracted $\ord(\aem^{b+1})$
coefficient $d\hsig^{[1]}$, which may not be readily available in
the context of a massive-lepton computation. In order to eliminate it,
we can exploit eq.~(\ref{fct1}), to arrive at:
\beq
d\delta_S^{[2]}=d\delta_{Se}^{[2]}-d\delta_{Sc}^{[2]}\,,
\label{fctdel2Semc}
\eeq
with
\beqn
d\delta_{Se}^{[2]}&=&\Big(
\Gamma^{[1]}\star\Gamma^{[0]}
+\Gamma^{[0]}\star\Gamma^{[1]}\Big)\star d\bsig^{[1]}\,,
\label{fctdel2Se}
\\*
d\delta_{Sc}^{[2]}&=&\Big(
2\,\Gamma^{[1]}\star\Gamma^{[1]}
+\big(\Gamma^{[1]}\star\Gamma^{[1]}\big)\star\Gamma^{[0]}
+\Gamma^{[0]}\star\big(\Gamma^{[1]}\star\Gamma^{[1]}\big)\Big)
\star d\hsig^{[0]}.
\label{fctdel2Sc}
\eeqn
In eq.~(\ref{fctdel2Se}) only unsubtracted massive-lepton matrix elements 
appear; however, there are now double logarithms, which are cancelled by
those in eq.~(\ref{fctdel2Sc}). It should be clear that while we have
understood logarithms of $m^2/\mu^2$, when a cutoff is present we can 
actually have the more involved argument that appears on the r.h.s.~of 
eq.~(\ref{lgwc}), 

It is appropriate to comment explicitly on the physical meaning of the various
contributions that appear in eqs.~(\ref{fctdel2D}), (\ref{fctdel2Se}), 
and~(\ref{fctdel2Sc}): for example \mbox{$\Gamma^{[1]}\star\Gamma^{[1]}$} 
stems from two $1\to 2$ branchings, one for each incoming leg, while
\mbox{$\Gamma^{[0]}\star\Gamma^{[2]}$} stems from one $1\to 3$ branching
in the leg incoming from the right, and
\mbox{$\Gamma^{[0]}\star\big(\Gamma^{[1]}\star\Gamma^{[1]}\big)$} is its
strongly-ordered analogue, i.e.~one that stems from two consecutive
$1\to 2$ branchings from the same right leg. This also renders it clear that 
only terms such as \mbox{$\Gamma^{[1]}\star\Gamma^{[1]}$} are possibly 
associated with VBF configurations; diagrammatically, 
\mbox{$\Gamma^{[0]}\star\Gamma^{[2]}$}
can be seen to emerge from situations analogous to that depicted on
the right panel of fig.~\ref{fig:div}, i.e.~from what has been called
an anti-collinear configuration.

The latter observation is key in view of the fact that eq.~(\ref{fct2}) 
can be used in the same way as what was suggested for eq.~(\ref{fct1}), 
namely as a tool to check the mutual consistency  between massive- and 
massless-lepton matrix element computations. Clearly, some attention is
required, since at relative ${\cal O}(\aem^2)$ we restrict ourselves
to considering tree-level matrix elements, whereas the quantities
$d\bsig^{[i]}$ and $d\hsig^{[i]}$ are supposed to be the complete
cross sections. However, as was  already observed, the emergence of
mass logarithms is chiefly driven by tree-level matrix elements,
and this can indeed be verified explicitly. Thus, for the case of 
interest for this paper, we shall identify the l.h.s.~of eq.~(\ref{fct2}) 
with the massive-lepton tree-level matrix element of the $2\to 2+m$ process, 
while its r.h.s.~will be given by eq.~(\ref{final00}), with the photon PDFs 
there ($\PDF{\gamma}{\lep}$) replaced by their $\ord(\aem)$ coefficients
\mbox{($\aem/(2\pi)\PDF{\gamma}{\lep}^{[1]}$)}. It should then be 
clear that this operation does lead to a potential mismatch -- specifically, 
contributions corresponding to terms such as 
\mbox{$\Gamma^{[0]}\star\Gamma^{[2]}$} and
\mbox{$\Gamma^{[0]}\star\big(\Gamma^{[1]}\star\Gamma^{[1]}\big)$} 
will feature double and single mass logarithms in the massive computation,
and logarithms of the cutoff used to impose the technical cuts in the
massless computation. There are therefore two things that can be
done. One first puts the massive- and massless-lepton results on the 
same footing, by introducing the same cutoff for the anti-collinear 
configurations in the two cases, whereby the formal replacement of
eq.~(\ref{lgwc}) becomes relevant. Then one can either compare the
results when \mbox{$m^2/\mu^2\ll 1-y_c$}, where an agreement is
expected if the procedure that leads to eq.~(\ref{final00}) is
correct. Or one can compare the cutoff-ed massless calculation with
the massive one where no cutoff is applied -- any significant difference
found between the two would signal a possible bias on the
predictions of eq.~(\ref{final00}) due to the cutoff choice.
We shall discuss these two options, as well as a number of other
issues relevant to the technical cuts, in sect.~\ref{sec:cuts}.

\subsection{Cuts\label{sec:cuts}}
As is discussed both in sect.~\ref{sec:meaning} (see in particular 
fig.~\ref{fig:div} there) and at the beginning of sect.~\ref{sec:tech}, 
eq.~(\ref{final00}) requires technical cuts lest it diverge, 
since its $\dNLO$- and $\dNNLO$-type contributions are not fully-fledged  
$\dNLO$ and $\dNNLO$ computations, and thus feature unsubtracted
singularities. While the advocated replacement of the $\dNLO$-type terms with 
a complete $\dNLO$ computation will address the former case, for the latter 
one cuts will still be necessary. By construction, such cuts will be
applied on the outgoing leptons. A preliminary question then arises, namely
whether and how these cuts will affect not only the contribution on the
first line of eq.~(\ref{final00}), but also the three other ones (where
one or both of the outgoing leptons are remnants collinear to the parent 
beam line, and do not formally appear in the final state). Note that the
latter three terms have essentially the same forms, and thus pose the same 
problems, as the counterevents in the former.

As to the ``whether'', the answer appears to be positive. Physically,
this can be argued as follows. All contributions to eq.~(\ref{final00}) 
emerge from manipulating the matrix elements of eq.~(\ref{VBFproc}) by
means of kinematical transformations, enforced by identities such as
those of eqs.~(\ref{ybe2}) and~(\ref{ybe3}). Then, if one multiplies the 
matrix elements from the very beginning by a string of $\stepf$ functions
that impose the cuts, these kinematical transformations will 
affect the arguments of such $\stepf$ functions (for example, by turning
a hard configuration into a collinear one), but will not eliminate them
(of course, this does not prevent some of these $\stepf$ functions from being 
null for certain kinematical configurations). In other words, the various 
contributions to eq.~(\ref{final00}) are related to each other, in spite
of the fact that one is entitled to deal with each of them independently.
More formally, these relationships among contributions can be seen as 
follows. One starts from the distribution identity:
\beq
\left(\frac{1}{1-y}\right)_+ =
\left(\frac{1}{1-y}\right)_{\deltaI}+
\log\frac{\deltaI}{2}\,\delta(1-y)\,,
\label{ypvsyd}
\eeq
with $0<\deltaI\le 2$, and having defined:
\beq
\langle \left(\frac{1}{1-y}\right)_{\deltaI},f\rangle =
\int_{-1}^1\frac{dy}{1-y}\,\Big[f(y)-f(1)\stepf\big(y\ge 1-\deltaI\big)\Big].
\eeq
By using the massless limits of the formulae given in sect.~\ref{sec:VBF}
one arrives at the usual factorised expression, e.g.~for emissions from 
leg~1:
\beqn
&&\log\frac{\deltaI}{2}\,\delta(1-y_1)\,(1-y_1)
\ampsq^{(n+1)}(p_1,p_2)\,d\phi_{n+1}(p_1,p_2)=
\nonumber\\*&&\phantom{aaaaaaa}
\frac{\aem}{2\pi}\,\log\frac{\deltaI}{2}\,
P(z_1)\ampsq^{(n)}(z_1p_1,p_2)\,dz_1\,d\phi_{n}(z_1p_1,p_2)\,,
\label{collM}
\eeqn
where $P$ is the relevant Altarelli-Parisi kernel.
By employing eq.~(\ref{ypvsyd}) to eliminate all of the plus distributions
that appear in eq.~(\ref{final00}), and by subsequently using 
eq.~(\ref{collM}) to manipulate all of the emerging terms that feature
a $\delta(1-y_{1,2})$ distribution, it is matter of simple algebra to
arrive at a result which is identical to that of eq.~(\ref{final00}),
bar for the formal replacements:
\beqn
\left(\frac{1}{1-y_{1,2}}\right)_+ \;\;&\longrightarrow&\;\;
\left(\frac{1}{1-y_{1,2}}\right)_{\deltaI}\,,
\\
{\cal Q}_{\gamma\lep}^\prime(z)\;\;&\longrightarrow&\;\;
{\cal Q}_{\gamma\lep}^{(\deltaI)^\prime}(z)\,,
\eeqn
where
\beq
{\cal Q}_{\gamma\lep}^{(\deltaI)^\prime}(z)=
\frac{1+(1-z)^2}{z}\left(
\log\frac{s\deltaI}{2\mu^2}+2\log(1-z)\right) +z - K_{\gamma\lep}(z)\,.
\label{Qpresdel}
\eeq
Note that the result of eq.~(\ref{Qpresdel}) coincides with that
of sect.~4.3 of ref.~\cite{Frederix:2009yq}, again in keeping with
a fact remarked in sect.~\ref{sec:meaning} --  
${\cal Q}_{\gamma\lep}^{(\deltaI)^\prime}$ is the kernel relevant to 
the FKS degenerate $(n+1)$-body contributions. It now becomes apparent
that the various terms of eq.~(\ref{final00}) are related to
each other, since the dependence on $\deltaI$ only cancels in their
sum, and not in each of them individually. 

The procedure above also helps clarify how the technical cuts must
be implemented on the second, third, and fourth contributions to 
eq.~(\ref{final00}), as well as in the counterevents of the first
contribution there, since it underscores how the terms that 
feature the $\PDF{\gamma}{\lep}(z_{1,2})$ and 
${\cal Q}_{\gamma\lep}^{(\deltaI)^\prime}(z_{1,2})$ kernels have
emerged from the strictly collinear $\lep\to\gamma\lep$ branching 
enforced by a $\delta(1-y_{1,2})$ factor; thus, the lepton which does
not explicit appear in the final state travels along the beamline
with momentum equal\footnote{This is strictly true only at $\ord(\aem)$,
i.e.~for ${\cal Q}_{\gamma\lep}^{(\deltaI)^\prime}(z_{1,2})$ but not
necessarily for $\PDF{\gamma}{\lep}(z_{1,2})$, which in general contains
terms to all orders. By associating all of the remnant momentum with the
lepton we make the simplest choice, which is still compatible with the
perturbative order we are working at.} to \mbox{$(1-z_{1,2})p_{1,2}$}.

Finally, as far as the technical cuts proper are concerned, and in reference
to the process of eq.~(\ref{VBFproc}) and to the reduced ones emerging
from it according to the various contributions to eq.~(\ref{final00}):
the anticollinear divergences are avoided by imposing the following
conditions (hemisphere cuts)\footnote{In other words: the cutoff 
parameters in eqs.~(\ref{hemic}), (\ref{massc1}) and~(\ref{massc2})
play the role in the actual computations of what has been generically
denoted by $y_c$ in eq.~(\ref{lgwc}).}:
\beq
\stepf_{H_1}\stepf_{H_2}\;\equiv\;
\stepf\left(\eta_{\lm(k_1)}\ge -\eta_c\right)
\stepf\left(\eta_{\lp(k_2)}\le \eta_c\right)
\label{hemic}
\eeq
(or equivalent ones given in terms of angles), with $\eta_{\lep(k)}$ the
pseudorapidity of lepton $\lep$ with momentum $k$, and $\eta_c$ an 
arbitrary quantity, understood to be positive. If either (or both) lepton(s)
is (are) in the respective anticollinear region, the product in 
eq.~(\ref{hemic}) vanishes.
Conversely, its collinear limits exist, and therefore these cuts are
collinear safe. In addition to them, the requirement that the
$\gamma\to\lp\lm$ and $Z\to\lp\lm$ branchings (the latter when
the $Z$ width is set equal to zero) do not induce singularities is
equivalent to using
\beqn
\stepf_\gamma&=&\stepf\left(M_{\lm(k_1)\lp(k_2)}\ge\delta_{M_1}\right)\,,
\label{massc1}
\\
\stepf_Z&=&1-\stepf\left(M_Z-\delta_{M_2}\le M_{\lm(k_1)\lp(k_2)}
\le M_Z+\delta_{M_3}\right)\,,
\label{massc2}
\eeqn
for some $\delta_{M_1}$, $\delta_{M_2}$, and $\delta_{M_3}$; these cuts
are manifestly soft and collinear safe. The overall technical cuts
amount to imposing
\beq
\stepf_{H_1}\stepf_{H_2}\stepf_\gamma\stepf_Z\,.
\label{allc}
\eeq
This implies that the outgoing leptons must be in their respective hemisphere
(which includes the case when either of them is a beam remnant). There, they
must also pass the invariant mass cuts, that force the pair to not be
on the $\gamma$ and $Z$ mass shells. If either lepton is outside of its
hemisphere the event is discarded, and the invariant mass cuts are
irrelevant.

We note that there is some redundancy in the cuts of eq.~(\ref{allc}).
For example, in the anticollinear region the contribution due to the 
graph on the left panel of fig.~\ref{fig:div} has both\footnote{Note
that for this to happen the role of the beam-remnant lepton in defining
the cuts is crucial.} $\stepf_{H_2}=0$ and $\stepf_\gamma=0$. Still, 
it seems preferable (owing to its larger flexibility, with only a very 
minor increase of complexity) to implement both the hemisphere and the 
invariant-mass cuts in the context of phenomenology studies.

\subsection{Factorisation scheme dependence\label{sec:scheme}}
Lepton PDFs can be computed in perturbative QED only after making
some assumptions; these, we collectively call a ``scheme''. At the LL 
the choice of a scheme (which is {\em not} a factorisation scheme, 
since it is not associated with the subtraction of a singularity) 
typically involves quantities which are beyond accuracy -- 
see e.g.~eqs.~(A.14)--(A.19) in ref.~\cite{Bertone:2022ktl}.
Note that differences among LL schemes may or may not be parametric, and
as such can be very large and difficult to quantify; for example, some 
LL PDFs (notably, those which have been used historically) do not have 
any photon content, others do. Conversely, at the NLL any scheme 
{\em is} a factorisation scheme\footnote{There is actually a
dependence on the renormalisation scheme as well, which is of a
different nature w.r.t.~that associated with the factorisation scheme.
This point has been discussed at length in ref.~\cite{Bertone:2022ktl},
and will not be repeated here, where we shall always consider the 
renormalisation scheme as given, choosing the $G_\mu$ one for 
numerical predictions.}, that essentially defines the 
non-divergent part of the subtraction of the collinear singularities 
of bare PDFs. It is convenient to parametrise factorisation
schemes in terms of the $K_{ij}(z)$ functions, given that these also enter
the FKS short-distance cross sections (specifically, in the $(n+1)$-body
degenerate contributions). 

Since we emphasise the importance of accuracy and of the ability to
sensibly assess theoretical uncertainties, we discuss here in the
main text the case of NLL PDFs; some facts of relevance to their
LL counterparts are given in appendix~\ref{sec:factLL}.

The basic idea is that the factorisation theorem, eq.~(\ref{factsimp}),
is scheme-independent up to higher-order terms (i.e.~those not included 
in the short-distance cross sections). Thus, in eq.~(\ref{factsimp}) 
the scheme dependence of the PDFs $\Gamma$ cancels that of the
cross sections $d\hsig$. At the NLO this is understood in minute
details; mnemonically, one bears in mind that the $K_{ij}$ functions
enter linearly with a plus sign in the expansions of the PDFs (see 
e.g.~eqs.~(\ref{Ggesol2}) and~(\ref{G1sol2})), and with a minus sign in 
the kernels of the $(n+1)$-body degenerate contributions (see 
e.g.~ref.~\cite{Frederix:2009yq}), so that the physical cross sections 
(after a perturbative expansion in $\aem$) are $K_{ij}$-independent.
At the NNLO the principle is the same, but there are no explicit studies
of this matter for leptonic collisions. We point out that the 
$\ord(\aem^{b+2})$ contributions we are considering here only account for 
some terms of a proper $\dNNLO$ result. However, we can observe that in 
the combinations
\beq
\PDF{\gamma}{\lep}(z)+
\aemotpi\,{\cal Q}_{\gamma\lep}^\prime(z)\equiv
\aemotpi\,{\cal Q}_{\gamma\lep}(z)+\ord(\aem^2)
\label{GmQ}
\eeq
that appear in eq.~(\ref{final00}) the dependence 
on $K_{\gamma\lep}(z)$ drops out at this order after expanding the PDF
(see eqs.~(\ref{Ggesol2}) and~(\ref{Qpres})). While this is true by 
the construction of ${\cal Q}_{\gamma\lep}^\prime$, which inherits the 
$K_{\gamma\lep}$ dependence from $d\hsig_{\gamma\ord(\aem)}^{(n)}$, 
i.e.~from the expansion of the photon PDF itself, it will become 
a useful cross-check tool when the two terms on the l.h.s.~of
eq.~(\ref{GmQ}) will be associated with different contributions to
a cross section.

These observations have implications for what has been discussed at the
beginning of sect.~\ref{sec:tech}. Note, in particular, that $d\delta^{[1]}$, 
$d\delta_D^{[2]}$, and $d\delta_S^{[2]}$ are scheme-dependent through 
$\Gamma^{[1]}$ and $\Gamma^{[2]}$\footnote{The former point renders it 
clear that when employing the WW function in the
definition of the subtraction terms one introduces uncertainties which
are difficult to quantify, since there is no known scheme in which the
first-order coefficient in the expansion of the photon PDF coincides
with the WW function.}. Further to this observation, we note that the
r.h.s.~of eqs.~(\ref{fct1}) and~(\ref{fct2}) are truly scheme independent
only if $d\hsig^{[1]}$ and $d\hsig^{[2]}$, respectively, feature
collinear-remainder contributions (the $(n+1)$-degenerate terms of FKS),
and not only real-emission ones. This is because the subtractions of
collinear singularities of a massless matrix element is inherently
defined up to finite terms, whose arbitrariness is compensated by
that of the collinear remainders. In the context of the massive- vs
massless-lepton checks that we have discussed in sect.~\ref{sec:cuts}
the actual cancellation of all arbitrary terms of collinear origin
is automatically taken care of by the presence of the scale- and
scheme-independent linear combination on the l.h.s.~of eq.~(\ref{GmQ}).
Explicit results that show the extent of such a cancellation are
given in sect.~\ref{sec:resfact}, and in ref.~\cite{Frixione:2025xxx}.

\subsection{Two-side vs one-side VBF topologies\label{sec:sides}}
The main result of this paper will stem from eq.~(\ref{final00}). 
In order to arrive from the latter to the improved prediction we seek,
we need to investigate a couple of potential issues on top of those that
have already been considered. Firstly, how
eq.~(\ref{final00}) emerges by ``iterating'' the procedure that
leads to eq.~(\ref{final0A}). Secondly, the identification of certain
terms with $\dNLO$-type contributions (according to the FKS formalism)
is based on the fact that the former and the latter have the same
{\em functional} form. However, this does not mean that they are identical. 
This is because the angular variables $y_{1,2}$  of eq.~(\ref{final00})
are defined in the rest frame of \mbox{$p_1+p_2$}, while FKS would
require $y_1$ ($y_2$) to be defined in the rest frame of 
\mbox{$p_1+z_2p_2$} (\mbox{$z_1p_1+p_2$}), since this is the c.m.~frame of 
the parton pair that initiates the hard process. As we shall see in the
following, these two items are related to each other to a certain
extent; we shall start our discussion from the former.

By ``iterating'' the procedure that leads to eq.~(\ref{final0A}) we
literally mean to apply it separately to the following two quantities:
\beqn
&&\left(\frac{1}{1-y_1}\right)_+\!
\left((1-y_1)\ampsq_{\lep\lep}^{(m+2)}\right)
d\phi_{m+2}\big(p_1,p_2\big)\,,
\label{source1}
\\*&&
\left(\PDF{\gamma}{\lep}(z_1)+
\aemotpi\,{\cal Q}_{\gamma\lep}^\prime(z_1)\right)
\ampsq_{\gamma\lep}^{(m+1)}\,
d\phi_{m+1}\big(z_1p_1,p_2\big)\,dz_1\,.
\label{source2}
\eeqn
It is apparent that the former generates the first and the second 
contributions in eq.~(\ref{final00}), and the latter the third
and the fourth. However, eq.~(\ref{final0A}) has been obtained
by starting from a massive matrix element ($\bampsq_{\lep\lep}^{(m+2)}$),
and a massive $(m+2)$-body phase space -- which is not what appears
in eqs.~(\ref{source1}) and~(\ref{source2}). The key observation is
the factorised structure of the VBF-like contributions we are 
considering. Such a structure implies that the characteristics of
the vertex where the photon branches from $\lp(p_2)$ are not affected
by the form of the contributions to the matrix elements relevant
to the vertex where the photon branches from $\lm(p_1)$, which here 
give rise the $y_1$- and $z_1$-dependent prefactors in eqs.~(\ref{source1}) 
and~(\ref{source2}), respectively. Because of this, we can easily
restore the mass dependence of the \mbox{$\lp(p_2)\to\gamma\lp(k_2)$}
vertex in the $\ampsq_{\lep\lep}^{(m+2)}$ and $\ampsq_{\gamma\lep}^{(m+1)}$ 
matrix elements -- this is sufficient to give one the collinear-dominant 
contributions. Likewise, the phase spaces can again be taken to be
massive. In this way, the only remaining difference w.r.t.~the original
starting point is the presence, in eq.~(\ref{source2}) of an
$(m+1)$-body phase space, rather than of an $(m+2)$-body one. In fact,
the implicit assumption is that such a phase space is written as
follows:
\beq
d\phi_{m+1}(z_1p_1,p_2)dz_1=
\left(\frac{s\xi_1}{4}\frac{1}{2(2\pi)^2}\right)^{-1}\!\!
\delta(1-y_1)\,d\phi_{m+2}(p_1,p_2)\,.
\label{phsprepar}
\eeq
Again, the prefactors on the r.h.s.~of this equation do not have
any bearings on the features relevant to the second step of the 
procedure considered here, since the latter are essentially limited to
parametrising $p_2\mydot k_2$. By employing eq.~(\ref{phsprepar})
one also still works in the rest frame of \mbox{$p_1+p_2$}.
In this way, one finally arrives at eq.~(\ref{final00}), where
by construction all of the angular variables are defined in the rest 
frame of \mbox{$p_1+p_2$}.

An alternative and more rigorous way to obtain eq.~(\ref{final00}) is 
that of observing that the two collinear limits of interest, namely 
$p_1\parallel k_1$ and $p_2\parallel k_2$, are independent of each
other -- the double-collinear limit is incoherent:
\beqn
\bampsq_{\lep\lep}^{(n+2)}
&\stackrel{p_1\parallel k_1}{\longrightarrow}&
\frac{e^2}{p_1\mydot k_1-m^2}\,P_{\gamma^\star\lep}^{<}\left(z_1\right)\,
\ampsq_{\gamma}^{(n+1)}(zp_1,p_2)
\label{dcll}
\\*
&\stackrel{p_2\parallel k_2}{\longrightarrow}&
\frac{e^2}{p_1\mydot k_1-m^2}\,P_{\gamma^\star\lep}^{<}\left(z_1\right)\,
\frac{e^2}{p_2\mydot k_2-m^2}\,P_{\gamma^\star\lep}^{<}\left(z_2\right)\,
\ampsq_{\gamma\gamma}^{(n)}(z_1p_1,z_2p_2)\,.
\nonumber
\eeqn
Thus, the identities of eqs.~(\ref{ybe2}) and~(\ref{ybe3}), written
in terms of 
\beq
\rho_1=2\frac{(1-\xi_1)^2}{\xi_1^2}\frac{m^2}{s}\,,
\;\;\;\;\;\;\;\;\;
\rho_2=2\frac{(1-\xi_2)^2}{\xi_2^2}\frac{m^2}{s}\,,
\eeq
can be exploited simultaneously for the two incoming legs, as a factorised
series expansion in $(\rho_1,\rho_2)$.

What was done above, and in particular eq.~(\ref{phsprepar}), allows us
to discuss the second of the items introduced at the beginning of this
section. Here, we can formulate it as follows: while eq.~(\ref{phsprepar})
gives one a way to obtain the sought result, it is by no means 
{\em necessary} to use it. Then, what happens if one tries and works
in the rest frame of \mbox{$z_1p_1+p_2$}? This is an equally
valid approach, since the massification of the matrix elements and of
the phase space is as before.

In order to follow the details of this discussion, the reader will 
need some information about the reference frames relevant to the problem, 
and to exploit an identity among distributions, which is proven in 
appendix~\ref{sec:distr}. As far as the former ones are concerned, we 
present all of the material about them in appendix~\ref{sec:frames};
here, we limit ourselves to introduce the notation with which we
identify the frames that we shall employ in the following, namely:
\beqn
&&F:\;\;\;\;{\rm rest~frame~of}\;p_1+p_2\,,
\\*
&&F_1^\prime:\;\;\;\;{\rm rest~frame~of}\;z_1p_1+p_2\,.
\eeqn
With the help of appendix~\ref{sec:frames} and of appendix~\ref{sec:distr},
we are now able to consider the iteration of our procedure
starting from the quantity in eq.~(\ref{source2}) but, at variance
with what has been done before, we shall not employ eq.~(\ref{phsprepar})
but work directly in frame $F_1^\prime$. By using the notation
of eqs.~(\ref{phinpo}) and~(\ref{tphin}) this leads to:
\beqn
&&\left(\PDF{\gamma}{\lep}(z_1)+
\aemotpi\,{\cal Q}_{\gamma\lep}^\prime(z_1;s)\right)
\left(\frac{1}{1-\yb_2}\right)_+\!
\left((1-\yb_2)\ampsq_{\gamma\lep}^{(m+1)}(z_1p_1,p_2)\right)
\nonumber\\*&&\phantom{aaaaaaaa}\times
d\tilde{\phi}_n(p_1,p_2)
\frac{1}{2(2\pi)^3}\,\frac{z_1s}{4}\,\xib_2 d\xib_2\,d\yb_2\,d\varphi_2\,dz_1
\nonumber\\*&+&
\left(\PDF{\gamma}{\lep}(z_1)+
\aemotpi\,{\cal Q}_{\gamma\lep}^\prime(z_1;s)\right)
\left(\PDF{\gamma}{\lep}(z_2)+
\aemotpi\,{\cal Q}_{\gamma\lep}^\prime(z_2;z_1s)\right)
\nonumber\\*&&\phantom{aaaaaaaaa}\times
\ampsq_{\gamma\gamma}^{(m)}(z_1p_1,z_2p_2)\,
d\phi_{m}\big(z_1p_1,z_2p_2\big)\,dz_1\,dz_2\,,\phantom{aaa}
\label{tmp3}
\eeqn
which must be compared to the last two lines of eq.~(\ref{final00}).
There are two obvious differences emerging from such a comparison,
namely: the plus prescription is expressed in terms of the $\yb_2$
variable (as opposed to $y_2$), which is due to working in frame
$F_1^\prime$ rather than in $F$; as a consequence of this, the integration
variables are also those relevant to the former frame. And secondly,
the rightmost ${\cal Q}_{\gamma\lep}^\prime$ term in the last line of
eq.~(\ref{tmp3}) features a $(z_1s)$ factor as argument of the logarithm
that multiplies the Altarelli-Parisi splitting kernel, rather than $s$; 
in view of this, the notation has been modified.
In order to proceed, in the first term of eq.~(\ref{tmp3}) one uses
eq.~(\ref{plvsplb2F1}) and changes variables\footnote{Note that part
of the jacobian of this transformation is already included in
eq.~(\ref{plvsplb2F1}), and thus one need only take into account
$d\xib_2/d\xi_2$; this, thanks to the fact that $\yb_2$ is independent
of $\xi_2$, and thus the jacobian is the product of two terms.} 
\mbox{$(\yb_2,\xib_2)\to (y_2,\xi_2)$}; furthermore, one also observes that:
\beq
1-\yb_2=\frac{4}{z_1s\xib_2}\,(p_2\mydot k_2)=
\frac{\xi_2}{z_1\xib_2}\,(1-y_2).
\eeq
After some algebra, the first term of eq.~(\ref{tmp3}) thus transformed
results in the sum of two contributions, the first of which is identical
to the third line of eq.~(\ref{final00}), while the second of which reads:
\beqn
&&\left(\PDF{\gamma}{\lep}(z_1)+
\aemotpi\,{\cal Q}_{\gamma\lep}^\prime(z_1;s)\right)
\left(-\aemotpi\log z_1\, P_{\gamma\lep}(z_2)\right)
\nonumber\\*&&\phantom{aaaaaaaaa}\times
\ampsq_{\gamma\gamma}^{(m)}(z_1p_1,z_2p_2)\,
d\phi_{m}\big(z_1p_1,z_2p_2\big)\,dz_1\,dz_2\,,\phantom{aaa}
\label{tmp4}
\eeqn
where use has been made of eq.~(\ref{collM}) to simplify the purely
collinear term\footnote{Note that at $y_2=1$ we have $\xi_2=\xib_2$,
and therefore there is no need to introduce a variable $\bar{z}_2$.}
stemming from $\delta(1-y_2)$. Then, in the sum of eq.~(\ref{tmp4}) plus
the second term of eq.~(\ref{tmp3}) one finds the linear combination:
\beq
-\log z_1 P_{\gamma\lep}(z_2)+
{\cal Q}_{\gamma\lep}^\prime(z_2;z_1s)\equiv
{\cal Q}_{\gamma\lep}^\prime(z_2;s).
\label{Qpextrad}
\eeq
This is identical to what appears in the last line of eq.~(\ref{final00}),
thus finally proving that, regardless of whether the second step of the
iterative procedure is carried out in $F$ or $F_1^\prime$, the result
is always the same. We point out that this also shows that the final
result is independent of whether one starts with the $p_1\parallel k_1$
or the $p_2\parallel k_2$ collinear configuration.

\section{The NNLO-improved NLO predictions\label{sec:eq35}}
We are finally in the position to firmly establish the connection between
the various terms in eq.~(\ref{final00}) and their counterparts in 
NNLO, NLO, and LO cross sections, and thus to achieve our stated goal
of replacing some of the former with some of the latter. We start 
by re-writing eq.~(\ref{final00}) as follows (see page~\pageref{page:names}
for the notation used here):
\beq
\bampsq_{\lep\lep}^{(m+2)} d\phi_{m+2}\longrightarrow
d\hat{\Sigma}_{\dNNLOG}+d\hat{\Sigma}_{\dNLOG}+d\Sigma_{\LOG}\,,
\label{Msplit}
\eeq
where:
\beqn
d\hat{\Sigma}_{\dNNLOG}(p_1,p_2)&=&
\left(\frac{1}{1-y_1}\right)_+\!
\left(\frac{1}{1-y_2}\right)_+\!
\left((1-y_1)(1-y_2)\ampsq_{\lep\lep}^{(m+2)}\right)
d\phi_{m+2}\big(p_1,p_2\big)
\nonumber\\*&+&
\aemotpi\,{\cal Q}_{\gamma\lep}^\prime(z_2)
\left(\frac{1}{1-y_1}\right)_+\!
\left((1-y_1)\ampsq_{\lep\gamma}^{(m+1)}\right)
d\phi_{m+1}\big(p_1,z_2p_2\big)\,dz_2
\nonumber\\*&+&
\aemotpi\,{\cal Q}_{\gamma\lep}^\prime(z_1)
\left(\frac{1}{1-y_2}\right)_+\!
\left((1-y_2)\ampsq_{\gamma\lep}^{(m+1)}\right)
d\phi_{m+1}\big(z_1p_1,p_2\big)\,dz_1
\nonumber\\*&+&
\left(\aemotpi\right)^2\,
{\cal Q}_{\gamma\lep}^\prime(z_1)\,
{\cal Q}_{\gamma\lep}^\prime(z_2)
\ampsq_{\gamma\gamma}^{(m)}\,
d\phi_{m}\big(z_1p_1,z_2p_2\big)\,dz_1\,dz_2\,,\phantom{aaa}
\label{final00NNLO}
\\*
d\hat{\Sigma}_{\dNLOG}(p_1,p_2)&=&
\PDF{\gamma}{\lep}(z_2)
\left(\frac{1}{1-y_1}\right)_+\!
\left((1-y_1)\ampsq_{\lep\gamma}^{(m+1)}\right)
d\phi_{m+1}\big(p_1,z_2p_2\big)\,dz_2
\nonumber\\*&+&
\PDF{\gamma}{\lep}(z_1)
\left(\frac{1}{1-y_2}\right)_+\!
\left((1-y_2)\ampsq_{\gamma\lep}^{(m+1)}\right)
d\phi_{m+1}\big(z_1p_1,p_2\big)\,dz_1
\nonumber\\*&+&
\aemotpi
\Big({\cal Q}_{\gamma\lep}^\prime(z_1)\PDF{\gamma}{\lep}(z_2)+
\PDF{\gamma}{\lep}(z_1){\cal Q}_{\gamma\lep}^\prime(z_2)\Big)
\nonumber\\*&&\phantom{aaaaaaaaa}\times
\ampsq_{\gamma\gamma}^{(m)}\,
d\phi_{m}\big(z_1p_1,z_2p_2\big)\,dz_1\,dz_2\,,\phantom{aaa}
\label{final00NLO}
\\*
d\Sigma_{\LOG}(p_1,p_2)&=&
\PDF{\gamma}{\lep}(z_1)
\PDF{\gamma}{\lep}(z_2)
\ampsq_{\gamma\gamma}^{(m)}\,
d\phi_{m}\big(z_1p_1,z_2p_2\big)\,dz_1\,dz_2\,,\phantom{aaa}
\label{final00LO}
\eeqn
are, as the notation suggests, the $\dNNLO$-, $\dNLO$-, and LO-like 
contributions, respectively, due to $\gamma\gamma$ fusion.
These three quantities are independent of one another\footnote{In the
sense that they are separately IR-finite. One can still exchange finite terms 
among them, which is what has been anticipated at page~\pageref{page:two} 
and the reason why the various identifications have to be done with utmost
care. We shall return to this point later\label{ft:tt}.}, and will be
thus manipulated independently. Starting with eq.~(\ref{final00NNLO}),
by exploiting eq.~(\ref{ypvsyd}) one arrives at:
\beqn
d\hat{\Sigma}_{\dNNLOG}(p_1,p_2)&=&
\left(\frac{1}{1-y_1}\right)_{\deltaI}\!
\left(\frac{1}{1-y_2}\right)_{\deltaI}\!
\left((1-y_1)(1-y_2)\ampsq_{\lep\lep}^{(m+2)}\right)
d\phi_{m+2}\big(p_1,p_2\big)
\nonumber\\*&+&
\aemotpi\,{\cal Q}_{\gamma\lep}^{(\deltaI)^\prime}(z_2)
\left(\frac{1}{1-y_1}\right)_{\deltaI}\!
\left((1-y_1)\ampsq_{\lep\gamma}^{(m+1)}\right)
d\phi_{m+1}\big(p_1,z_2p_2\big)\,dz_2
\nonumber\\*&+&
\aemotpi\,{\cal Q}_{\gamma\lep}^{(\deltaI)^\prime}(z_1)
\left(\frac{1}{1-y_2}\right)_{\deltaI}\!
\left((1-y_2)\ampsq_{\gamma\lep}^{(m+1)}\right)
d\phi_{m+1}\big(z_1p_1,p_2\big)\,dz_1
\nonumber\\*&+&
\left(\aemotpi\right)^2\,
{\cal Q}_{\gamma\lep}^{(\deltaI)^\prime}(z_1)\,
{\cal Q}_{\gamma\lep}^{(\deltaI)^\prime}(z_2)
\ampsq_{\gamma\gamma}^{(m)}\,
d\phi_{m}\big(z_1p_1,z_2p_2\big)\,dz_1\,dz_2\,.\phantom{aaa}
\label{final00NNLO2}
\eeqn
By construction, this short-distance cross section is independent
of the arbitrary parameter $\deltaI$. It can be used to define the 
relative $\ord(\aem^2)$ improvement we have been seeking, thus:
\beq
\Delta d\hat{\sigma}_{\dNNLOG}(p_1,p_2)=
d\zeta_1\,d\zeta_2\,\PDF{\lep}{\lep}(\zeta_1)\,\PDF{\lep}{\lep}(\zeta_2)\,
d\hat{\Sigma}_{\dNNLOG}(\zeta_1 p_1,\zeta_2 p_2)\,,
\label{sNNLOfinal}
\eeq
where we have convoluted the short-distance cross section of 
eq.~(\ref{final00NNLO2}) with the lepton PDFs relevant to each incoming leg.
We remind the reader that the matrix elements that appear in the
four lines of eq.~(\ref{final00NNLO2}) are the complete ones (i.e.~they
are not limited to the contributions of the graphs that feature $t$-channel
vector-boson exchanges) relevant to the processes of eqs.~(\ref{VBFproc}),
(\ref{mugaWW}), (\ref{gamuWW}), and~(\ref{gagaWW}), respectively.

Equations~(\ref{final00NNLO2}) and~(\ref{sNNLOfinal}) constitute
a physical result as long as the two rightmost terms in 
eq.~(\ref{Msplit}) can be identified with the relevant contributions
to an NLO and a LO cross section, respectively. This is now easy
to do. Specifically, as far as eq.~(\ref{final00NLO}) is concerned, 
one starts by rewriting it in the natural frames (according to the FKS
subtraction formalism) suggested by the arguments of the various phase
spaces. By employing again the results of appendix~\ref{sec:frames} and of
appendix~\ref{sec:distr} (that is, ultimately by performing an operation 
which is the inverse of that of sect.~\ref{sec:sides}), and the identity 
of eq.~(\ref{ypvsyd}) we obtain:
\beqn
d\hat{\Sigma}_{\dNLOG}(p_1,p_2)&=&
\PDF{\gamma}{\lep}(z_2)
\left(\frac{1}{1-\yb_1}\right)_{\bdeltaI}\!
\left((1-\yb_1)\ampsq_{\lep\gamma}^{(m+1)}\right)
d\phi_{m+1}\big(p_1,z_2p_2\big)\,dz_2
\nonumber\\*&+&
\PDF{\gamma}{\lep}(z_1)
\left(\frac{1}{1-\yb_2}\right)_{\bdeltaI}\!
\left((1-\yb_2)\ampsq_{\gamma\lep}^{(m+1)}\right)
d\phi_{m+1}\big(z_1p_1,p_2\big)\,dz_1
\nonumber\\*&+&
\aemotpi
\Big({\cal Q}_{\gamma\lep}^{(\bdeltaI)^\prime}(z_1;z_2s)
\PDF{\gamma}{\lep}(z_2)+
\PDF{\gamma}{\lep}(z_1)
{\cal Q}_{\gamma\lep}^{(\bdeltaI)^\prime}(z_2;z_1s)\Big)
\nonumber\\*&&\phantom{aaaaaaaaa}\times
\ampsq_{\gamma\gamma}^{(m)}\,
d\phi_{m}\big(z_1p_1,z_2p_2\big)\,dz_1\,dz_2\phantom{aaa}
\label{final00NLO2}
\\*&\equiv&
d\zeta_1\,\PDF{\gamma}{\lep}(\zeta_1)\,
d\hat{\Sigma}_{\dNLOG}^{(1)}(\zeta_1 p_1,p_2)
+d\zeta_2\,\PDF{\gamma}{\lep}(\zeta_2)\,
d\hat{\Sigma}_{\dNLOG}^{(2)}(p_1,\zeta_2 p_2)\,,
\eeqn
where in the rightmost side we have introduced\footnote{We point out that 
we always define $s=(p_1+p_2)^2$, with $p_1$ and $p_2$ the arguments of the
function we are considering. Ultimately, we want $s$ to be the
partonic c.m.~energy, i.e.~the quantity to be rescaled by the 
product of the appropriate Bjorken $x$'s, $\zeta_i$.}:
\beqn
d\hat{\Sigma}_{\dNLOG}^{(1)}(p_1,p_2)&=&
\left(\frac{1}{1-\yb_2}\right)_{\bdeltaI}\!
\left((1-\yb_2)\ampsq_{\gamma\lep}^{(m+1)}\right)
d\phi_{m+1}\big(p_1,p_2\big)
\nonumber\\*&+&
\aemotpi
{\cal Q}_{\gamma\lep}^{(\bdeltaI)^\prime}(z_2;s)\,
\ampsq_{\gamma\gamma}^{(m)}\,
d\phi_{m}\big(p_1,z_2p_2\big)\,dz_2\,,\phantom{aaa}
\label{dSNLO1}
\\*
d\hat{\Sigma}_{\dNLOG}^{(2)}(p_1,p_2)&=&
\left(\frac{1}{1-\yb_1}\right)_{\bdeltaI}\!
\left((1-\yb_1)\ampsq_{\lep\gamma}^{(m+1)}\right)
d\phi_{m+1}\big(p_1,p_2\big)
\nonumber\\*&+&
\aemotpi{\cal Q}_{\gamma\lep}^{(\bdeltaI)^\prime}(z_1;s)
\ampsq_{\gamma\gamma}^{(m)}\,
d\phi_{m}\big(z_1p_1,p_2\big)\,dz_1\,,\phantom{aaa}
\label{dSNLO2}
\eeqn
and have renamed $z_1\to\zeta_1$ ($z_2\to\zeta_2$) in the second and
forth (first and third) terms on the r.h.s.~of eq.~(\ref{final00NLO2}),
while including such terms in eqs.~(\ref{dSNLO1}) and~(\ref{dSNLO2}), 
respectively.
The two quantities in eqs.~(\ref{dSNLO1}) and~(\ref{dSNLO2}) are
separately independent of the arbitrary parameter $\bdeltaI$; note the 
augmented dependence of ${\cal Q}_{\gamma\lep}^\prime$, which is due to
the same mechanism as that which leads to eq.~(\ref{Qpextrad}). This fact, 
the functional forms, the variables used, render it now easy to see that the 
two terms of the following expression:
\beqn
&&d\zeta_1 d\zeta_2\,\PDF{\lep}{\lep}(\zeta_1)\,\PDF{\gamma}{\lep}(\zeta_2)\,
d\hat{\Sigma}_{\dNLOG}^{(2)}(\zeta_1 p_1,\zeta_2 p_2)
\nonumber\\*&+&
d\zeta_1 d\zeta_2\,\PDF{\gamma}{\lep}(\zeta_1)\,\PDF{\lep}{\lep}(\zeta_2)\,
d\hat{\Sigma}_{\dNLOG}^{(1)}(\zeta_1 p_1,\zeta_2 p_2)
\label{ourNLO}
\eeqn
coincide with the FKS $\dNLO$ real-correction contributions for the 
processes whose tree-level graphs are associated with 
eqs.~(\ref{mugaWW}) and~(\ref{gamuWW}), respectively.

Finally, it is self-evident that eq.~(\ref{final00LO}) is
already in the form of the LO cross section for the process
of eq.~(\ref{gagaWW}). Therefore, we have established that the
decomposition of eq.~(\ref{Msplit}) gives rise to a genuine 
(i.e.~that does not double count) $\dNNLO$ contribution ultimately in
the form of eq.~(\ref{sNNLOfinal}), which can therefore be safely
summed to a complete NLO-accurate cross section. We write this
as follows, by employing the notation of eq.~(\ref{fact0}):
\beqn
d\sigma(p_1,p_2)&=&\sum_{ij}\PDF{i}{\lep}(\zeta_1)\,\PDF{j}{\lep}(\zeta_1)
\left(d\hsig_{ij}^{[0]}(\zeta_1 p_1,\zeta_2 p_2)+
\aemotpi\,d\hsig_{ij}^{[1]}(\zeta_1 p_1,\zeta_2 p_2)\right)
d\zeta_1\,d\zeta_2
\nonumber\\*&+&
\Delta d\hat{\sigma}_{\dNNLOG}(p_1,p_2)\,.
\label{xsNNLOimp}
\eeqn
Note that the sums in the first line of eq.~(\ref{xsNNLOimp}) extend
to all parton types, and are not restricted to leptons of the same 
flavours as the incoming ones. This is in keeping with what the 
factorisation theorem dictates, and shows explicitly how 
eq.~(\ref{xsNNLOimp}) is valid everywhere in the phase space, where
it is at least NLO-accurate, with the accuracy increasing effectively
to NNLO in the $\gamma\gamma$-dominated regions, thanks to the rightmost 
term on the r.h.s.. In fact, eq.~(\ref{xsNNLOimp}) can be re-written in
a form where its being {\em the} factorisation-theorem formula, albeit
featuring only a subset of the NNLO contributions, is more manifest. This
can be done by observing that the matrix elements which appear
in eq.~(\ref{final00NNLO2}) are such that:
\beq 
\ampsq_{ab}^{(m+q)}\;\propto\;\aem^b\aem^q\,,\;\;\;\;\;\;\;\;
q=0,1,2\,.
\label{Mapref}
\eeq
By also taking into account the explicit $\aem$-dependent prefactors
in eq.~(\ref{final00NNLO2}), eq.~(\ref{Mapref}) motivates the definition 
of the NNLO analogue of the LO and NLO short distance cross sections which 
appear in the first line of eq.~(\ref{xsNNLOimp}), thus:
\beq
\left(\aemotpi\right)^2\!d\hsig_{\dNNLOG}^{[2]}(p_1,p_2)=
d\hat{\Sigma}_{\dNNLOG}(p_1,p_2)\,,
\label{dsig2Gdef}
\eeq
which leads us to:
\beqn
d\sigma(p_1,p_2)&=&
d\zeta_1\,d\zeta_2
\sum_{ij}\PDF{i}{\lep}(\zeta_1)\,\PDF{j}{\lep}(\zeta_1)
\nonumber
\\*&&\phantom{a}\times
\Bigg(d\hsig_{ij}^{[0]}(\zeta_1 p_1,\zeta_2 p_2)+
\aemotpi\,d\hsig_{ij}^{[1]}(\zeta_1 p_1,\zeta_2 p_2)
\nonumber
\\*&&\phantom{aaaaa}+
\delta_{i\lep}\delta_{j\lep}
\left(\aemotpi\right)^2\!d\hsig_{\dNNLOG}^{[2]}(\zeta_1 p_1,\zeta_2 p_2)
\Bigg),
\label{xsNNLOimprep3}
\eeqn
which is the highly symmetric expression we were seeking.

We conclude by returning to the comment made in footnote~\ref{ft:tt}. By 
introducing the quantities which appear on the r.h.s.~of eq.~(\ref{Msplit}),
and by treating them independently from one another (in particular as
far as the integration over the phase space is concerned), the {\em local}
cancellations between the two terms on the l.h.s.~of eq.~(\ref{GmQ}) does
not occur any longer. Such cancellations are relevant to two quantities:
the hard scale $\mu$, and the change-of-scheme function $K_{\gamma\lep}$.
Rather than being a problem, this is actually an opportunity to check
that the sum of the terms on the r.h.s.~of eq.~(\ref{xsNNLOimprep3}) is more
stable w.r.t.~changes of scales and scheme than its individual summands.
Note, in particular, that in the context of a strict NLO-accurate computation 
(e.g.~one where only the first two terms within the round brackets on the 
r.h.s.~of eq.~(\ref{xsNNLOimprep3}) are retained) the scheme dependence of 
$\gamma\gamma$-channel contributions is beyond accuracy, being
of relative $\ord(\aem^2)$ (each $K_{\gamma\lep}$ function is multiplied
by an $\aem$ factor, and there are no $\ord(\aem^0)$ contributions
to the photon PDF). We have shown that such scheme-dependent terms are 
exactly cancelled, at least in the $\gamma\gamma$-dominated regions, by the
cross section in the second line of eq.~(\ref{xsNNLOimp}). We shall
give an actual numerical evidence of this fact in sect.~\ref{sec:resfact}.

\subsection{Results\label{sec:res}}
In this section we present a few sample results which document the
impact of the technical cuts (sect.~\ref{sec:rescuts}) and of
the choice of factorisation scheme (sect.~\ref{sec:resfact}) 
on actual observables.

The $\dNNLOG$ cross section of eq.~(\ref{final00NNLO2}) and (where relevant) 
the $\dNLOG$ one of eq.~(\ref{ourNLO}) have been implemented in a 
non-automated manner in a dedicated code (some details of which are 
given in appendices~\ref{sec:impl} and~\ref{sec:azicorr}) for the two
processes we have considered in this paper, namely $t\bt$ and $W^+W^-$
production. Aside from this implementation, the LO and NLO cross sections 
are automatically obtained by employing
{\tt MadGraph5\_aMC@NLO}~\cite{Alwall:2014hca}, and specifically
its EW-corrections capabilities~\cite{Frederix:2018nkq,
Bertone:2022ktl}\footnote{We remind the reader that while the first 
automation of EW corrections in {\tt MadGraph5\_aMC@NLO} had been 
presented in ref.~\cite{Frederix:2018nkq}, it did not apply to lepton
collisions, but only to hadron ones. The relevant extension has been 
achieved in ref.~\cite{Bertone:2022ktl}.}.
The basic parameters used for our simulations are as follows:
\beqn
m_W&=&80.419~\GeV\,,
\\*
m_Z&=&91.188~\GeV\,,
\\*
m_t&=&173.3~\GeV\,,
\\*
m_H&=&125~\GeV\,.
\eeqn
All widths are set equal to zero. Beyond-LO results are obtained
in the context of the $G_\mu$ renormalisation scheme~\cite{Sirlin:1980nh},
with
\beq
G_f=1.16639\mydot 10^{-5}~\GeV^{-2}
\;\;\;\;\Longrightarrow\;\;\;\;
\aem=1/132.5070\,.
\eeq
The hard scale is generally set equal to the pair invariant mass of the 
massive system produced (i.e.~either $t\bt$ or $W^+W^-$), with additional
non-dynamical choices employed in sect.~\ref{sec:rescuts}. We consider 
$\mu^+\mu^-$ collisions, with the muon NLL PDFs taken from 
refs.~\cite{Frixione:2023gmf,Bonvini:2025xxx}.

\subsubsection{Impact of the technical cuts\label{sec:rescuts}}
The results concerning the impact of the technical cuts on 
observables which we present in this section are in keeping
with what has been discussed at the end of sect.~\ref{sec:tech}
and in sect.~\ref{sec:cuts}. We have considered $t\bt$ and $W^+W^-$
production in $\mu^+\mu^-$ collisions at different collider energies
$\sqrt{S}$. Lest there should be a proliferation of plots, and given that 
the conclusions we shall reach are essentially process-independent, we
show here the results for $t\bt$ production at $\sqrt{S}=3$~TeV and
$10$~TeV, while those relevant to $W^+W^-$ production are collected
in appendix~\ref{sec:extrapl}. In both cases, we have chosen the pair
invariant mass and the transverse momentum of the positively-charged
outgoing heavy particle as representative observables; as is the case
for the production process, the conclusions would not change if other 
observables (such as pair and single-inclusive rapidity and pseudorapidity,
or pair transverse momentum, all of which we have studied) were shown.

The essence of the idea discussed in sect.~\ref{sec:tech} which we shall
employ here is that of a direct comparison between massive- and 
massless-lepton predictions. The strictest version of such a comparison
at the level of $2\to 2+m$ tree-level matrix elements is the one that uses 
eq.~(\ref{final00}), with its l.h.s.~as is (this is the massive-lepton 
calculation), and its r.h.s.~with the photon PDFs replaced by their 
$\ord(\aem)$ coefficients \mbox{($\aem/(2\pi)\PDF{\gamma}{\lep}^{[1]}$)},
and no convolution with muon PDFs\footnote{We point out that this is
equivalent to convoluting with the first-order term of the perturbative
expansion of the muon PDF, i.e.~with $\delta(1-z)$. This is in keeping
with the strict fixed-order comparison strategy we are following here.}
(this is the massless-lepton computation; as its massive counterpart, 
it is then exactly of $\ord(\aem^{2+m})$). In other words, the latter
is $\NNLOG$ with the PDFs replaced by their $\ord(\aem)$ coefficients;
with abuse of notation, we still denote it by $\NNLOG$. Conversely, the
results of the former are denoted by MG, having been obtained 
by means of an {\tt MadGraph5\_aMC@NLO} branch which allows one to keep
the light-lepton masses different from zero. In the text, these two
simulations will be loosely referred to as ``massive'' and ``massless''.

The technical cuts have been applied in the same manner to the massive
and massless predictions, with the outgoing leptons dealt with as was
explained in sect.~\ref{sec:cuts}. We have considered several combinations
of cuts. As far as the lepton-pair invariant mass is concerned, and in order
to magnify any possible cut dependence, we have chosen (see 
eqs.~(\ref{massc1}) and~(\ref{massc2})):
\beqn
&&\delta_{M_1}=120~~{\rm or}~~200~\GeV\,,
\\
&&\delta_{M_2}=\delta_{M_3}=0\,.
\eeqn
As hemisphere cuts, we have considered (see eq.~(\ref{hemic}))
\beq
\eta_c=2.5~~{\rm or}~~5\,,
\eeq
as well as the rather draconian choice of eliminating at the level 
of amplitudes the $s$-channel resonant diagrams, i.e.~those in which
either the $\gamma$ or the $Z$ splits into a $\mu^+\mu^-$ pair.
In this way, we have obtained six different cut combinations, the bottom
line of which is that the hemisphere cuts have a negligible effect
when these invariant-mass cuts are also applied (in other words, the latter
also act very effectively on anti-collinear configurations). Because of
this, we shall only show (as representatives of worst-case scenarios)
the predictions obtained with the two invariant mass cuts and no hemisphere
cuts, and with the loosest hemisphere cuts ($\eta_c=5$), and the removal
of the $s$-channel resonant diagrams\footnote{For many processes, including 
but not limited to those studied in this paper, such a removal can be 
achieved in a gauge-invariant way by considering the production of the
relevant final state as initiated by a fictitious $\mu^\pm e^\mp$ pair,
with the electron mass set equal to the muon mass.}, calling them as follows:
\beqn
&&{\rm scenario}~1:\;\;\;m(\mu^+\mu^-)\ge 120~\GeV\,,
\label{cs1}
\\*
&&{\rm scenario}~2:\;\;\;m(\mu^+\mu^-)\ge 200~\GeV\,,
\label{cs2}
\\*
&&{\rm scenario}~3:\;\;\;\eta_c=5\,,\;\;\;{\rm resonance~removal}\,.
\label{cs3}
\eeqn
In order to be clear: we do not necessarily advocate such a diagram removal 
(nor, to a lesser extent, invariant mass cuts without hemisphere cuts) 
as a solution to be adopted when computing cross sections of 
phenomenological interest. Here, we use it precisely because we are 
not concerned about phenomenology but rather by the theoretical viability 
of the strategy proposed in this work, which we are trying to break 
by means of extreme setups.
In keeping with this principle, for each cut scenario we have considered
the following three hard-scale choices for the massless calculation:
\beq
\mu=10~\GeV;\;\;\;\;\;\;
\mu=1000~\GeV;\;\;\;\;\;\;
\mu=m(t\bt)~~{\rm or}~~m(W^+W^-)\,.
\label{scalech}
\eeq
Among these, the only one of phenomenological relevance is the third,
while the first one is clearly an outlier at a multi-TeV collider.

\begin{figure}[htb]
  \begin{center}
  \includegraphics[scale=1.0,width=0.48\textwidth]{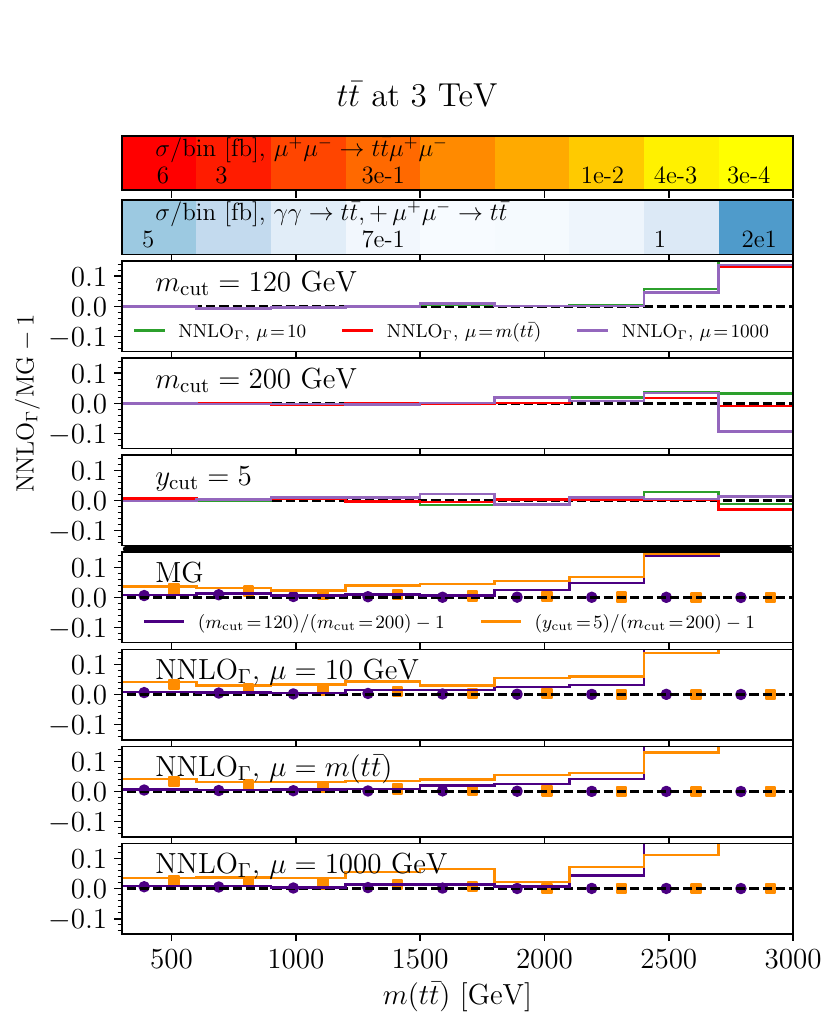}
$\phantom{a}$
  \includegraphics[scale=1.0,width=0.48\textwidth]{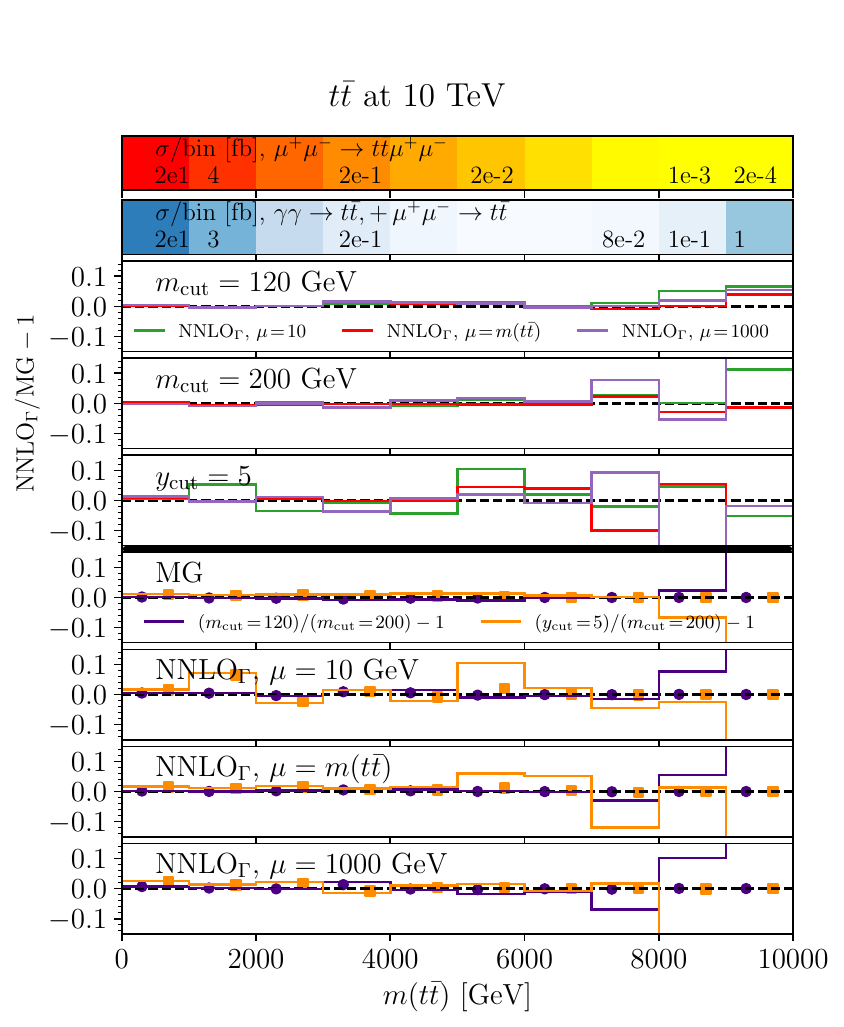}
\caption{\label{fig:tcuts1}
Impact of technical cuts on the $t\bt$ invariant mass, at $\sqrt{S}=3$~TeV
(left panel) and 10~TeV (right panel). See the text for details.
}
  \end{center}
\end{figure}
\begin{figure}[htb]
  \begin{center}
  \includegraphics[scale=1.0,width=0.48\textwidth]{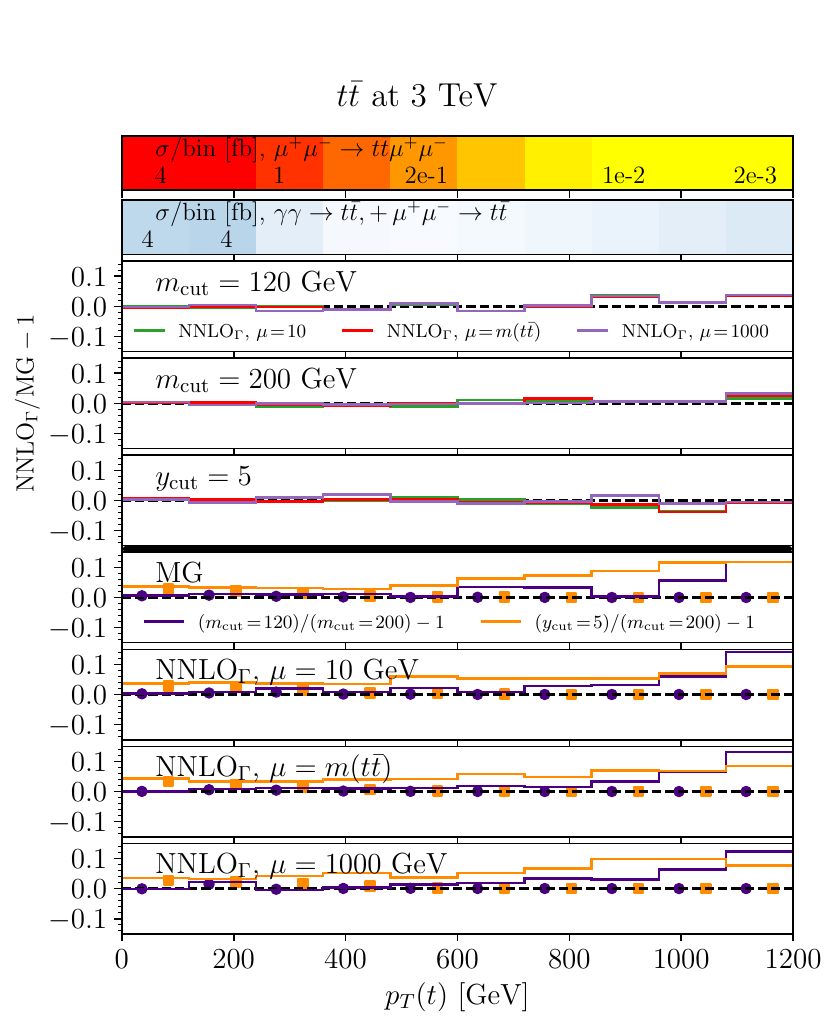}
$\phantom{a}$
  \includegraphics[scale=1.0,width=0.48\textwidth]{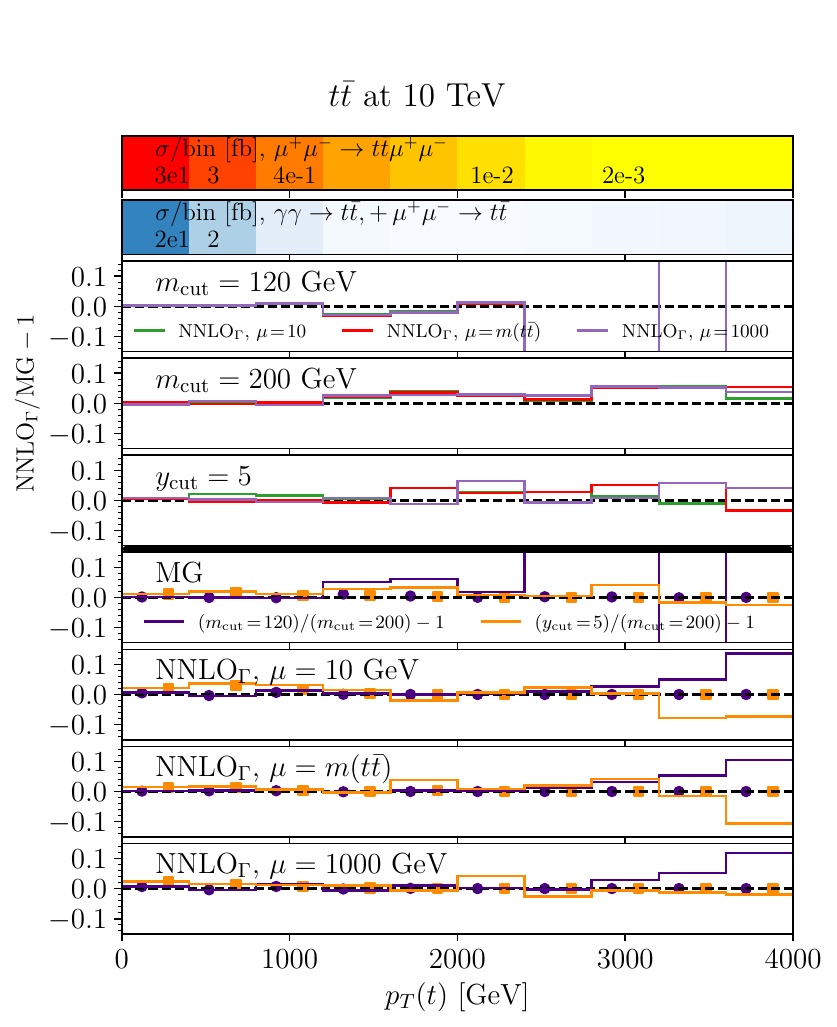}
\caption{\label{fig:tcuts2}
As in fig.~\ref{fig:tcuts1}, for the transverse momentum of the top quark.
}
  \end{center}
\end{figure}

We now turn to discussing the results we have obtained; as was said 
above, we shall comment here on $t\bt$ production (figs.~\ref{fig:tcuts1}
and~\ref{fig:tcuts2} for the pair invariant mass and the single-inclusive
$\pt$, respectively), with additional plots relevant to $W^+W^-$ production 
presented in appendix~\ref{sec:extrapl} (figs.~\ref{fig:tcuts3}
and~\ref{fig:tcuts4}). Irrespective of the production process and
of the observable, all of the figures have the same layout. At the very
top there are two bands, which give one a rough indication of the differential
rates, in fb per bin, predicted by: {\em a)} the massive $2\to 4$ 
computation\footnote{In turn essentially identical, as we shall see, to 
the massless one emerging from of eq.~(\ref{final00}) modified as was 
explained above.} (upper band), and: {\em b)} the Born-level simulation 
of $t\bt$ production stemming from the sum of the $\gamma\gamma$ and
$\mu^+\mu^-$ channels (lower band). In both bands, darker (lighter) hues 
correspond to larger (smaller) cross sections; the upper band uses red tones,
the lower one blue tones, so that the same cross section in absolute value
is associated with two different colours in the two bands. The actual value
of the cross section is superimposed to the bands for a few selected bins.

The main frame of the figures is composed of two parts, separated by a 
thick black line, both of which display ratios of cross sections. In the 
upper part of the main frame there are three ratio plots, with the ratios 
obtained by dividing a massless prediction by the corresponding massive 
one, and by subtracting one from the result thus obtained (therefore, an 
horizontal line at zero would mean perfect agreement between the two 
predictions). These three ratio plots are obtained with the 
technical-cut scenarios of eqs.~(\ref{cs1}), (\ref{cs2}), and~(\ref{cs3})
for the upper, middle, and lower plot, respectively. Each plot shows three 
histograms, each of which is the ratio computed according to what has been 
just explained, after choosing the hard scale as is indicated in 
eq.~(\ref{scalech}) (with the green and violet histograms stemming from
the first and second options, and the red histograms from the third one).
Conversely, the lower part of the main frame presents four ratio plots, where 
again we subtract one from the actual ratio, so as to have a line at zero in 
the case of a perfect agreement between the numerator and the denominator. 
In each plot, the dark blue (orange) histogram corresponds to the ratio of the
predictions obtained by dividing the result of cut scenario~1 (scenario~3) 
by that of cut scenario~2. The symbols are obtained from the histograms
of the same colour (with circles and boxes used in the case of the blue
and orange colours, respectively), by locally (i.e.~bin-by-bin) multiplying
the latter by the ratio of the Born cross section that originates from the
$\gamma\gamma$ channel over the total Born cross section (which is the
sum of those emerging from the $\gamma\gamma$ and $\mu^+\mu^-$ channels) --
as will be explained later, the symbols have a more direct connections with
physical observables w.r.t.~to the corresponding histograms, and will be
crucial to establish the fact that the technical cuts are ultimately harmless.
Roughly speaking, in a given bin this ratio of Born cross sections is close 
to one when the two coloured bands at the top of the figure have a similar
(dark or light) hue in that bin, and close to zero when the upper coloured 
band has a much lighter hue than the lower band. The layout of the four
plots in the lower part of the main frame that we have just described
is identical -- they differ in that they correspond to different types
of simulations, namely: the upper plot is obtained with the massive
calculation in both the numerator and denominator, whereas the three lower 
plots feature the massless predictions at the numerator and denominator,
and differ only by the hard-scale choice.

We start by commenting the results for the pair invariant mass of
fig.~\ref{fig:tcuts1}, with the left and right panel there obtained
with $\sqrt{S}=3$~TeV and $\sqrt{S}=10$~TeV, respectively. As far
as the cross sections are concerned, from the upper coloured band we
see that the $\gamma\gamma$-induced one falls by about four (five)
orders of magnitude at 3~TeV (10~TeV) when moving from threshold
(the left-hand side of the plot) to the kinematic limit (the right-hand side
of the plot). Conversely, the total Born cross section (lower coloured
band) has a global minimum at about 2~TeV at $\sqrt{S}=3$~TeV (7~TeV
at $\sqrt{S}=10$~TeV), with similar values at the threshold and
at the kinematic limit. This implies that, on the whole range, both
production mechanisms are essential to obtain sensible predictions,
with the $\gamma\gamma$ channel dominant at the threshold (its importance
increasing with the c.m.~energy), and the $\mu^+\mu^-$ one dominant at 
the kinematic limit. The three plots in the upper part of the main frame
are, up to statistical fluctuations, basically indistinguishable from one 
another, and all of the histograms there are compatible with being equal to 
zero. Therefore, we conclude that, regardless of the cut scenario and 
hard-scale choice, the behaviours of the massive and massless calculations are 
identical, which is a direct indication that eq.~(\ref{final00}) is correct.

As far as the four plots in the lower part of the main frame are concerned,
the overarching conclusion is that their patterns are identical -- again,
the massive and the massless calculations behave the same, regardless of
the choice of the hard scale in the latter. Having said that, we remark
that such patterns, in the case of the histograms at $\sqrt{S}=3$~TeV, 
are less trivial than those of the plots in the upper part of the main 
frame. Specifically, we see that both histograms move away from zero
when moving towards the kinematic limit, and that the orange histograms
also differ from zero (by $1-2$\%) near threshold. This is the signal
of a dependence on the cut scenario adopted. While it has been stressed
how these scenarios are rather extreme, this is still a potentially
disturbing fact, since it renders it difficult to exclude technical-cut
dependences even for sensible choices of such cuts. Fortunately, the
dependence we observe is totally irrelevant at the level of physical
observables -- this is borne out by the fact that the symbols
are all compatible with being exactly equal to zero. This can only 
happen because when the histograms move away from zero (while the symbols
are equal to zero), they do so in regions where the $\gamma\gamma$ 
production channel that underpins them become negligible w.r.t.~the 
$\mu^+\mu^-$ one. Conversely, when the $\gamma\gamma$-channel
contributions are non-negligible or dominant, we observe a complete
independence of the technical cuts (this is particularly striking
when considering the differences between the predictions at 
$\sqrt{S}=3$~TeV and $\sqrt{S}=10$~TeV, with the histograms in the
latter case being very close to zero almost on the entire range).
Taken together, these two facts give one a crucial message: namely,
that while technical cuts are always irrelevant, they are so for
basic physics reasons that are process- and observable independent.
Namely, their impact is negligible in those kinematic regions where
the underlying Born process (from which the would-be singular branching
emerges) is initiated by $\gamma\gamma$ fusion. When that impact
is visible, the underlying Born process is a $\mu^+\mu^-$-annihilation one.
That being the case, the corresponding contribution will be completely
swamped by the contributions of genuine $\mu^+\mu^-$ processes,
which are both convoluted with the muon PDFs and are perturbatively
dominant (i.e.~relatively enhanced by a coupling-constant factor).

A corroborating evidence of what we have just said emerges from 
considering different observables (see e.g.~fig.~\ref{fig:tcuts2})
and different processes (see~appendix~\ref{sec:extrapl}).
The same discussion as the one we presented for fig.~\ref{fig:tcuts1}
could be repeated in the case of fig.~\ref{fig:tcuts2}. What is 
interesting is to remark that the same conclusions emerge from
different mechanisms, in the sense that the $\gamma\gamma$-fusion
vs the $\mu^+\mu^-$ annihilation balance is different in the case of the 
transverse momentum of the top quark, presented in fig.~\ref{fig:tcuts2},
w.r.t.~to the case of the pair invariant mass discussed before.
This is {\em a fortiori} true for longitudinally-dominated observables
such as rapidities, which also follow the same pattern as the one
shown explicitly above.

Therefore, the take-home message that should be retained is that
the technical cuts are harmless, physics-wise -- this renders
the procedure advocated in this paper viable, and under firm 
theoretical control. From the purely technical viewpoint we can
also extract another message, namely: in the regions where the
technical cuts have a visible effect on the subdominant $\gamma\gamma$
channel, the driving mechanism is that associated with the final-state
$s$-channel branchings, over that stemming from initial-state
anti-collinear configurations. This gives one the possibility of
addressing the problem in a relatively easy way, since $s$-channel
resonances are more straightforward to address w.r.t.~anti-collinear
singularities (which have a genuine NNLO topology). It remains true
that a solution of this kind will not have any impact on physics
predictions\footnote{By being sufficiently perverse one can always find 
a region of the phase space where this is not true any longer. However, 
as we have shown, such a region will be associated with a cross sections 
many orders of magnitude smaller than the one at the peak(s).}.
Finally, we remark that what has been done here constitutes a mutual
consistency check of the massive-lepton branch of {\tt MadGraph5\_aMC@NLO},
and of the non-automated implementation of eqs.~(\ref{final00NNLO2}) 
and~(\ref{xsNNLOimprep3}) achieved for the first time during the course
of this work.

\subsubsection{Choice of factorisation scheme\label{sec:resfact}}
We now turn to showing the impact of the choice of factorisation
scheme, following the formal discussion given in sect.~\ref{sec:scheme}. 
In order to allow for a direct connection with the results shown 
previously, we shall consider a subset of the same observables as those 
already analysed in sect.~\ref{sec:rescuts}. Since the issue of scheme
dependence has a very direct bearing on phenomenological predictions,
the interested reader is urged to check the discussion on this matter 
which we give in ref.~\cite{Frixione:2025xxx}.

The factorisation-scheme dependence is assessed by comparing
predictions computed in two different factorisation schemes, namely 
$\MSb$~\cite{Bardeen:1978yd} and $\Delta$~\cite{Frixione:2021wzh}.
At variance with what was done in sect.~\ref{sec:rescuts}, such
predictions will always be obtained by means of our master formula,
eq.~(\ref{xsNNLOimp}), i.e.~with a massless calculation which is of
NLO, and is NNLO-improved at its most accurate -- in other words, the evolved
PDFs are employed, rather than their first-order coefficients, in order
to obtain physical predictions. Since the factorisation-scheme 
dependence grows with energy, we present here the results obtained
with \mbox{$\sqrt{S}=10$~TeV} in order to show the largest effects among
those we have studied.

\begin{figure}[htb]
  \begin{center}
  \includegraphics[scale=1.0,width=0.48\textwidth]{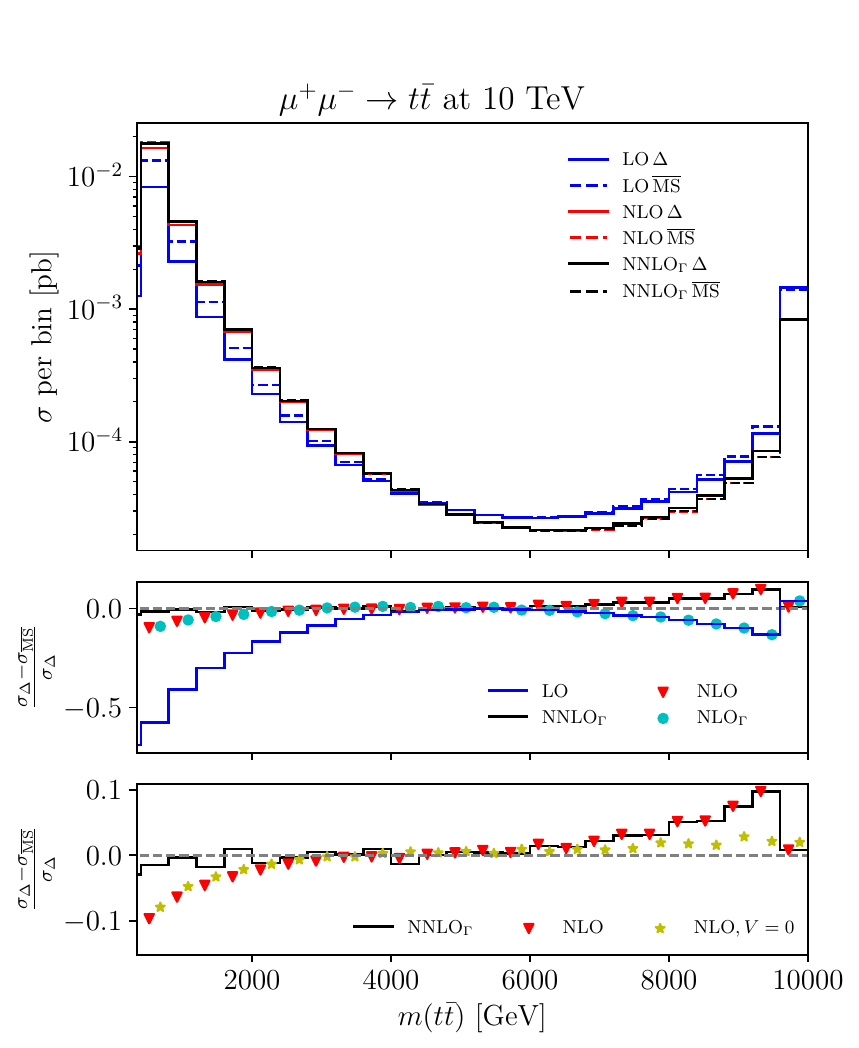}
$\phantom{a}$
  \includegraphics[scale=1.0,width=0.48\textwidth]{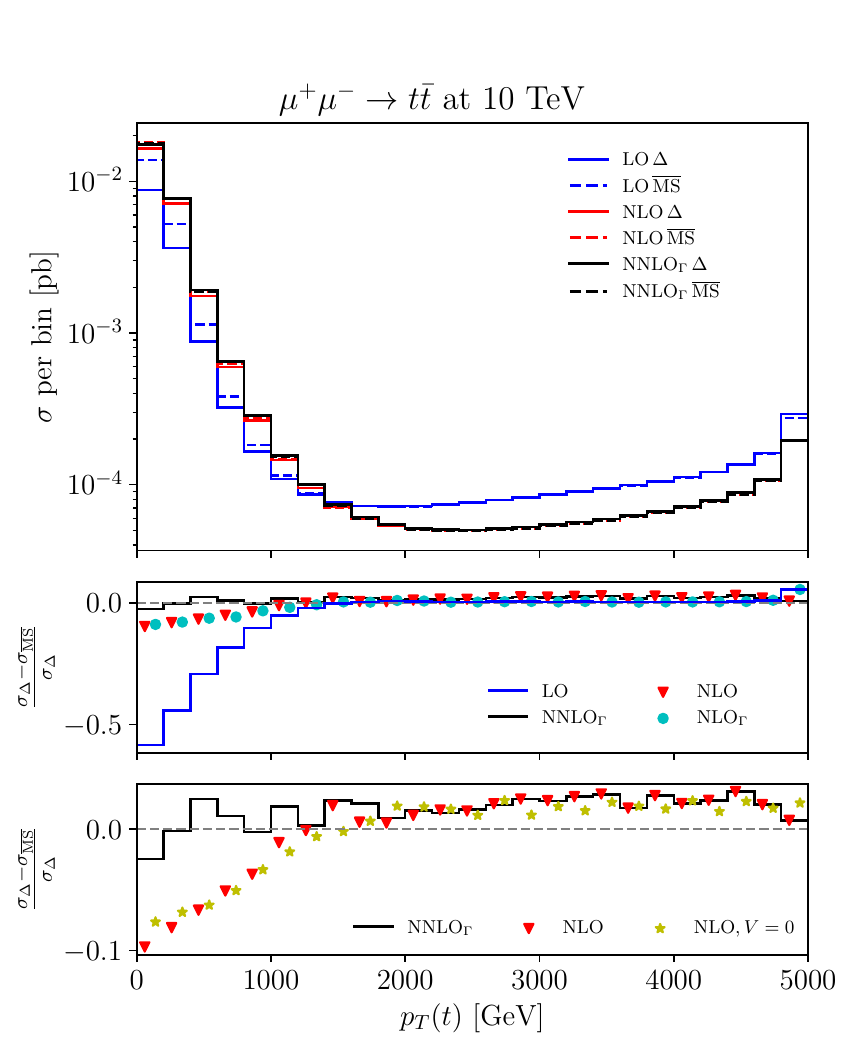}
\caption{\label{fig:scheme1}
Impact of factorisation-scheme choice on the $t\bt$ invariant mass
(left panel), and the transverse momentum of the top quark (right panel),
at $\sqrt{S}=10$~TeV. See the text for details.
}
  \end{center}
\end{figure}
The results for the pair invariant mass and the single-inclusive transverse
momentum in $t\bt$ production are presented in the left and right panel
of fig.~\ref{fig:scheme1}, respectively, while analogous observables 
in $W^+W^-$ production are reported in appendix~\ref{sec:extrapl} 
(fig.~\ref{fig:scheme3}). All of the plots in these figures have 
the same layout, as follows. In the main frame the
differential cross section (in pb per bin) is shown in absolute value.
There are six histograms, which correspond to an increasing level of
perturbative accuracy (LO: blue histograms; NLO: red histograms; 
NNLO: black histograms), and to the two different factorisation-scheme
choices (solid histograms: $\Delta$ scheme; dashed histograms: $\MSb$
scheme). Below the main frame there are two insets, where the relative
differences associated with the factorisation-scheme choice are displayed
according to the following definition:
\beq
\frac{d\sigma_{\Delta}-d\sigma_{\MSb}}{d\sigma_{\Delta}}\,,
\label{relfrac}
\eeq
which must be understood in a local sense (i.e.~the ratio is computed
bin-by-bin). The insets have almost identical contents, but the lower
one has a reduced range on the $y$ axis, and thus constitutes a zoomed-in
view of the upper inset. The results that appear in the insets are
relevant to an increasing level of perturbative accuracy, as those in the
main frame; specifically, they are all computed by means of 
eq.~(\ref{relfrac}), where the two quantities $d\sigma_{\Delta}$ and 
$d\sigma_{\MSb}$ are the solid and dashed histograms, respectively, in the 
main frame which have the same colour as either the histograms (black 
for NNLO, and blue for LO) or the symbols (red triangles for NLO) in the 
insets\footnote{Symbols rather than histograms are employed in order to 
improve visibility, given the proximity of the NLO and NNLO predictions.}. 
Furthermore, there is an additional set of symbols in the upper (blue
circles) and lower (yellow stars) insets. The former is equal to $\NLOG$
(i.e.~it is computed with eq.~(\ref{ourNLO}) for the NLO corrections), while 
the latter is identical to the NLO result except for the fact that the 
finite part of the virtual contribution, defined according to the conventions 
of ref.~\cite{Frixione:1995ms}, is set equal to zero.

We begin by commenting on the predictions for the pair invariant mass (left
panel of fig.~\ref{fig:scheme1}). The defining feature of the main frame is the
fact that the distribution has two peaks, at the threshold\footnote{The fact 
that the leftmost bin is lower than the bin to its right is simply due to 
the fact that its left border is smaller than the actual $t\bt$ mass 
threshold.}  and at the kinematic limit which, as was already 
pointed out in sect.~\ref{sec:rescuts},
are the results of the dominance of the $\gamma\gamma$ and of the
$\mu^+\mu^-$ channel, respectively. We observe that the global maximum 
(i.e.~the highest of the two peaks) changes from that at the right to that 
at the left when increasing the c.m.~energy (e.g.~at $\sqrt{S}=3$~TeV
is the one at the right which is highest) -- as is known,
larger c.m.~energies allow the hard process to probe increasingly smaller
Bjorken $x$'s, where the photon density grows fast. The relative heights
of the two peaks also change with the perturbative precision at a
given c.m.~energy, with the one at threshold (at the kinematic limit)
higher (lower) at the NLO(NNLO) w.r.t.~the LO result. In general, the 
changes induced by the inclusion of higher-order corrections are {\em very} 
significant, and such corrections are therefore essential in order to obtain
sensible physical predictions. This is reflected on the topic which is
our primary interest here, namely the factorisation-scheme dependence,
whose actual size is best seen in the insets. At the LO, such a fractional 
dependence is extremely large, and grows with the c.m.~energy. It is largest 
at the threshold, where the ratio of eq.~(\ref{relfrac}) is equal to $-30$\% 
and $-60$\% at $\sqrt{S}=3$~TeV and $\sqrt{S}=10$~TeV, respectively. The 
reduction of these fractions when including the $\dNLO$ and the approximate 
$\dNNLO$ corrections is impressive. Specifically, at the NLO one obtains 
$-5$\% and $-10$\% at the threshold at the two c.m.~energies, and this is 
further reduced at the NNLO to being statistically compatible with zero -- 
this is entirely due to the cancellation mechanism explained in 
sect.~\ref{sec:scheme}, and it is a very direct confirmation that our 
NNLO-improved cross section of eq.~(\ref{xsNNLOimprep3}) works as we envisage 
it to do in the regions dominated by the $\gamma\gamma$ channel. Such
a dominance can be immediately understood by observing the almost perfect 
agreement between the NLO and $\NLOG$ predictions (red triangles and blue 
circles, respectively) on the left side of the figure. Conversely, we note 
that in the kinematic-limit region a visible factorisation-scheme dependence
remains. In that respect, we observe that, firstly, our NNLO improvement
has no impact there, since that region is a $\mu^+\mu^-$-dominated one;
and, secondly, that this effect is almost entirely due to the virtual
corrections, which are very large -- one must bear in mind that NLO-driven
factorisation-scheme dependences (as is the case of the virtuals here)
can only be cancelled at the NNLO, and in the kinematic-limit region
we do not include any sizeable contribution of that order. Indeed, this is 
shown clearly by the fact that when such virtual corrections are set equal
to zero, one finds an almost perfect cancellation of factorisation-scheme
effects also when approaching the kinematic limit (the yellow symbols
are close to zero there). While of course switching virtual corrections
off is an unphysical operation, 
we point out that the bulk of the NNLO scheme-compensating terms
can in fact be predicted (essentially, they are equal to the virtual
matrix elements convoluted with the $K_{ij}$ functions), and if need
be they could be included in order to reduce the theoretical systematics.
This is not an entirely satisfactory solution, since it leaves out other
terms that are perturbatively on the same footing as those one would
include. A better approach is to use the theoretical argument that,
in the kinematic regions dominated by large Bjorken $x$'s, the
$\Delta$ scheme~\cite{Frixione:2021wzh} is superior to the $\MSb$ one --
for example, $\Delta$ never introduces the double logarithms of $(1-z)$ 
which are present in $\MSb$ in both the PDFs and the short-distance cross 
sections, and which are eventually cancelled in physical observables. 
An unpleasant by-product of this fact is that, in order to
obtain distributions of comparable smoothness in $\MSb$- and $\Delta$-based
runs, one must use a statistical sample several times larger in the former 
case w.r.t.~the latter one (this factor being approximately equal to five 
at $\sqrt{S}=10$~GeV). The interested reader can find extensive discussions 
on this point in ref.~\cite{Bertone:2022ktl}, but should note that here
the problem is much more severe, owing to the vastly larger c.m.~energies
w.r.t.~those of relevance to the studies of ref.~\cite{Bertone:2022ktl}.

Most of what has been said about the pair invariant mass applies to the case 
of the top-quark transverse momentum distribution, shown in right panel of
fig.~\ref{fig:scheme1}. We remark again the change in the prominence
of the peak, with the one at the left end of the spectrum increasingly
dominant with increasing c.m.~energy w.r.t.~the one at the right end of 
the spectrum. That is, one passes from a large-$x$, $\mu^+\mu^-$-dominated 
configuration to a small-$x$, $\gamma\gamma$-dominated one. Again there is 
an extremely large factorisation-scheme dependence at the LO, which is 
significantly reduced when $\dNLO$, and especially approximated $\dNNLO$, 
corrections are included -- as was the case for the pair invariant mass,
this is most evident on the left side of the distribution. Conversely,
the residual scheme dependence in the kinematic-limit region is smaller 
than in the case of the pair invariant mass, and this is due to the fact 
that the distribution here is less peaked than there, which dilutes the 
impact of the virtual corrections; this is demonstrated by the fact
that setting these corrections equal to zero has a much smaller effect
than for the pair invariant mass at the kinematic limit.

\section{Conclusions\label{sec:concl}}
The main achievement of this paper is the definition of a process- and 
observable-independent contribution of NNLO, which can be added to any
NLO-accurate prediction emerging from the collinear factorisation theorem,
and which improves the latter in regions dominated by vector-boson
fusion, thus including those associated with VBF topologies. Specifically,
such an NNLO contribution is given in eq.~(\ref{final00NNLO2}), 
which must be used as is prescribed by eqs.~(\ref{dsig2Gdef}) 
and~(\ref{xsNNLOimprep3}); in the latter equation, these NNLO results 
are summed to those of an NLO computation of the $2\to m$ lepton-collision 
process of interest. In order for this final section to be as self-contained 
as is possible, we re-write those key equations here: our physical predictions 
correspond to the l.h.s.~of eq.~(\ref{xsNNLOimprep3})
\beqn
d\sigma(P_1,P_2)&=&
d\zeta_1\,d\zeta_2
\sum_{ij}\PDF{i}{\lep}(\zeta_1)\,\PDF{j}{\lep}(\zeta_1)
\nonumber
\\*&&\phantom{a}\times
\Bigg(d\hsig_{ij}^{[0]}(\zeta_1 P_1,\zeta_2 P_2)+
\aemotpi\,d\hsig_{ij}^{[1]}(\zeta_1 P_1,\zeta_2 P_2)
\nonumber
\\*&&\phantom{aaaaa}+
\delta_{i\lep}\delta_{j\lep}
\left(\aemotpi\right)^2\!d\hsig_{\dNNLOG}^{[2]}(\zeta_1 P_1,\zeta_2 P_2)
\Bigg),
\label{xsNNLOimprep4}
\eeqn
where the momenta of the beam muons are now denoted by uppercase symbols 
($P_i$), and the contribution of NNLO that we have obtained in this work is
normalised as follows:
\beq
\left(\aemotpi\right)^2\!d\hsig_{\dNNLOG}^{[2]}(p_1,p_2)=
d\hat{\Sigma}_{\dNNLOG}(p_1,p_2)\,,
\label{dsig2Gdef2}
\eeq
where its short-distance cross section is defined thus:
\beqn
d\hat{\Sigma}_{\dNNLOG}(p_1,p_2)&=&
\left(\frac{1}{1-y_1}\right)_{\deltaI}\!
\left(\frac{1}{1-y_2}\right)_{\deltaI}\!
\left((1-y_1)(1-y_2)\ampsq_{\lep\lep}^{(m+2)}\right)
d\phi_{m+2}\big(p_1,p_2\big)
\nonumber\\*&+&
\aemotpi\,{\cal Q}_{\gamma\lep}^{(\deltaI)^\prime}(z_2)
\left(\frac{1}{1-y_1}\right)_{\deltaI}\!
\left((1-y_1)\ampsq_{\lep\gamma}^{(m+1)}\right)
d\phi_{m+1}\big(p_1,z_2p_2\big)\,dz_2
\nonumber\\*&+&
\aemotpi\,{\cal Q}_{\gamma\lep}^{(\deltaI)^\prime}(z_1)
\left(\frac{1}{1-y_2}\right)_{\deltaI}\!
\left((1-y_2)\ampsq_{\gamma\lep}^{(m+1)}\right)
d\phi_{m+1}\big(z_1p_1,p_2\big)\,dz_1
\nonumber\\*&+&
\left(\aemotpi\right)^2\,
{\cal Q}_{\gamma\lep}^{(\deltaI)^\prime}(z_1)\,
{\cal Q}_{\gamma\lep}^{(\deltaI)^\prime}(z_2)
\ampsq_{\gamma\gamma}^{(m)}\,
d\phi_{m}\big(z_1p_1,z_2p_2\big)\,dz_1\,dz_2\,.\phantom{aaa}
\label{final00NNLO2rep}
\eeqn
The contribution of eq.~(\ref{final00NNLO2rep}) to eq.~(\ref{xsNNLOimprep4})
constitutes the improvement of NNLO mentioned at the beginning of this 
section, while the NLO predictions appear in the second line of the latter 
equation. As was said, we expect this improvement to be particularly 
accurate in the kinematic regions dominated by VBF-like 
configurations or, more generally, by underlying partonic collisions 
stemming from the $\gamma\gamma$ channel. Equation~(\ref{final00NNLO2rep}) 
features the $2\to 2+m$ matrix elements that allow one to retain exactly 
both logarithmic and power-suppressed effects in the mass of the $Z$ boson, 
as well as all of those due to the $Z/\gamma$ interference. At the same time, 
this formalism resums all light-fermion mass effects through the usage of 
lepton PDFs, at the logarithmic accuracy to which such PDFs are available. 
The PDFs must not include weak massive vector bosons in their partonic 
contents.

While of NNLO, eq.~(\ref{final00NNLO2rep}) is {\em not} a complete NNLO
cross section, as is also underscored by the restrictions on the sums over 
parton types in eq.~(\ref{xsNNLOimprep4}) which are enforced by the
Kronecker delta's. It cannot thus have the same level of accuracy as that 
of complete NNLO predictions, but it has the advantage of being fairly 
easy to calculate, essentially implying the same level of computational 
complexity as the corresponding NLO cross section. Therefore, it can easily 
be automated, although this is something that we have not yet considered. 
More importantly, it gives one a viable, and theoretically vastly superior, 
alternative to the usage of weak-boson PDFs, which is correctly 
defined in the whole of the phase space, with a straightforward 
inclusion of higher logarithmic and fixed-order contributions, a 
well-defined theoretical systematics, and the complete control over 
theoretical uncertainties.

A number of immediate phenomenological consequences of the results
presented here are discussed in a companion paper~\cite{Frixione:2025xxx}.

\section*{Acknowledgements}
We thank Giovanni Stagnitto for having provided us with a still-unreleased
NLL- and NLO-accurate version of the muon PDFs of ref.~\cite{Frixione:2023gmf}.
DP~and MZ~acknowledge the financial support by the MUR (Italy), with 
funds of the European Union (NextGenerationEU), through the PRIN2022
grant 2022EZ3S3F; likewise FM, through the PRIN2022 grant 2022RXEZCJ.
SF~thanks the TH division of CERN for the hospitality during the course 
of this work.

\appendix
\section{Additional plots\label{sec:extrapl}}
In this appendix we present the analogues of the plots shown in
sect.~\ref{sec:rescuts} and~\ref{sec:resfact}, obtained when considering
$W^+W^-$ production rather than $t\bt$ production as was done in the
main text.
\begin{figure}[htb]
  \begin{center}
  \includegraphics[scale=1.0,width=0.48\textwidth]{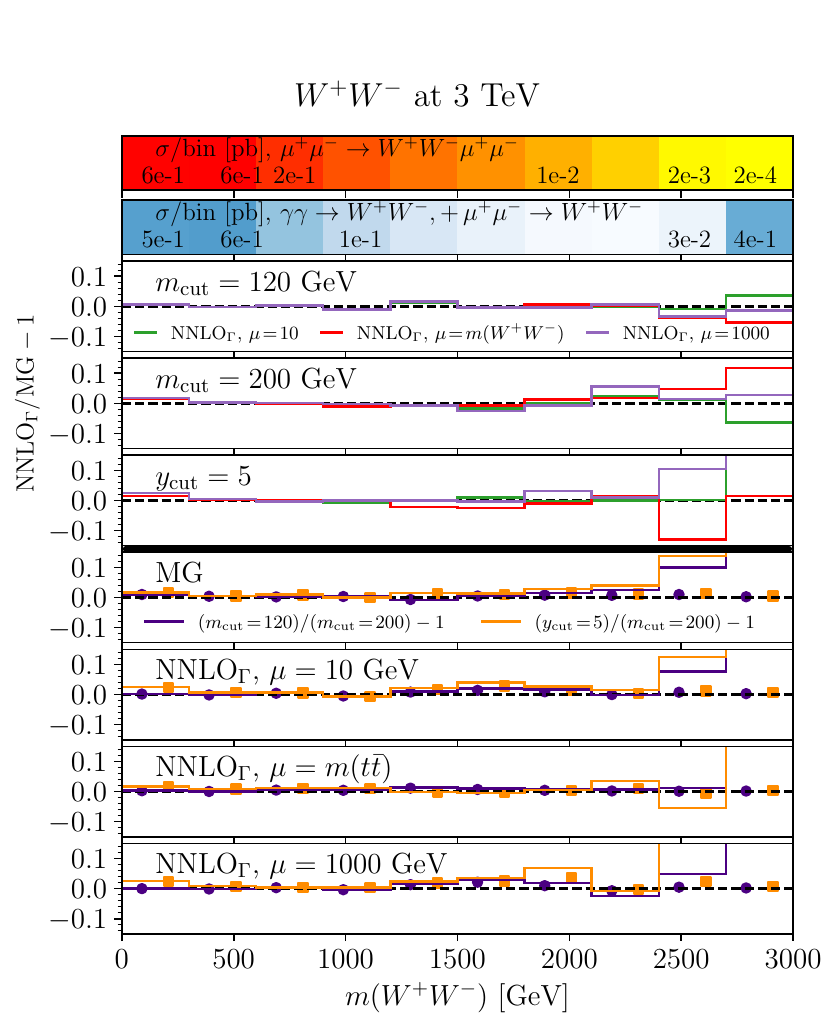}
$\phantom{a}$
  \includegraphics[scale=1.0,width=0.48\textwidth]{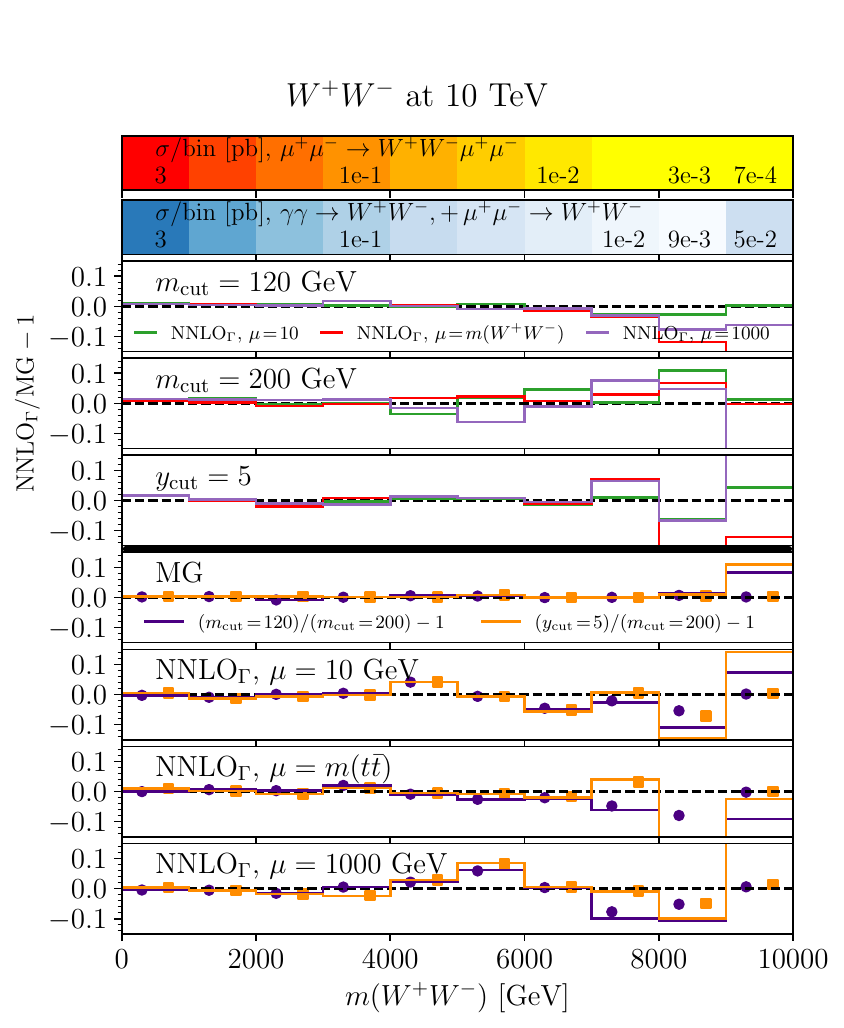}
\caption{\label{fig:tcuts3}
As in fig.~\ref{fig:tcuts1}, for $W^+W^-$ production.
}
  \end{center}
\end{figure}
\begin{figure}[htb]
  \begin{center}
  \includegraphics[scale=1.0,width=0.48\textwidth]{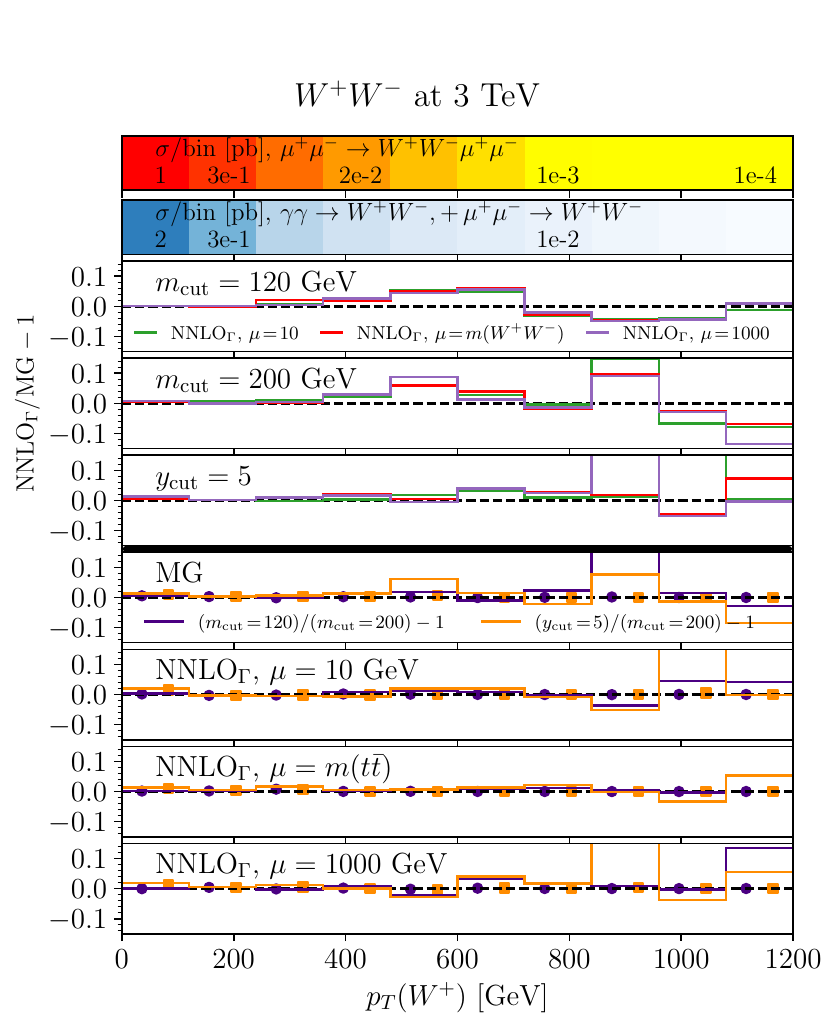}
$\phantom{a}$
  \includegraphics[scale=1.0,width=0.48\textwidth]{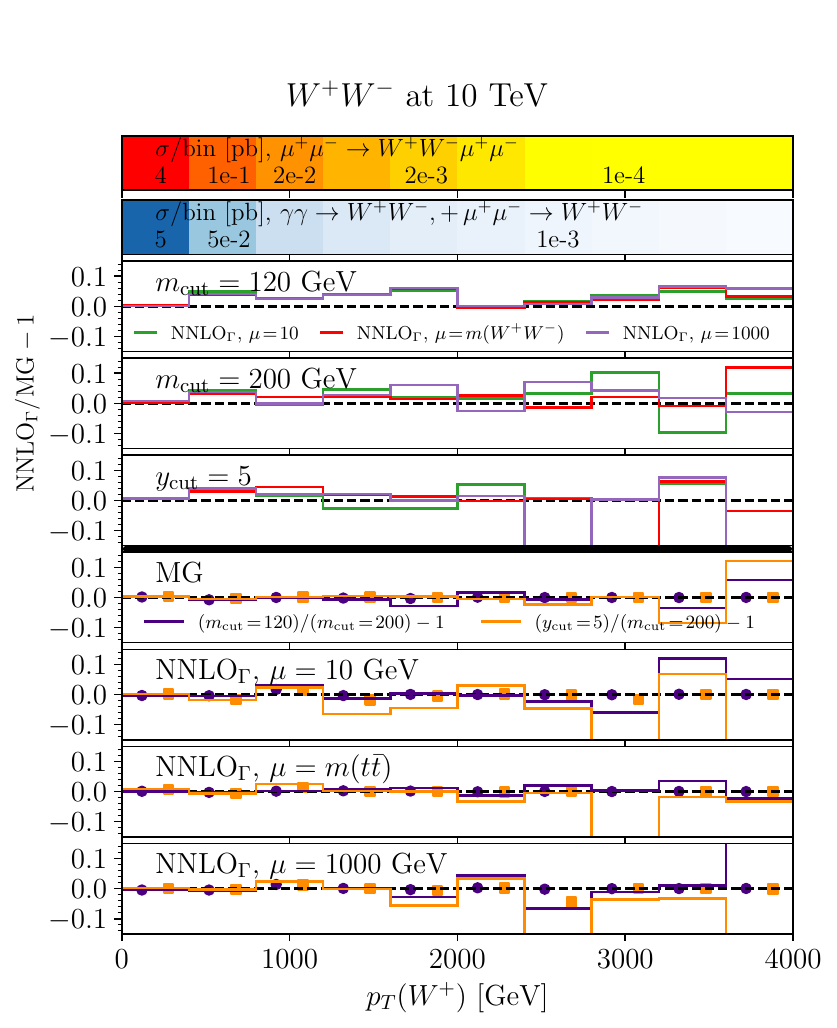}
\caption{\label{fig:tcuts4}
As in fig.~\ref{fig:tcuts2}, for $W^+W^-$ production.
}
  \end{center}
\end{figure}

The take-home message is that, despite the significant differences in
the production mechanisms of $W^+W^-$ and $t\bt$ pairs,
the conclusions about the dependences on the technical cuts
and on the factorisation-scheme choice are exactly the same. As we
have argued in the text, this is far from being accidental, and is
rather the consequence of the physics which underpins the procedure
we have constructed (in addition to constituting a check that such a 
procedure has been correctly implemented in a computer code).
\begin{figure}[htb]
  \begin{center}
  \includegraphics[scale=1.0,width=0.48\textwidth]{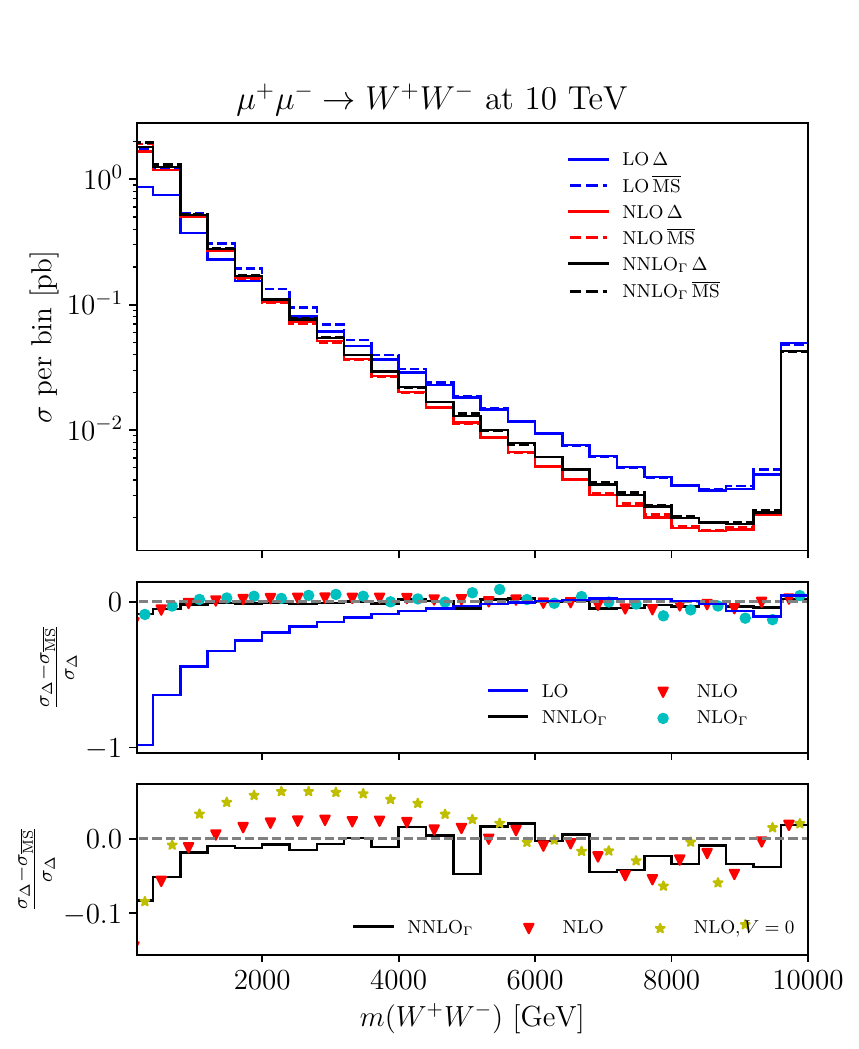}
$\phantom{a}$
  \includegraphics[scale=1.0,width=0.48\textwidth]{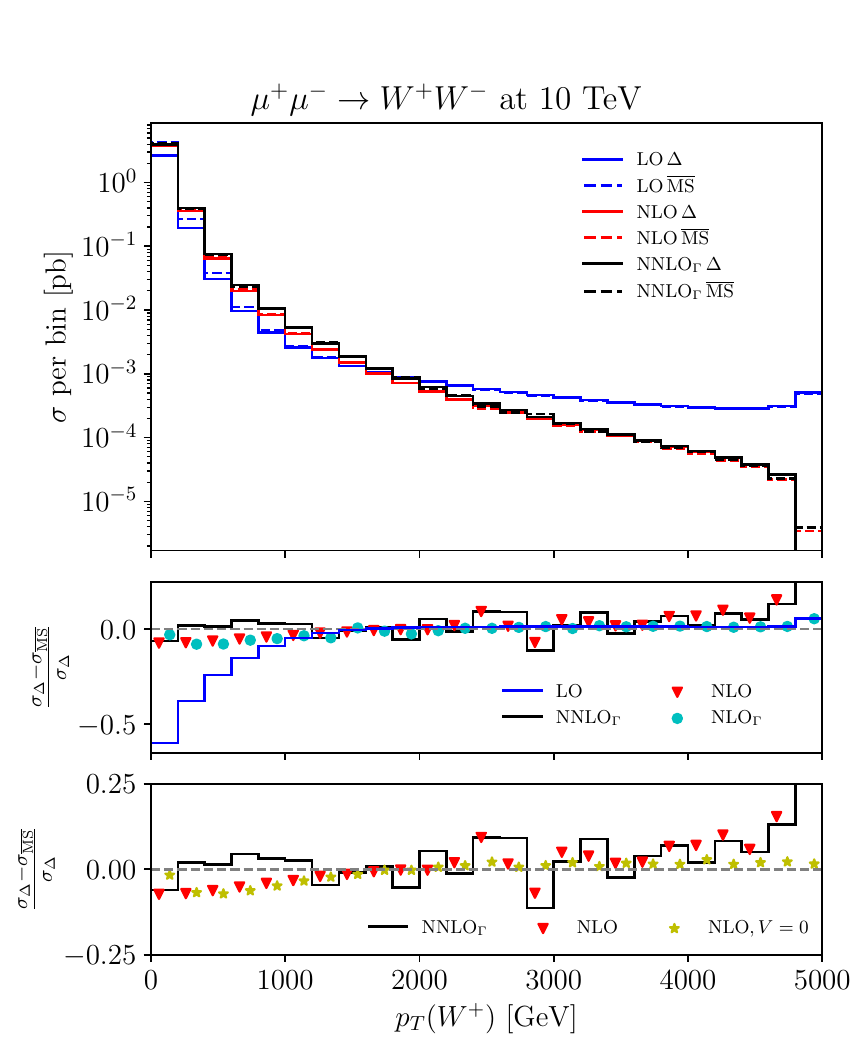}
\caption{\label{fig:scheme3}
As in fig.~\ref{fig:scheme1}, for $W^+W^-$ production.
}
  \end{center}
\end{figure}

\section{An FKS-type phase space for double-real contributions
\label{sec:impl}}
The implementation of eq.~(\ref{final00NNLO2rep}) is independent of
that of the NLO cross section used to obtain phenomenological
predictions, since the two are now uncorrelated. Having said that, it is 
convenient for us to adopt also in the case of eq.~(\ref{final00NNLO2rep})
a phase-space parametrisation analogous to that we employ in our
NLO computations, namely one based on the FKS formalism. 
With the momentum assignments of eq.~(\ref{VBFproc}), we can 
write\footnote{The particles of the set of the tagged objects and
their momenta are denoted by the same symbols ($T_i$) here, since no
confusion is possible.}:
\beqn
&&d\phi_{m+2}\equiv
d\phi_{m+2}\left(p_1+p_2;k_1,k_2,\{T_i\}_{i=1}^m\right)
\label{phsp1}
\\*
&&\phantom{aa}=
(2\pi)^4\delta\!\left(p_1+p_2-k_1-k_2-\sum_{i=1}^m T_i\right)
\prod_{j=1}^2\frac{d^3 k_j}{(2\pi)^3 2k_j^0}\,
\prod_{i=1}^m\frac{d^3 T_i}{(2\pi)^3 2T_i^{0}}\,.
\phantom{aa}
\nonumber
\eeqn
With
\beq
s=(p_1+p_2)^2
\eeq
and the FKS-type parametrisations in the $\lp\lm$ cm 
frame\footnote{According to eq.~(\ref{final00}), \mbox{$k_2\parallel p_2$}
when $y_2=1$, whence the sign of the third spatial component in 
eq.~(\ref{kjdef}).}:
\beq
k_j=\frac{\sqrt{s}}{2}\xi_j\left(1,\sqrt{1-y_j^2}\vec{e}_{Tj},
(-)^{j-1}y_j\right)\,,
\label{kjdef}
\eeq
where
\beq
\vec{e}_{Tj}=\left(\sin\varphi_j,\cos\varphi_j\right).
\eeq
Then:
\beqn
d\phi_{m+2}&=&\left(\frac{s}{8(2\pi)^3}\right)^2
\xi_1d\xi_1dy_1d\varphi_1\xi_2d\xi_2dy_2d\varphi_2
\nonumber
\\&&\phantom{aaaaaa}\times
d\phi_{m}\left(p_1+p_2-k_1-k_2;\{T_i\}_{i=1}^m\right)\,.
\label{phsp2}
\eeqn
The quantity on the second line of eq.~(\ref{phsp2}) is a proper
$m$-body phase space, where the four vector \mbox{$p_1+p_2-k_1-k_2$}
is thought to be given; as such, it can be defined as it is most
convenient. The variables $y_{1,2}$ coincide with their namesakes in 
eq.~(\ref{final00}), while we shall take $z_{1,2}=1-\xi_{1,2}$ (bear in 
mind that both of these apply solely to the NNLO-type contributions).
It is a matter of simple algebra to arrive at:
\beq
\left(\sum_{i=1}^m T_i\right)^2\equiv
\left(p_1+p_2-k_1-k_2\right)^2=
s+\frac{s}{2}\xi_1\xi_2(1-\Omega)-s(\xi_1+\xi_2)\ge\minf^2\,,
\label{xiboundeq}
\eeq
with $\minf^2$ the lower bound on the invariant mass squared of the
$T$ system, and
\beq
\Omega=\sqrt{1-y_1^2}\sqrt{1-y_2^2}\,\vec{e}_{T1}\mydot\vec{e}_{T2}-y_1y_2
\;\;\;\;\Longrightarrow\;\;\;\;
-1\le\Omega\le 1\,.
\eeq
Equation~(\ref{xiboundeq}) leads to
\beq
0\le\xi_1\le 1-\hminf^2\,,\;\;\;\;\;\;\;\;
0\le\xi_2\le\frac{2(1-\hminf^2-\xi_1)}{2-(1-\Omega)\xi_1}\,,
\label{xibound}
\eeq
where we have defined:
\beq
\hminf^2=\frac{\minf^2}{s}\,.
\eeq
Note that:
\beq
\frac{2(1-\hminf^2-\xi_1)}{2-(1-\Omega)\xi_1}\le 1-\hminf^2\,.
\eeq
A possible procedure is then the following:
\begin{itemize}
\item Generate $-1\le y_{1,2}\le 1$ and $0\le\varphi_{1,2}\le 2\pi$.
\item Generate $\xi_{1,2}$ in the ranges of eq.~(\ref{xibound}).
\item With the above, construct $k_1$ and $k_2$, and thus
$p_1+p_2-k_1-k_2$.
\item In the rest frame of $p_1+p_2-k_1-k_2$ generate the $m$ $T_i$
momenta; boost them back to the $p_1+p_2$ rest frame.
\end{itemize}
The generation of $\xi_{1,2}$ requires the prior knowledge of $\minf^2$.
This is typically a constant (e.g.~$4m_W^2$ for $W^+W^-$ production).
However, for processes with $m\ge 3$, it is conceivable that one of
the variables that parametrise $d\phi_m$ be equal to the invariant
mass squared of the $T$ system. If so, one can generate this
variable before the generation of $\xi_{1,2}$, and then proceed
as is written above; the last step would entail the generation of
the remaining $3m-5$ variables for $d\phi_m$.

The parametrisation defined above works also for a one-body system
(i.e.~for $m=1$). However, in that special case the inequality in 
eq.~(\ref{xiboundeq}) becomes an equality, and $\minf^2=m_{T_1}^2$,
with $m_{T_1}$ the invariant mass of the final state particle
(e.g.~the Higgs). This equality is formally enforced by the fact that now:
\beqn
&&d\phi_{1}\left(p_1+p_2-k_1-k_2;T_1\right)=
(2\pi)^4\delta\!\left(p_1+p_2-k_1-k_2-T_1\right)
\nonumber\\*&&
\phantom{d\phi_{1}\left(p_1+p_2-k_1-k_2;T_1\right)=aaa}\times
\frac{d^4 T_1}{(2\pi)^3}\,
\delta\!\left(T_1^{2}-m_{T_1}^2\right)
\\*&&\phantom{aaaaaaaa}=(2\pi)
\delta\!\left(s+\frac{s}{2}\xi_1\xi_2(1-\Omega)-s(\xi_1+\xi_2)-
m_{T_1}^2\right)\,.
\label{delmeq1}
\eeqn
Equation~(\ref{delmeq1}) allows one to trivially integrate out either
$\xi_1$ or $\xi_2$, and to restore the correct counting of independent
phase-space variables (with a one-body system, there must be 5 independent
variables, but eq.~(\ref{phsp2}) has a 6-dimensional measure). Instead
of eq.~(\ref{xibound}) we thus have:
\beq
0\le\xi_1\le 1-m_{T_1}^2/s\,,\;\;\;\;\;\;\;\;
\xi_2=\bar{\xi}_2\equiv\frac{2(1-m_{T_1}^2/s-\xi_1)}
{2-(1-\Omega)\xi_1}\,,
\label{xibound2}
\eeq
and
\beq
d\phi_{1}\left(p_1+p_2-k_1-k_2;T_1\right)=
\frac{2(2\pi)}{s(2-(1-\Omega)\xi_1)}\,\delta(\xi_2-\bar{\xi_2})\,.
\label{1dphsp}
\eeq
The slightly unpleasant feature of eq.~(\ref{xibound}), and of
eqs.~(\ref{xibound2}) and~(\ref{1dphsp}), is the asymmetric role played
by $\xi_1$ and $\xi_2$. Although this does not necessarily imply a poor
numerical behaviour, one can possibly interchange the roles of these
two variables in a random manner. Alternatively, one can proceed by
following one of the two approaches we are now going to describe.

In the first approach, we define
\beq
\varsigma=\xi_1+\xi_2\,,\;\;\;\;\;\;\;\;
\delta=\xi_1-\xi_2\,,
\label{sdvar}
\eeq
so that
\beq
\xi_1=\half\big(\varsigma+\delta\big)\,,\;\;\;\;\;\;\;\;
\xi_2=\half\big(\varsigma-\delta\big)\,,
\eeq
and
\beq
d\xi_1 d\xi_2 = \half d\varsigma d\delta\,.
\eeq
The region of integration in the $\langle\varsigma,\delta\rangle$ plane
can be found be imposing \mbox{$\xi_i\ge 0$}, \mbox{$\xi_i\le 1$}, and the 
condition of eq.~(\ref{xiboundeq}). The borders of these regions are depicted
\begin{figure}[htb]
  \begin{center}
  \includegraphics[width=0.65\textwidth]{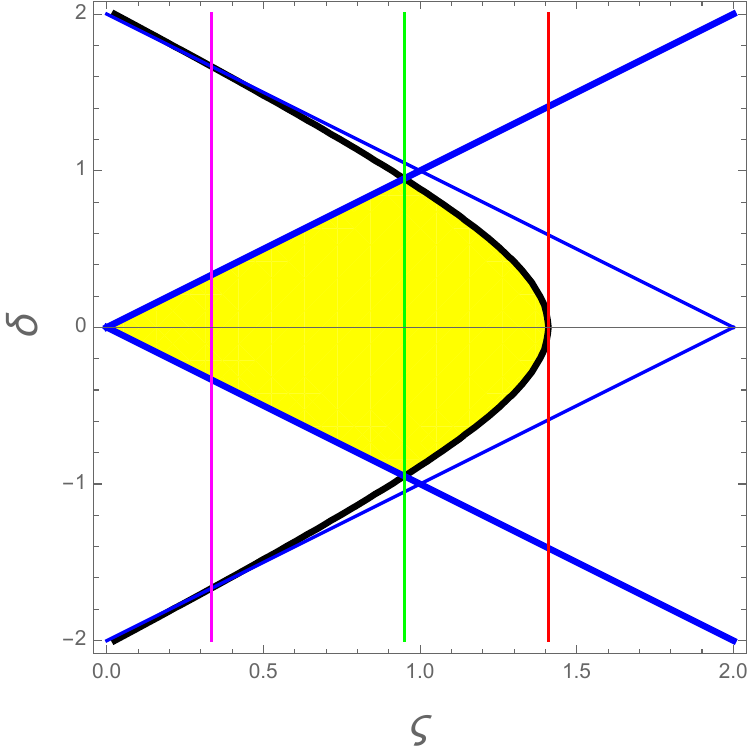}
\caption{\label{fig:vbfphsp1} 
Integration region for the variables of eq.~(\ref{sdvar}),
having set $\Omega=-0.85$ and $\hminf^2=0.05$.
}
  \end{center}
\end{figure}
in fig.~\ref{fig:vbfphsp1} as thick blue lines, thin blue lines,
and a thick black line respectively; the resulting overlapping region,
i.e.~the sought region of integration, is shown in yellow. It turns out
to be convenient to first integrate in $\delta$ and then in $\varsigma$.
Thus, the upper branch ($\delta\ge 0$) of the black line of 
fig.~\ref{fig:vbfphsp1} is given by
\beq
\Delta(\varsigma)=\sqrt{
\frac{8(1-\hminf^2)-8\varsigma+(1-\Omega)\varsigma^2}
{1-\Omega}}\,.
\eeq
As is shown in the figure, the function will be relevant in
\mbox{$\varsigma_1\le\varsigma\le\varsigma_2$}, where
\beqn
\varsigma_1&=&1-\hminf^2\,,
\\
\varsigma_2&=&\frac{4-2\sqrt{2}\sqrt{1+\Omega+\hminf^2(1-\Omega)}}
{1-\Omega}\,,
\eeqn
whose values are shown as green and red lines in fig.~\ref{fig:vbfphsp1}.
By construction:
\beqn
\Delta(\varsigma_1)&=&\varsigma_1\,,
\\
\Delta(\varsigma_2)&=&0\,.
\eeqn
We note that $\varsigma_2\ge\varsigma_1$, and that:
\beq
\lim_{\Omega\to 1}\varsigma_2=\varsigma_1\,,
\eeq
which implies that the integral of any function regular in 
\mbox{$\varsigma_1\le\varsigma\le\varsigma_2$} will vanish for
$\Omega\to 1$. Furthermore, by defining $\varsigma_0$ so that
\beq
\Delta(\varsigma_0)=2-\varsigma_0\,,
\eeq
(whose value is shown as a magenta line in fig.~\ref{fig:vbfphsp1})
we have
\beq
\varsigma_0<\varsigma_1<1\,,\;\;\;\;\;\;\;\;
\varsigma_2<2\,,
\eeq
for any $\hminf^2>0$. This implies that the conditions stemming from
requiring \mbox{$\xi_i\le 1$} are never relevant for the determination
of the integration region. So finally we have that:
\beqn
&&\!\!\!\!\!\!\!\!\int d\xi_1 \stepf\!\left(\xi_2\le 1-\hminf^2\right)
\int d\xi_2
\stepf\!\left(\xi_2\le\frac{2(1-\hminf^2-\xi_1)}{2-(1-\Omega)\xi_1)}\right)
f(\xi_1,\xi_2)
\label{volsd}
\\
&=&\half\int_0^{\varsigma_1} d\varsigma \int_{-\varsigma}^\varsigma d\delta
f\left(\frac{\varsigma+\delta}{2},\frac{\varsigma-\delta}{2}\right)
+\half\int_{\varsigma_1}^{\varsigma_2} d\varsigma 
\int_{-\Delta(\varsigma)}^{\Delta(\varsigma)} d\delta
f\left(\frac{\varsigma+\delta}{2},\frac{\varsigma-\delta}{2}\right)
\nonumber
\eeqn
for any function $f(\xi_1,\xi_2)$. We have verified eq.~(\ref{volsd})
numerically for different choices of~$f$, and analytically for $f\equiv 1$.

We now turn to discussing the second approach, based on two variables
that we denote by $\lambda$ and $t$, which are such that:
\beq
\xi_1=\lambda t\,,\;\;\;\;\;\;\;\;
\xi_2=\lambda (1-t)\,,
\label{lamtvar}
\eeq
with
\beq
d\xi_1 d\xi_2 = \lambda d\lambda dt\,.
\eeq
As was done before, the region of integration in the 
$\langle\lambda,t\rangle$ plane
can be found be imposing \mbox{$\xi_i\ge 0$}, \mbox{$\xi_i\le 1$}, and the 
condition of eq.~(\ref{xiboundeq}). The borders of these regions are depicted
\begin{figure}[htb]
  \begin{center}
  \includegraphics[width=0.65\textwidth]{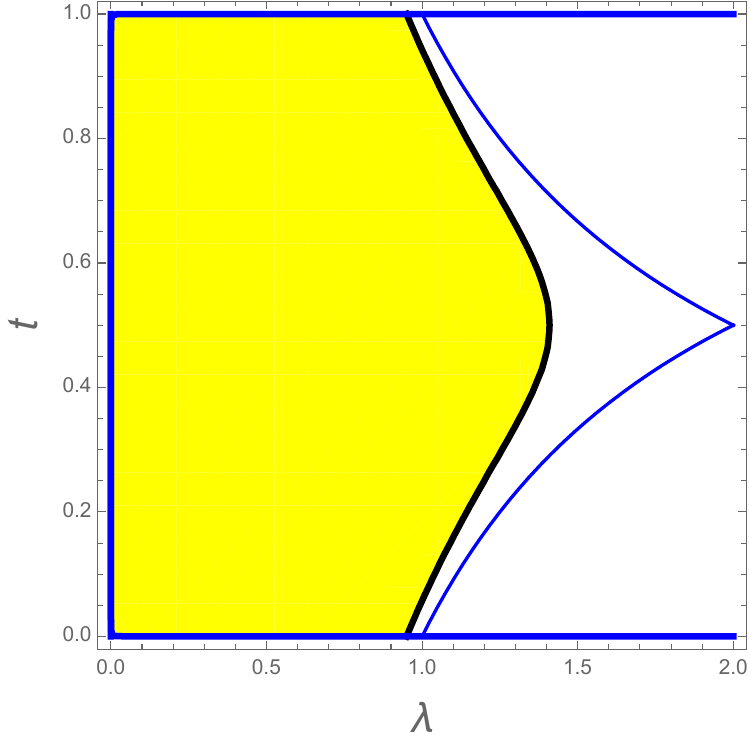}
\caption{\label{fig:vbfphsp2} 
Integration region for the variables of eq.~(\ref{lamtvar}),
having set $\Omega=-0.85$ and $\hminf^2=0.05$.
}
  \end{center}
\end{figure}
in fig.~\ref{fig:vbfphsp1} as thick blue lines, thin blue lines,
and a thick black line respectively; the resulting overlapping region,
i.e.~the sought region of integration, is shown in yellow. As was already
the case, the conditions stemming from \mbox{$\xi_i\le 1$} are irrelevant,
and the only non-trivial constraint is due to eq.~(\ref{xiboundeq}).
In the present case, the best strategy is that of integrating first in 
$\lambda$ and then in $t$; thus, the black line of fig.~\ref{fig:vbfphsp2} 
is given by:
\beq
\Lambda(t)=\frac{1-\sqrt{1-2(1-\hminf^2)(1-\Omega)(1-t)t}}
{(1-\Omega)(1-t)t}.
\eeq
We note that:
\beq
\lim_{t\to 0}\Lambda(t)=
\lim_{t\to 1}\Lambda(t)=
\lim_{\Omega\to 1}\Lambda(t)=1-\hminf^2\,.
\eeq
When $t\simeq 0$, and/or $t\simeq 1$, and/or $\Omega\simeq 1$ a numerical 
$0/0$ result in the evaluation of $\Lambda(t)$ can be avoided by using 
the following series expansion:
\beq
\Lambda(t)=(1-\hminf^2)\sum_{k=0}^\infty\widehat{\Lambda}_k R^k\,,
\;\;\;\;\;\;\;\;
R=(1-\hminf^2)(1-\Omega)(1-t)t\,,
\eeq
with
\beq
\widehat{\Lambda}_k=1,\,\half,\,\half,\,\frac{5}{8},\,
\frac{7}{8},\,\frac{21}{16},\,\ldots
\eeq
This expansion is quite accurate: by retaining its first six terms
whose coefficients are given here, the relative difference between
$\Lambda(t)$ and its truncated expansion is smaller than $10^{-5}$
for $R\le 0.1$. Finally:
\beqn
&&\!\!\!\!\!\!\!\!\int d\xi_1 \stepf\!\left(\xi_2\le 1-\hminf^2\right)
\int d\xi_2
\stepf\!\left(\xi_2\le\frac{2(1-\hminf^2-\xi_1)}{2-(1-\Omega)\xi_1)}\right)
f(\xi_1,\xi_2)
\nonumber
\\
&=&\int_0^1 dt \int_0^{\Lambda(t)} d\lambda\,\lambda\,
f\left(\lambda t,\lambda (1-t))\right)\,.
\label{vollamt}
\eeqn
We have verified eq.~(\ref{vollamt}) numerically for $f\equiv 1$
and several values of $\Omega$ and $\hminf^2$.

\section{Azimuthal correlations\label{sec:azicorr}}
As is known, a local collinear counterterm requires an additional 
contribution w.r.t.~the standard Altarelli-Parisi kernel when the off-shell 
parton that participates in the branching has spin one; such a contribution
has a non-trivial dependence on the azimuthal angle, and because of
this it is essential to obtain a stable numerical behaviour, in spite
of the fact that it integrates to zero -- more details on this in
the context of the FKS subtraction can be found in 
refs.~\cite{Frixione:1995ms,Frederix:2009yq}.

VBF topologies feature two $t$-channel photon exchanges, and therefore
there will be two azimuthal kernels (one for each of the photons) to be
included. As was stressed multiple times in the main text, in such
topologies the collinear limits are incoherent, and one can thus arrive
at the final result by iterating the one relevant to a single branching.

The quantity of interest is the matrix element of the process of 
eq.~(\ref{VBFprocred}), in the collinear limits \mbox{$p_1\parallel k_1$} 
and \mbox{$p_2\parallel k_2$}. Following the procedure of appendix~B
of ref.~\cite{Frixione:1995ms}, we write
\beqn
\ampsq_{\lep\lep}^{(m+2)}\!\!\!\!&=&\!\!\!\!
\sum_{h_1g_1}\sum_{h_2g_2}\abs{\amp_{\lep\lep}^{(m+2)}(h_1,g_1;h_2,g_2)}^2\,,
\\
\amp_{\lep\lep}^{(m+2)}(h_1,g_1;h_2,g_2)
&\stackrel{p_1\parallel k_1,p_2\parallel k_2}{\longrightarrow}&
e^2\sum_{f_1f_2}\amp_{\gamma\gamma}^{(m)}(f_1;f_2)
S_{\gamma\lep}^{f_1h_1g_1}(z_1)S_{\gamma\lep}^{f_2h_2g_2}(z_2)\,.
\label{ampsplit}
\eeqn
We have denoted by $h_i$ ($g_i$) the helicity of the incoming (outgoing)
lepton with momentum $p_i$ ($k_i$), and by $f_i$ the polarisation index
of the photon emitted in the branching of leg $i$. In other words,
the amplitude relevant to the branching process
\beq
\lep(p_i,h_i)\,\longrightarrow\,\lep(k_i,g_i)+\gamma(p_i-k_i,f_i)
\equiv\lep((1-z_i)p_i,g_i)+\gamma(z_ip_i,f_i)\,,
\eeq
has been denoted by
\beq
S_{\gamma\lep}^{f_ih_ig_i}(z_i)\,.
\eeq
By squaring the r.h.s.~of eq.~(\ref{ampsplit}) we simplify the result
by using the relationships between the branching amplitudes and the
splitting kernels, per ref.~\cite{Frixione:1995ms}:
\beqn
\sum_{h_ig_i}\abs{S_{\gamma\lep}^{+h_ig_i}(z_i)}=
\sum_{h_ig_i}\abs{S_{\gamma\lep}^{-h_ig_i}(z_i)}&=&
\frac{1}{2p_i\mydot k_i}\,P_{\gamma\lep}^<(z_i)\,,
\\*
\sum_{h_ig_i}S_{\gamma\lep}^{+h_ig_i}(z_i)
\left(S_{\gamma\lep}^{-h_ig_i}(z_i)\right)^\star&=&
\frac{1}{4p_i\mydot k_i}\,\frac{\langle p_ik_i\rangle}{[p_ik_i]}\,
Q_{\gamma^\star\lep}(z_i)\,,
\eeqn
whence we obtain:
\beqn
\ampsq_{\lep\lep}^{(m+2)}
&\stackrel{p_1\parallel k_1,p_2\parallel k_2}{\longrightarrow}&
\frac{(4\pi\aem)^2}{(p_1\mydot k_1)(p_2\mydot k_2)}
\Bigg[
P_{\gamma\lep}^<(z_1)\,P_{\gamma\lep}^<(z_2)\,
\ampsq_{\gamma\gamma}^{(m)}
+P_{\gamma\lep}^<(z_1)\,Q_{\gamma^\star\lep}(z_2)\,
\widetilde{\ampsq}_{\gamma\gamma^\star}^{(m)}
\nonumber
\\*&&\phantom{aaaa}
+Q_{\gamma^\star\lep}(z_1)\,P_{\gamma\lep}^<(z_2)\,
\widetilde{\ampsq}_{\gamma^\star\gamma}^{(m)}
+Q_{\gamma^\star\lep}(z_1)\,Q_{\gamma^\star\lep}(z_2)\,
\widetilde{\ampsq}_{\gamma^\star\gamma^\star}^{(m)}
\Bigg],
\label{dcazi}
\eeqn
where $\ampsq_{\gamma\gamma}^{(m)}$ is the matrix element for the
process of eq.~(\ref{gagaWW}), which is ubiquitous in the text, and
we have defined:
\beqn
\widetilde{\ampsq}_{\gamma\gamma^\star}^{(m)}&=&
\Re\left[\sum_{f_1}\amp_{\gamma\gamma}^{(m)}(f_1;+)
\left(\amp_{\gamma\gamma}^{(m)}(f_1;-)\right)^*
\frac{\langle p_2k_2\rangle}{[p_2k_2]}
\right]\,,
\label{Mtil1}
\\*
\widetilde{\ampsq}_{\gamma^\star\gamma}^{(m)}&=&
\Re\left[\sum_{f_2}\amp_{\gamma\gamma}^{(m)}(+;f_2)
\left(\amp_{\gamma\gamma}^{(m)}(-;f_2)\right)^*
\frac{\langle p_1k_1\rangle}{[p_1k_1]}
\right]\,,
\label{Mtil2}
\\*
\widetilde{\ampsq}_{\gamma^\star\gamma^\star}^{(m)}&=&
\half\,\Re\left[\amp_{\gamma\gamma}^{(m)}(+;+)
\left(\amp_{\gamma\gamma}^{(m)}(-;-)\right)^*
\frac{\langle p_1k_1\rangle}{[p_1k_1]}\,
\frac{\langle p_2k_2\rangle}{[p_2k_2]}
\right]
\nonumber
\\*&+&
\half\,\Re\left[\amp_{\gamma\gamma}^{(m)}(+;-)
\left(\amp_{\gamma\gamma}^{(m)}(-;+)\right)^*
\frac{\langle p_1k_1\rangle}{[p_1k_1]}\,
\frac{[p_2k_2]}{\langle p_2k_2\rangle}
\right]\,.
\label{Mtil12}
\eeqn
We point out that the matrix elements of eqs.~(\ref{Mtil1}) and~(\ref{Mtil2})
coincide with the standard $\widetilde{\ampsq}$ of FKS implementations; the
new notation only helps one understand which of the two legs that undergo a
collinear splitting induces azimuthal correlations.

If one adopts the language of dipole subtraction~\cite{Catani:1996vz}
instead, for a single collinear lepton splitting one finds:
\beq
\abs{\amp_\lep|}^2= 
\frac{8\pi\alpha}{-k^2}
\left(\amp_{\gamma}\amp_{\gamma}^\star \right)^{\rho\sigma}
\left[- g_{\rho\sigma} z + 
\frac{4(1-z)}{z} \hat k^\perp_\rho \hat k^\perp_\sigma\right]\,.
\label{eq:collpwg}
\eeq
Here, $k$ is the momentum of the photon that emerges from the lepton
branching. With FKS variables,  $k^2=-s \xi (1-y)/2$, and $z=1-\xi$. 
Furthermore, $\left(\amp_{\gamma}\amp_{\gamma}^\star \right)^{\rho\sigma}$ 
would be the squared amplitude for the process where the lepton line is 
replaced by an on-shell photon, if it were summed over photon polarisations.
Finally, $\hat k^\perp = \left(0,\cos \phi, \sin \phi,0\right)$.
With this, we can iterate eq.~(\ref{eq:collpwg}), and obtain the
analogue of eq.~(\ref{dcazi}), which reads as follows:
\beqn
\ampsq_{\lep\lep}^{(m+2)}
&\stackrel{p_1\parallel k_1,p_2\parallel k_2}{\longrightarrow}&
\frac{(4\pi\aem)^2}{(p_1\mydot k_1)(p_2\mydot k_2)}
\left(\amp_{\gamma\gamma}
\amp_{\gamma\gamma}^\star \right)^{\rho_1\sigma_1\rho_2\sigma_2}
\label{diazicorr}
\\*&&\times\Bigg[
g_{\rho_1\sigma_1}g_{\rho_2\sigma_2} z_1 z_2 
-g_{\rho_1\sigma_1} z_1 
\frac{4(1-z_2)}{z_2} \hat k^{\perp2}_{\rho_2} \hat k^{\perp2}_{\sigma_2}
\nonumber
\\*&&
-g_{\rho_2\sigma_2} z_2 
\frac{4(1-z_1)}{z_1} \hat k^{\perp1}_{\rho_1} \hat k^{\perp1}_{\sigma_1} +
\frac{4(1-z_1)}{z_1} \hat k^{\perp1}_{\rho_1} \hat k^{\perp1}_{\sigma_1}
\frac{4(1-z_2)}{z_2} \hat k^{\perp2}_{\rho_2} \hat k^{\perp2}_{\sigma_2}
\Bigg].
\nonumber
\eeqn
We have used eq.~(\ref{diazicorr}) in our implementation of the
$\dNNLOG$ corrections, since in doing so we have benefitted from the fact 
that squared amplitudes contracted with an arbitrary four-vector can 
be obtained by means of {\tt MadNkLO}~\cite{Lionetti:2018gko,Hirschi:2019fkz,
Becchetti:2020wof,Bonciani:2022jmb,Bertolotti:2022ohq}.

\section{Reference frames\label{sec:frames}}
The natural reference frame to obtain eq.~(\ref{final00}) is:
\beq
F:\;\;\;\;{\rm rest~frame~of}\;p_1+p_2\,,
\eeq
which is what has been assumed in appendix~\ref{sec:impl} in order to
parametrise the four momenta; in particular, we have
\beq
p_{1,2}=\frac{\sqrt{s}}{2}\left(1,0,0,\pm 1\right)
\eeq
and eq.~(\ref{kjdef})
\beq
k_j=\frac{\sqrt{s}}{2}\xi_j\left(1,\sqrt{1-y_j^2}\vec{e}_{Tj},
(-)^{j-1}y_j\right)
\label{kjdefbis}
\eeq
for the momenta involved in the collinear splittings of interest.

We also need to consider the c.m.~frame of the parton pair that
initiates the hard collision -- the following is the case where a 
photon PDF is convoluted on the leg incoming from the left:
\beq
F_1^\prime:\;\;\;\;{\rm rest~frame~of}\;z_1p_1+p_2
\;\;\;\;\longleftrightarrow\;\;\;\;
\delta(1-y_1)\,,
\eeq
where the Dirac $\delta$ here reminds one that this frame is
associated with the collinear configuration $p_1\parallel k_1$.
The boost $B_1$ from frame $F$ to frame $F_1^\prime$ is given in terms
of the boost rapidity $\eta_B$. By imposing
\beq
B_1\left(z_1p_1+p_2\right)=\frac{\sqrt{z_1s}}{2}\left(1,0,0,0\right)\,,
\eeq
we obtain
\beq
\exp(\eta_B)=\sqrt{z_1}\,,
\eeq
and we can thus verify that\footnote{Here and elsewhere, boosted
momenta will be denoted by means of an overline.}
\beqn
B_1\left(z_1p_1\right)&\equiv&
\overline{(z_1p_1)}=\frac{\sqrt{z_1s}}{2}\left(1,0,0,1\right)\,,
\\*
B_1\left(p_2\right)&\equiv&
\pb_2=\frac{\sqrt{z_1s}}{2}\left(1,0,0,-1\right)\,.
\eeqn
In $F_1^\prime$, we parametrise $k_2$ as follows:
\beq
\bk_2=
\frac{\sqrt{z_1s}}{2}\,
\xib_2\left(1,\sqrt{1-\yb_2^2}\vec{e}_{T2},-\yb_2\right),
\label{bk2def}
\eeq
thus introducing the FKS variables $\xib_2$ and $\yb_2$. By equating
$\bk_2$ of eq.~(\ref{bk2def}) with what we obtain from $B_1(k_2)$ we
arrive at:
\beqn
\yb_2&=&-\frac{1-z_1-y_2(1+z_1)}{1+z_1-y_2(1-z_1)}\,,
\label{yb2def}
\\
\xib_2&=&\frac{\xi_2}{2z_1}\left(1+z_1-y_2(1-z_1)\right)\,.
\label{xib2def}
\eeqn
These variables have the expected properties
\beqn
\xi_2(1-y_2)&=&z_1\xib_2(1-\yb_2)\,,
\\
\xi_2 d\xi_2 dy_2 &=& z_1\xib_2 d\xib_2 d\yb_2\,,
\eeqn
stemming from the invariance of $p_2\mydot k_2$ and $d\phi_1(k_2)$,
respectively. One can also verify that the volume of the one-body
phase space is correctly computed by the two sets of variables, 
namely that (see appendix~\ref{sec:impl}):
\beqn
&&\int_{-1}^1 dy_2\int d\xi_2\,\xi_2\,
\stepf\!\left(\xi_2\le\frac{2(z_1-\hminf^2)}{2-(1+y_2)(1-z_1)}\right)
\nonumber\\*&&=
z_1\int_{-1}^1 d\yb_2\int d\xib_2\,\xib_2\,
\stepf\!\left(\xib_2\le 1-\hminf^2/z_1\right)=
z_1\left(1-\frac{\hminf^2}{z_1}\right)^2\,.
\label{F1pvol}
\eeqn
The upper limit of the $\xi_2$ integration on the l.h.s.~of eq.~(\ref{F1pvol})
is that of eq.~(\ref{xibound}) in the collinear configuration $y_1=1$; that
on the $\xib_2$ integration is the same one, after $y_2$ and $\xi_2$ are
given in terms of $\yb_2$ and $\xib_2$ by inverting eqs.~(\ref{yb2def})
and~(\ref{xib2def}). This gives another check of self-consistency, since
the upper limit on $\xib_2$ is manifestly the correct one for that FKS
variable.

The other frame of interest (where a photon PDF is convoluted on the 
leg incoming from the right) is:
\beq
F_2^\prime:\;\;\;\;{\rm rest~frame~of}\;p_1+z_2p_2
\;\;\;\;\longleftrightarrow\;\;\;\;
\delta(1-y_2)\,.
\eeq
The boost from $F$ to $F_2^\prime$ is obtained as before in terms
of the boost rapidity:
\beq
\exp(\eta_B)=1\big/\sqrt{z_2}\,.
\eeq
The analogue of eq.~(\ref{bk2def}) reads:
\beq
\bk_1=\frac{\sqrt{z_2s}}{2}\xib_1\left(1,\sqrt{1-\yb_1^2}\vec{e}_{T1},
\yb_1\right).
\label{bk1def}
\eeq

\section{A distribution identity\label{sec:distr}}
Let $y$ be a variable with domain $-1\le y\le 1$, and $\yb$ a variable
with domain $-1\le\yb\le 1$, which can be related to each other by
means of a function $g$:
\beq
y=g(\yb)\,.
\label{yvsgyb}
\eeq
We assume $g(\yb)$ to have the following properties: monotonicity
in the whole domain (so that eq.~(\ref{yvsgyb}) constitute a legit
change of integration variable $y\leftrightarrow\yb$), and:
\beqn
&&g(\pm 1)=\pm 1\,,\;\;\;\;\;\;\;\;
g(\yb)=1-(1-\yb)h(\yb)\,,
\label{gdef}
\\*&&
h(\yb)=h_0+h_1(1-\yb)+\ord\left((1-\yb)^2\right)\,,\;\;\;\;\;\;\;\;
h_0\ne 0
\label{hprop}
\eeqn
whence
\beq
h_0=\frac{dg}{d\yb}(1)>0\,.
\label{h0prop}
\eeq
It is easy to see that the variables $y_{1,2}$ and $\yb_{1,2}$ introduced
in appendix~\ref{sec:frames} fulfill these conditions.

We use the identity
\beq
(1-y)^{-1-\ep}=-\frac{2^{-\ep}}{\ep}\,\delta(1-y)
+\left(\frac{1}{1-y}\right)_+ + \ord(\ep).
\label{plusreg}
\eeq
In what follows, we shall systematically ignore terms of $\ord(\ep)$ or
higher, since the final result will be obtained in the $\ep\to 0$ limit;
in keeping with this, the $\ord(\ep)$ notation will be omitted.

The integral of the $y$ plus distribution (where we omit to denote
a test function $f(y)$, which will be understood in the following) is:
\beq
I=\int dy \left(\frac{1}{1-y}\right)_+=
\int d\yb\frac{dg}{d\yb}\left[
(1-\yb)^{-1-\ep}\left(h(\yb)\right)^{-1-\ep}
+\frac{2^{-\ep}}{\ep}\,\delta(1-g(\yb))\right],
\eeq
where in the r.h.s.~we have changed integration variable $y\to\yb$, and
expressed all of the dependences on $y$ by means of $g(\yb)$. By using
again eq.~(\ref{plusreg}) we obtain:
\beqn
I&=&\int d\yb\frac{dg}{d\yb}\Bigg\{
\left[-\frac{2^{-\ep}}{\ep}\,\delta(1-\yb)
+\left(\frac{1}{1-\yb}\right)_+\right]\left(h(\yb)\right)^{-1-\ep}
\nonumber\\*&&\phantom{aaaaaaa}
+\frac{2^{-\ep}}{\ep}\,\abs{\frac{dg}{d\yb}}^{-1}\delta(1-\yb)
\Bigg\},
\eeqn
whence
\beqn
I&=&\int d\yb\frac{dg}{d\yb}
\left(\frac{1-g(\yb)}{1-\yb}\right)^{-1}
\!\left(\frac{1}{1-\yb}\right)_+
\nonumber\\*&+&
\frac{2^{-\ep}}{\ep}\int d\yb\, h_0
\Big[-h_0^{-1-\ep}+h_0^{-1}\Big]\delta(1-\yb)\,.
\eeqn
Here, we have set $\ep=0$ in the term proportional to the plus distribution
(since the result coincides with its finite $\ep\to 0$ limit). By expanding 
in $\ep$ the coefficients of the terms proportional to the Dirac $\delta$ 
we see that the poles cancel, and we are left with a finite result. 
Thus we arrive at the sought result:
\beq
dy\left(\frac{1}{1-y}\right)_+ =
d\yb\Bigg\{\frac{dg}{d\yb}\left(\frac{1-g(\yb)}{1-\yb}\right)^{-1}
\!\left(\frac{1}{1-\yb}\right)_+ +
\log h_0\,\delta(1-\yb)\Bigg\}.
\label{plvsplb}
\eeq
Equation~(\ref{plvsplb}) can be further manipulated by employing
the identity of eq.~(\ref{ypvsyd}) on the plus distributions on
the two sides, introducing two arbitrary parameters $\deltaI$ and
$\bdeltaI$ to control the endpoint subtraction in the $y$ and $\yb$
variable, respectively. In this way:
\beq
dy\left(\frac{1}{1-y}\right)_{\deltaI} =
d\yb\Bigg\{\frac{dg}{d\yb}\left(\frac{1-g(\yb)}{1-\yb}\right)^{-1}
\!\left(\frac{1}{1-\yb}\right)_{\bdeltaI} +
\log\frac{h_0\bdeltaI}{\deltaI}\,\delta(1-\yb)\Bigg\}.
\label{plvsplb2}
\eeq
We can apply eq.~(\ref{plvsplb2}) to the $(y_2,\yb_2)$ pair of variables
we have used in frame $F_1^\prime$, and to the pair $(y_1,\yb_1)$ relevant
to frame $F_2^\prime$. A simple algebra leads to:
\beqn
dy_2\left(\frac{1}{1-y_2}\right)_{\deltaI} &\!\!=\!\!&
d\yb_2\Bigg\{\left(\frac{d\xi_2}{d\xib_2}\right)^{-1}
\!\left(\frac{1}{1-\yb_2}\right)_{\bdeltaI} +
\log\frac{z_1\bdeltaI}{\deltaI}\,\delta(1-\yb_2)\Bigg\},
\label{plvsplb2F1}
\\
dy_1\left(\frac{1}{1-y_1}\right)_{\deltaI} &\!\!=\!\!&
d\yb_1\Bigg\{\left(\frac{d\xi_1}{d\xib_1}\right)^{-1}
\!\left(\frac{1}{1-\yb_1}\right)_{\bdeltaI} +
\log\frac{z_1\bdeltaI}{\deltaI}\,\delta(1-\yb_1)\Bigg\}.\phantom{aaaa}
\label{plvsplb2F2}
\eeqn

\section{Factorisation scheme at the LL\label{sec:factLL}}
We begin this discussion by reminding the reader that while at 
this order the scheme is not a factorisation scheme, it has been shown
in appendix~A of ref.~\cite{Bertone:2022ktl} how $K_{ij}$ functions may be 
introduced here as well, as compensating factors which correct for the 
fact that the first-order coefficients in the expansion of the PDFs do not 
coincide with the correct results computed in ref.~\cite{Frixione:2019lga}.
However, it is crucial to bear in mind that such compensating factors
are {\em solely} associated with $\ord(\aem^{b+1})$ cross sections, and
not with PDFs. In other words, the relevant $K_{ij}$ functions must only
be included in the $(n+1)$-body degenerate contributions to $d\hsig^{[1]}$
(and, by extension, to their analogues at higher orders), and not in the 
expansions of the PDFs, where they simply do not appear. In the present
context, by construction ${\cal Q}_{\gamma\lep}^\prime$, $d\delta^{[1]}$, 
$d\delta_D^{[2]}$, and $d\delta_S^{[2]}$ are matrix-level quantities
and, therefore, must feature the LL $K_{ij}$ functions introduced above.
For example, this is the only way in which eq.~(\ref{GmQ}) can be
fulfilled.

In order to be more definite, we report some explicit results relevant to
the LL PDFs. We limit ourselves to considering the so-called collinear
and running schemes, introduced in ref.~\cite{Bertone:2019hks}, since
in the standard ones (namely, beta, eta, and mixed) the photon is equal
to zero. For the quantities we are interested in, the collinear and
running scheme results coincide. We have:
\beqn
\PDF{\lep}{\lep}(z)&=&\delta(1-z)+\aemotpi\left(\frac{1+z^2}{1-z}\right)_+
\log\frac{\mu^2}{\muz^2}+\ord(\aem^2)\,,
\label{NSGamLL}
\\
\PDF{\gamma}{\lep}(z)&=&\aemotpi\,\frac{1+(1-z)^2}{z}\,
\log\frac{\mu^2}{\muz^2}+\ord(\aem^2)\,.
\label{gaGamLL}
\eeqn
The $\ord(\aem)$ coefficients above can be compared to their NLL-PDF
counterparts in \EWeq{4.121} and \EWeq{4.189} (eq.~(\ref{Ggesol2}) here).
Note that at $\ord(\aem)$ the LL PDFs depend on the starting scale $\muz$;
such a dependence drops out at the NLL (in practice, however, there is no good
reason to choose $\muz\ne m$). The compensating $K_{\lep\lep}$ function 
stemming 
from eq.~(\ref{NSGamLL}) are given in appendix~A  of 
ref.~\cite{Bertone:2022ktl}, and will not be repeated here. 
The one stemming from eq.~(\ref{gaGamLL}) is:
\beq
K_{\gamma\lep}=\frac{1+(1-z)^2}{z}
\left(2\log z+1-\log\frac{\muz^2}{m^2}\right).
\label{LLKgmu}
\eeq
By means of explicit calculations we can verify that:
\beq
{\rm eq}.\,(\protect\ref{Ggesol2})
\big[K_{\gamma\lep}(z)={\rm eq}.\,(\protect\ref{LLKgmu})\big]=
\frac{2\pi}{\aem}\,\times\,{\rm eq}.\,(\protect\ref{gaGamLL})\,,
\eeq
and:
\beq
{\cal Q}_{\gamma\lep}^\prime(z)
\big[K_{\gamma\lep}(z)={\rm eq}.\,(\protect\ref{LLKgmu})\big]+
\frac{2\pi}{\aem}\,\times\,{\rm eq}.\,(\protect\ref{gaGamLL})=
{\cal Q}_{\gamma\lep}(z)\,.
\eeq

\phantomsection
\addcontentsline{toc}{section}{References}
\bibliographystyle{JHEP}
\bibliography{lepvbf}

\end{document}